\documentclass[sn-aps,iicol]{sn-jnl}

\jyear{2022}%

\usepackage{combelow}

\usepackage[utf8]{inputenc}
\usepackage{bm}
\usepackage{bbold}
\usepackage{mathtools}

\usepackage{graphicx}
\usepackage{subfigure}
\usepackage{scalerel,amssymb}
\DeclareMathAlphabet{\pazocal}{OMS}{zplm}{m}{n}

\usepackage{braket}
\usepackage{tensor}
\usepackage{slashed}
\usepackage{xcolor}
\usepackage{ulem}
\usepackage{xr}
\usepackage{textgreek}

\definecolor{purple}{rgb}{0.8,0,0.6}

\def\AVW{{\mathfrak{a}}}

\def\bp{{\mathbf{p}}}
\def\bk{{\mathbf{k}}}

\usepackage{fancyhdr}
\usepackage{lastpage}

\newcommand{\beqn}{\begin{eqnarray}}
\newcommand{\eeqn}{\end{eqnarray}}
\newcommand{\beqs}{\begin{subequations}}
\newcommand{\eeqs}{\end{subequations}\\[-2mm]\noindent}

\newcommand{\bs}{\boldsymbol}

\begin{document}

\title{Vortical waves in a quantum fluid with vector, axial and helical charges. II. Dissipative effects}

\author[1]{\fnm{Sergio} \sur{Morales-Tejera}}\email{sergio.morales@e-uvt.ro}

\author*[1]{\fnm{Victor E.} \sur{Ambru\cb{s}}}\email{victor.ambrus@e-uvt.ro}

\author*[1,2]{\fnm{Maxim N.} \sur{Chernodub}}\email{maxim.chernodub@univ-tours.fr}

\affil[1]{
\orgdiv{Department of Physics}, 
\orgname{West University of Timi\cb{s}oara},
\orgaddress{\street{Bd.~Vasile P\^arvan 4}, \city{Timi\cb{s}oara}, \postcode{300223}, \country{Romania}}}

\affil[2]{
\orgdiv{Institut Denis Poisson, CNRS UMR 7013},
\orgname{Universit\'e de Tours}, 
\orgaddress{\city{Tours}, \postcode{37200}, \country{France}}}

\abstract{
In this paper, we consider the effect of interactions on the local, average polarization of a quantum plasma of massless fermion particles characterized by vector, axial, and helical quantum numbers. Due to the helical and axial vortical effects, perturbations in the vector charge in a rotating plasma can lead to chiral and helical charge transfer along the direction of the vorticity vector. At the same time, interactions between the plasma constituents lead to the dissipation of the helical charge through helicity-violating pair annihilation (HVPA) processes and of the axial charge through the axial anomaly. We will discuss separately a QED-like plasma, in which we ignore background electromagnetic fields and thus the axial charge is approximately conserved, as well as a QCD-like plasma, where instanton effects lead to the violation of the axial charge conservation, even in the absence of background chromomagnetic fields. The non-conservation of helicity and chirality leads to a gapping of the Helical, Axial, and mixed Axial-Helical vortical waves that prevents their infrared modes from propagating. On the other hand, usual dissipative effects, such as charge diffusion, lead to significant damping of ultraviolet modes. We end this paper with a discussion of the regimes where these vortical waves may propagate.}


\maketitle

\section{Introduction}\label{sec:intro}

Anomalous breaking of continuous internal symmetries in chiral fluids leads to the emergence of a distinct class of gapless hydrodynamic modes that appear in nontrivial backgrounds such as classical electromagnetic fields or curved spacetimes~\cite{Kharzeev:2013jha}. A celebrated example of such modes is produced by the coherent interplay of two anomalous transport effects: the Chiral Magnetic Effect~\cite{Fukushima:2008xe, Vilenkin:1980fu, Alekseev:1998ds} and the Chiral Separation Effect~\cite{Son:2004tq, Metlitski:2005pr}. Both these transport phenomena, which appear as a result of the activation of the axial anomaly in the magnetic field background, generate a vector (chiral/axial) current in the regions with a nonvanishing chiral (vector) density. These effects yield a closed system of propagating vector-axial oscillations in currents and densities, known as the Chiral Magnetic Wave. This anomalous hydrodynamic excitation might manifest itself in the quark-gluon plasma and could potentially exhibit distinguishing characteristics that might be observed experimentally~\cite{ALICE:2023weh}. It is worth stressing that the word ``gapless'', originating from the condensed matter literature, implies that the excitation has no mass gap in its energy spectrum, making it somewhat similar to the acoustic sound waves.

Vortical backgrounds -- represented, for example, by whirlpools in fluids or plasmas -- host a similar class of hydrodynamic phenomenon known under the common name of Chiral Vortical Effects~\cite{Vilenkin:1978is, Vilenkin:1979ui}. These vortical transport effects generate intertwined oscillations of vector and axial currents, as well as their charges, propagating along the rotation axis~\cite{Jiang:2015cva, Gorbar:2017toh}. The resulting Chiral Vortical Waves, which are supported by the presence of the chiral anomaly and its mixed chiral-gravitational counterpart~\cite{Landsteiner:2011cp}, were suggested to exist in the rotating quark-gluon plasma created in highly-vortical noncentral heavy-ion collisions~\cite{Jiang:2015cva}.  

The picture described above traditionally represents a chiral fluid as a two-component system that includes only vector and axial degrees of freedom that propagate coherently. However, at this point, it is worth noticing that fermions possess a third type of quantum number associated with the fermion helicity, which is distinct from the vector and axial charges~\cite{Pal:2010ih}. Helicity and chirality are distinguishable physical properties of fermions that are often inaccurately identified with each other~\cite{Ambrus:2019ayb, Ambrus:2019khr}. The presence of this additional third characteristic of the fermion ensemble leads to the associated helical vortical transport phenomena that affect, in particular, the spectrum of the hydrodynamic modes~\cite{Ambrus:2019khr}.\footnote{Here, for completeness, we mention that similar, helicity-related nontrivial transport and associated hydrodynamic modes in the magnetic field background has been uncovered in Refs.~\cite{Ambrus:2023erf}. In this paper, we concentrate only on rotating systems.} 
Before proceeding further, it is worth mentioning that rotating media may also produce yet another, exotic ``zilch'' currents that are different from vector, chiral and helical charges. While the conserved zilch currents appear naturally in photonic gases~\cite{Chernodub:2018era, Prokhorov:2020npf}, they can also emerge in system of chiral fermions~\cite{Alexandrov:2020zsj}. Below, we concentrate on the vector-axial-helical triad of quantum numbers, leaving the zilch charge aside.

In our companion paper~\cite{Morales-Tejera:2024uzg}, the role of helical degrees of freedom in vortical chiral fluids is explored in an academic dissipationless limit when both axial and helical quantum numbers are conserved. This study has confirmed the presence of the Helical Vortical Wave~\cite{Ambrus:2019khr}, which corresponds to a gapless hydrodynamic excitation acting mainly in the helical and vector sectors of charge densities. Surprisingly, it also has uncovered yet another gapless wave in the system: the Axial Vortical Wave, which is driven by an axial imbalance accompanied, in the presence of vector (baryon) or helical (spin-polarized) background,  by coherent fluctuations in axial and helical/vector charges and their currents. At high vector (baryon) densities, where the fluid becomes degenerate,  both the Helical and the Axial Vortical Waves merge into a hydrodynamic excitation, the Axial-Helical Vortical Wave. The fascinating feature of the Axial Vortical Wave is its spatial non-reciprocity, which makes this hydrodynamic mode distinguishable from the usual acoustic waves: a pure Axial Vortical Wave propagates only opposite (along) the direction of the angular velocity for positive (negative) axial imbalance while being unable to travel in the opposite direction. 

This paper aims to study the fate of the hydrodynamic vortical excitations in realistic plasmas, which are characterized by axial and helical charge relaxation as well as kinetic damping effects. A similar analysis has been extensively applied to determine possible experimentally observable signatures~\cite{Ke:2012qb,STAR:2015wza} of a similar excitation, the Chiral Magnetic Wave, that emerges in the magnetic field background and incorporates oscillations of the vector and axial charges~\cite{Newman:2005hd, Kharzeev:2010gd}. 

In addition to the standard axial charge relaxation effects -- related to the fact that the axial charge is not a conserved quantity in quantum field theory -- the vector sector is also subjected to relaxation effects caused by the electromagnetic backreaction~\cite{Kharzeev:2010gd}. The latter phenomenon is supported by a finite electrical conductivity of the plasma that leads to an efficient screening of the local vector charge fluctuations carrying nonvanishing electric charge density. As a result, the Chiral Magnetic Wave is essentially ``overdamped'' in realistic environments such as quark-gluon plasma: the relaxation length of the wave appears to be shorter than the half-period of the wave oscillation~\cite{Shovkovy:2018tks}.\footnote{A recent work~\cite{Ahn:2024ozz} highlights the regimes, in which the Chiral Magnetic Wave can still be observed in a condensed matter setup as an underdamped collective mode in a Weyl semimetal which hosts relativistic fermionic quasiparticles similarly to quark-gluon plasma.} In the vector-chiral sector of rotating plasma, a related analysis has been performed in Ref.~\cite{Rybalka:2018uzh}, where various waves were analyzed, including the longitudinal sound wave, a circularly polarized vortical wave and diffusive modes that involve an oscillating vector charge density. 

Here, we include in the analysis the helical charge, which enriches the spectrum of the hydrodynamic excitations. In particular, the Helical Vortical Wave~\cite{Ambrus:2019khr} involves the oscillations of the helical and vector charge densities, and, therefore, our analysis gives a higher, conservative bound of the wave propagation length caused by the helical charge relaxation and kinetic dissipation effects. The electromagnetic backreaction of the vector charge density, which is not considered in this paper, will restrict this length even further. On the other hand, the Axial and Axial-Helical Vortical waves, found in the companion paper~\cite{Morales-Tejera:2024uzg}, are driven essentially by electrically neutral helical and axial charge densities. Therefore, they are not subjected to the backreaction coming from the electromagnetic sector associated with the vector charge fluctuations.

One  may argue that the helical degree of freedom is as good (bad) as the chiral charge of massive fermions: neither of them is conserved in the physically relevant theories, such as massive QED or QCD. Chirality is not conserved for massless fermions either, due to instanton-like effects that are present in the QCD medium, even in the absence of background chromomagnetic fields. In the relaxation time approximation, the pertinence of these quantum numbers for the dynamics of the system is determined by the relative order of magnitudes of the axial, $\tau_A$, and helical, $\tau_H$, relaxation times, which show how fast these charges dissolve in the system.  Similarly to the axial (chiral) charge~\cite{Ruggieri:2016asg, Astrakhantsev:2019zkr}, the relaxation of the helical charge might also be a rather fast process~\cite{Ambrus:2020oiw}. In addition to the charge relaxation times, the system is also characterized by the kinetic relaxation time $\tau_R$, which encodes how fast fluctuations in the thermodynamic characteristics of the system (such as pressure and energy density) relax towards the thermodynamic equilibrium. In general, all three relaxation times, $\tau_R$, $\tau_A$, and $\tau_H$, are independent of each other. An expression for $\tau_H$ was derived in Ref.~\cite{Ambrus:2019khr} for a neutral plasma. In this paper, we will extend this derivation for the case of a finite vector imbalance, encoding a difference between the number of particles and anti-particles in the system, which is necessary in order to assess its relevance in the limit of a degenerate (dense) system. A similar calculation for the axial relaxation time is beyond the scope of this work; as such, an analysis must be performed in fully non-perturbative QCD using, e.g., lattice methods.

Below, we concentrate on the theoretical questions related to the very existence of vortical modes, their spectrum, and their lifetimes. The discussion of their experimental signatures will be presented elsewhere.

The structure of the paper is as follows. In Sec.~\ref{sec:setup}, we briefly present the setup for the study of linear perturbations in the quantum plasma with vector/axial/helical degrees of freedom, following the approach in Ref.~\cite{Morales-Tejera:2024uzg}, which we extend to the case when the helical and/or axial charges are not conserved. Our focus is on the case of an unpolarized background plasma, in which both the helical and axial chemical potentials vanish. 

In Sec.~\ref{sec:QED}, we take into account the non-conservation of the helical charge stemming from local processes using a relaxation time approximation. Such a scenario is relevant for a massless QED plasma, where both vector and axial charges are conserved due to an unbroken U(1) gauge symmetry and the vector nature of inter-particle interactions, respectively. The Chiral Vortical Wave, as considered in \cite{Gorbar:2017toh}, is then recovered in the limit where the helical charge dissipates instantaneously. An interplay between Chiral and Helical Vortical Waves, as well as a distinct role played by the Axial Vortical Wave, are discussed. 

In Sec.~\ref{sec:QCD}, we study the spectra of collective excitations when both helical and axial charges are non-conserved. We show that the relaxation of helical and axial charge densities affects the spectrum of the wave in a different manner. 

Section~\ref{sec:diss} discusses in detail the effect of the kinetic dissipation, which appears at the level of corrections to the energy-momentum tensor and the charge currents. In a realistic fluid, this type of dissipation has different roots than the charge relaxation (for example, of axial or helical charge densities), as it occurs due to inter-particle interactions, which drive the system towards thermodynamic equilibrium. As anomalous transport also leads to charge and heat flow in the fluid rest frame, the kinetic dissipation has the effect of shifting the phase velocities by an imaginary quantity, leading to a gradual dissolution of the waves. The corresponding kinematic relaxation times are estimated. 

Finally, we briefly summarize in Sec.~\ref{sec:conc} our findings for the non-trivial modifications of the wave spectrum due to the non-vanishing lifetimes of the helical and axial charges.

In Appendix~\ref{apa}, we show that an alternative prescription to incorporate the relaxation of helical charge gives rise to an unphysical instability. Appendix~\ref{app:tauH} is devoted to the generalization of the expression given in Ref.~\cite{Ambrus:2019khr} for the helicity relaxation time $\tau_H$ to the presence of a finite vector imbalance.

\section{Hydrodynamic description of waves in V/A/H fluids}\label{sec:setup}

In this section, we briefly review the anomalous transport in the presence of rotation. In Subsec.~\ref{sec:setup:Landau}, we review the constitutive equations for the energy-momentum tensor and the charge currents for a Dirac fluid with vector, axial, and helical imbalance undergoing rigid rotation, as seen in the Landau (or energy) frame. Incorporating charge non-conservation is discussed in Subsec.~\ref{sec:setup:noncons}. We then summarize the treatment of small perturbations around this rigidly rotating state in Subsec.~\ref{sec:setup:linear}. 

\subsection{Landau frame decomposition for slow rotation} \label{sec:setup:Landau}

Let us consider a Dirac fluid characterized by the vector (V), axial (A) and helicity (H) quantum numbers undergoing rigid rotation with angular velocity $\Omega$ about the $z$ axis. In a region close to the rotation axis, where $\rho \Omega \ll 1$, and considering $\beta \Omega \ll 1$, the energy-momentum tensor and charge currents of this fluid can be approximated as 
\begin{align}
 &T^{\mu\nu} \simeq E_\beta u^\mu u^\nu - P_\beta \Delta^{\mu\nu} + u^\mu W_\beta^\nu + u^\nu W_\beta^\mu, \nonumber \\&
 J^\mu_\ell \simeq Q_{\ell;\beta} u^\mu + V_\beta^\mu,
\end{align}
where $W_\beta^\mu = \sigma^\omega_{\varepsilon;\beta} \omega^\mu$ and $V^\mu = \sigma^\omega_{\ell;\beta} \omega^\mu$ represent the heat flux and diffusion currents in the fluid rest frame, $\omega^\mu = \frac{1}{2} \varepsilon^{\mu\alpha\beta\gamma} u_\alpha \partial_\beta u_\gamma \simeq \Omega \delta^\mu{}_z$ is the vorticity four-vector and we neglected terms of quadratic or higher order in $\Omega$. In the above expression, $u^\mu \partial_\mu =\Gamma (\partial_t + \Omega \partial_\varphi)$ is the four-velocity of a rigidly-rotating fluid, which defines the so-called $\beta$ (thermometer) hydrodynamic frame. The energy density $E_\beta$ is related to the pressure $P_\beta$ through the ultrarelativistic equation of state, $E_\beta = 3P_\beta$. The charge densities $Q_{\ell;\beta}$ and the heat $\sigma^\omega_{\varepsilon;\beta}$ and charge $\sigma^\omega_{\ell;\beta}$ vortical conductivities, measured in the $\beta$ frame, can be obtained from the thermodynamic pressure $P_\beta$ via 
\begin{equation}
 Q_{\ell;\beta} = \frac{\partial P_\beta}{\partial \mu_\ell}, \ \ \
 \sigma^\omega_{\varepsilon;\beta} = Q_{A;\beta}, \ \ \
 \sigma^\omega_{\ell;\beta} = \frac{1}{2} \frac{\partial Q_{\ell;\beta}}{\partial \mu_A}.
\end{equation}

To linear order in $\Omega$, one can employ the Landau hydrodynamic frame, by which the four-velocity $u_L^\mu$ satisfies $T^\mu{}_\nu u_L^\nu = E u_L^\mu$, with $E$ being the energy density and $T^{\mu\nu}$ the energy-momentum tensor. The Landau and $\beta$-frame velocities are related, to leading order in $\Omega$, through $u^\mu_L = u^\mu + \sigma^\omega_\varepsilon \omega^\mu / (E + P)$. Applying a Lorentz boost along the $z$ axis brings the Landau frame velocity to the form corresponding to rigid rotation, such that the energy-momentum tensor and the charge currents read, to leading order in $\Omega$, as follows:
\begin{equation}
 T^{\mu\nu} = E u^\mu u^\nu - P \Delta^{\mu\nu}, \qquad 
 J^\mu_\ell = Q_\ell u^\mu + \sigma^\omega_\ell \omega^\mu.
 \label{eq:Landau_currents}
\end{equation}
The new Landau-frame vortical conductivities appearing above are related to the ones in the $\beta$ frame via
\begin{equation}
 \sigma^\omega_{\ell} = \sigma^{\omega}_{\ell;\beta} - \frac{Q_\ell Q_A}{E + P},
\end{equation}
where the last term represents the contribution from the $\beta$-frame heat conductivity, $\sigma^\omega_\varepsilon = Q_A$.

\subsection{Non-conserved charges}\label{sec:setup:noncons}

Our work is concentrated on hydrodynamic systems that possess non-conserved charges associated with axial and helical degrees of freedom. The non-conservation of axial charge is caused by the axial anomaly in both Abelian (like QED) and non-Abelian (QCD) theories. In the massless quark limit, the anomaly leads to topologically induced dissipation in QCD due to either inter-vacuum tunneling or thermal, sphaleron-induced processes related to the breaking of the axial symmetry by the axial anomaly~\cite{itzykson80,bertlmann96,Kharzeev:2013jha,buzzegoli20phd}. The conservation of axial charge is also spoiled by an eventual non-zero fermionic mass, which, however, provides a relatively small effect for light quarks in heavy-ion collisions. In addition, the conservation of helical charge is also violated, both in QED and QCD, through the so-called helicity-violating pair annihilation (HVPA) processes (see Sec.~5.2 in Ref.~\cite{Ambrus:2020oiw}).

The non-conservation of a charge $Q$ and the corresponding charge current $\bs J$ is often implemented in the relaxation time approximation:
\begin{equation}
\label{eq:dis1}
    \partial_{\mu}J^{\mu} \equiv \frac{\partial J^0}{\partial t} + {\bs \nabla} \cdot {\bs J} = - \frac{Q}{\tau}\,,
\end{equation}
where the relaxation rate is given by the appropriate relaxation time scale $\tau$. In order to preserve Lorentz symmetry, the charge on the right hand side of Eq.~\eqref{eq:dis1} must be obtained via contraction with the fluid four-velocity, $Q = u_\mu J^\mu$, thus indicating that the charge relaxation occurs only in the presence of matter and/or in thermal background.

It is understood that Eq.~\eqref{eq:dis1} holds in the vicinity of charge neutrality, where $Q \rightarrow \delta Q$ represents a small charge fluctuation around an otherwise uncharged background state. Large (persistent) departures from $Q = 0$ may be possible only in a far-from-equilibrium system, where the hydrodynamic description considered in this paper may not necessarily hold. Indeed, we are relying on a grand-canonical ensemble description of the system, where a chemical potential $\mu$ describes charge imbalance, and such a chemical potential only makes thermodynamic sense if the charge $Q$ is (approximately) conserved. Introducing the charge susceptibility, $\chi = \partial Q / \partial \mu$, Eq.~\eqref{eq:dis1} can be rewritten as
\begin{equation}
\label{eq:dis0}
    \partial_{\nu} J^\nu = - \frac{\chi}{\tau} \mu\,.
\end{equation}

The above prescription applies straightforwardly to the case when the system supports a single charge, which is approximately conserved.
In the presence of more than one charge, there are, however, two different ways how to generalize the charge non-conservation equation~\eqref{eq:dis1} in the relaxation time approximation, depending on whether we describe the chemical force that drives the charge relaxation as either the charge itself or the corresponding chemical potential. The difference between these approaches arises from the fact that in a system with multiple charges, the condition that a given charge $Q_\ell$ vanishes does not imply that its associated chemical potential disappears, and vice-versa. Therefore, it is important to discriminate between these two pictures of the charge relaxation as they lead to two physically different outcomes.

In the first approach, one assumes that the system relaxes towards a state with vanishing $Q_\ell$ charge.
The susceptibility becomes a matrix,
\begin{align}
 \delta Q 
 \rightarrow \delta Q_\ell =
 \chi_{\ell\ell'} \delta \mu_{\ell'} + \chi_{\ell T} \delta T\;,
    \label{eq_chi_ell}
\end{align}
where $\chi_{\ell\ell'} = \partial Q_\ell / \partial \mu_{\ell'}$ and $\chi_{\ell T} = \partial Q_\ell / \partial T$.
The variation of the charge density in the right-hand side of Eq.~\eqref{eq:dis1} should be taken with respect to the variation of all chemical potentials that exist in the system, as well as with respect to the temperature. 
In our case of the $V$, $A$, $H$ triad, the non-conservation~\eqref{eq:dis1} for the charge current $J^{\mu}_{\ell}$ thus gets generalized to the following form:
\begin{align}\label{eq:dis3}
    &\partial_{\mu} J^{\mu}_{\ell} = - \frac{1}{\tau_\ell} \left(\sum_{\ell' = V, A, H} \chi_{\ell\ell'}\delta \mu_{\ell'} + \chi_{\ell T} \delta T\right)\,, \nonumber \\ & \text{(no sum over $\ell = V, A, H$)}\,.
\end{align}
The conservation of the vector charge is encoded in the infinite vector relaxation time $\tau_V = \infty$. The other two relaxation times, $\tau_A$ and $\tau_H$ are, generally, finite quantities. Notice that despite the vector charge being a conserved quantity, in this prescription, its fluctuations affect the conservation of the other two charges so that the sum over $\ell'$ in the right-hand-side of Eq.~\eqref{eq:dis3} runs over all three charges in the $V$, $A$, $H$ triad. While the proposal in Eq.~\eqref{eq:dis3} is in principle viable, detailed analysis within our $V$, $A$, $H$ system indicates that it leads to instabilities appearing at the level of the linear modes governing hydrodynamic fluctuations (see Appendix~\ref{apa} for a more detailed discussion).

Our preferred approach is to work on the basis of the chemical potential, which implies that the dissipation of the charge is controlled not by the magnitude of the charge density on the right-hand side of Eq.~\eqref{eq:dis1} but rather by the magnitude of the corresponding chemical potential. One argument supporting this approach is based on the Chiral Magnetic Effect, giving rise to the Chiral Magnetic Wave. Indeed, the non-conservation of the axial charge appears due to the presence of the anomalous triangular diagram with the axial-vector-vector (AVV) vertex corners. The anomalous generation of the vector current (one V-corner) along the magnetic field (another V-corner) is given by the same diagram, whose remaining A-corner directly links the magnitude of the generated current with the axial chemical potential and not the axial charge density. In this sense, the charge densities serve as the auxiliary thermodynamic characteristics of the system. At the same time, the chemical potentials play a primary role in the grand-canonical description of a system, while charge densities are derived quantities. In addition, the fact that a chemical potential for non-conserved charges is not a well-defined quantity, in a thermodynamic sense, suggests that a system with non-conserved charges should be equilibrated to a state where such chemical potential is absent. Taking these considerations into account, we propose that Eq.~\eqref{eq:dis1} should be generalized as follows:
\begin{equation}\label{eq:dis2}
    \partial_{\mu} J^{\mu}_{\ell} = - \frac{\chi_\ell}{\tau_\ell} \delta \mu_{\ell} \,, \qquad \text{(no sum over $\ell = V, A, H$)}\,,
\end{equation}
which amounts to taking only the diagonal element in Eq.~\eqref{eq:dis3}. Note that the relaxation time approximation used to describe the non-conservation of charges is semi-phenomenological and \textit{a priori} disconnected from the relaxation time approximation of Anderson and Witting \cite{Anderson:1974nyl} in kinetic theory used to describe dissipative effects in Sec. \ref{sec:diss:RTA}. One may employ, e.g., the Shakhov extension of the RTA \cite{Ambrus:2024qsa} to allow for independent relaxation times, however we do not pursue this avenue in the present work.

In summary, while the two approaches are identical in the case of a single-charge system [{\it c.f.} Eqs.~\eqref{eq:dis1} and \eqref{eq:dis0}], they differ for systems that contain more than one charge. Since the charge dissipation is taken into account within an effective charge-relaxation-time formalism, we are free to choose between any of these prescriptions to proceed further. To this end, we notice that the first approach gives rise to unnatural unstable modes, as we point out in Appendix~\ref{apa}. We will, therefore, focus our discussion only on the second approach, summarized in Eq.\eqref{eq:dis2} above.

\subsection{Linear perturbations}\label{sec:setup:linear}

Following Ref.~\cite{Morales-Tejera:2024uzg}, we take small perturbations on top of the rigidly-rotating state. For definiteness, we consider only longitudinal perturbations along the vorticity vector. The detailed derivation of the equations for the linear perturbations is presented in Ref.~\cite{Morales-Tejera:2024uzg}, and for the sake of brevity, we do not repeat it here. Instead, we list the results:
\begin{subequations}\label{eq:conservation}
\begin{align}
 D E + (E + P) \theta &= 0, \label{eq:conservation_E}\\
 (E + P) Du^\mu - \nabla^\mu P &= 0,  \label{eq:conservation_umu}\\
 D Q_V + Q_V \theta + \omega^\mu \partial_\mu \sigma^\omega_V + 
 \sigma^\omega_V \partial_\mu \omega^\mu &= 0,\label{eq:conservation_JV}\\
 D Q_A + Q_A \theta + \omega^\mu \partial_\mu \sigma^\omega_A + 
 \sigma^\omega_A \partial_\mu \omega^\mu &= -\frac{T^2 \mu_A}{3\tau_A},\label{eq:conservation_JA}\\
 D Q_H + Q_H \theta + \omega^\mu \partial_\mu \sigma^\omega_H + 
 \sigma^\omega_H \partial_\mu \omega^\mu &= -\frac{T^2 \mu_H}{3\tau_H},
 \label{eq:conservation_JH}
\end{align}
\end{subequations}
where $D = u^\mu \partial_\mu$ represents the comoving derivative, $\nabla^\mu = \Delta^{\mu\nu} \partial_\nu$ is the spatial gradient in the fluid rest frame, while $\theta = \partial_\mu u^\mu$ is the expansion scalar. As we already discussed in Sec.~\ref{sec:setup:noncons} at the level of the charge currents, all three currents $J^\mu_{V/A/H}$ would be conserved in a quantum field theory of free (non-interacting) massless fermions. This property is certainly true also in the quantum case for the vector current when $\partial_\mu J^\mu_V = 0$. Contrary to Ref.~\cite{Morales-Tejera:2024uzg}, in this paper, we take into account the non-conservation of the axial and helical charges in the so-called relaxation-time approximation.

We now consider the Fourier decomposition 
\begin{align}
 &\bar{f} = f + \delta \bar{f}, \nonumber \\&
 \delta \bar{f} = \int_{-\infty}^\infty dk\, e^{ik z} \sum_\omega e^{-i \omega(k) t} \delta f_\omega(k),
 \label{eq_barf_Fourier}
\end{align}
where $f$ represents a background (possibly vanishing) constant value, and $\omega(k)$ represent the angular frequencies determined by the system of Eqs.~\eqref{eq:conservation}.\footnote{It is important not to confuse the angular frequency $\omega = \omega(k)$, which determines the time evolution of fluctuations~\eqref{eq_barf_Fourier}, with the vorticity $\omega^\mu$ which enters, for example, the system of equations~\eqref{eq:conservation} being associated with the local angular velocity of the vortical fluid.} 
In this paper, we take the approach of Ref.~\cite{Morales-Tejera:2024uzg} and consider at leading order only longitudinal perturbations. We note that, due to the vortical background, the amplitudes of the transverse velocity perturbations become non-vanshing at first-order with respect to $\Omega$. 

In order for linear perturbations to be applicable, we consider an unpolarized background state in which $\mu_A = \mu_H = 0$. Then, the right-hand sides of Eqs.~\eqref{eq:conservation_JA} and \eqref{eq:conservation_JH} are also linear in perturbations. 
The presence of the axial and helical relaxation times do not affect the conservation of energy and momentum. Therefore, the equations $\partial_\mu T^{\mu\nu} = 0$ lead to the usual acoustic modes, characterized by the dispersion relation
\begin{equation}
 \omega^\pm_{\rm ac.} = \pm c_s k, \qquad c_s = 1/\sqrt{3},
 \label{eq:acoustic_vpm}
\end{equation}
with $c_s$ being the speed of sound in an ultrarelativistic ideal fluid. Eq.~\eqref{eq:acoustic_vpm} is valid to $O(\Omega^2)$, as pointed out in Ref.~\cite{Morales-Tejera:2024uzg}, and it is consistent with the results derived in Ref.~\cite{Abbasi:2016rds}. Our result also coincides with that derived in Ref.~\cite{Gorbar:2017toh}, as we are considering the case of a vanishing background axial chemical potential. Note however that our result remains the same even at finite axial chemical potential, while that of Ref.~\cite{Gorbar:2017toh} exhibits a correction proportional to $\Omega Q_A$, due to two factors: 1) at finite $Q_A$, our reference frame is boosted with a velocity $V = - Q_A \Omega / 4P$ with respect to the laboratory frame considered in Ref.~\cite{Gorbar:2017toh}; 2) The amplitudes $\delta u^x_\omega$ and $\delta u^y_\omega$ of the transverse components of the velocity pick up an $(x,y)$ dependence at first order with respect to $\Omega$, which was neglected in Ref.~\cite{Gorbar:2017toh}, and which affects the dispersion relation.

The perturbations in the energy-momentum sector are not influenced by perturbations in the charges, $\delta Q_\ell$. For the charge-related modes, $\delta P_\omega = \delta u_\omega = 0$ (see discussion in Ref.~\cite{Morales-Tejera:2024uzg}), such that the equations for the charge currents become:
\begin{equation}
 \bigl(\omega \delta Q_{\ell;\omega} - k \Omega \delta \sigma^\omega_{\ell;\omega}\bigr){\biggl\rvert}_{\delta P_\omega = 0} 
 = -\frac{i T^2 \delta \mu_{\ell;\omega}}{3\tau_\ell}.
\label{eq:chargecons}
\end{equation}
The above equation involves the variations $\delta Q_{\ell;\omega}$ and $\delta \sigma^\omega_{\ell;\omega}$ of the charge densities and vortical conductivities at constant pressure. Thinking in terms of the grand canonical ensemble parameters, the constraint $\delta P_\omega = 0$, together with the thermodynamic relation $\delta P_\omega = s \delta T_\omega + \sum_\ell Q_\ell \delta \mu_{\ell;\omega}$ lead to the constraint
\begin{equation}
 \delta T_\omega = -\sum_\ell \frac{Q_\ell}{s} \delta \mu_{\ell;\omega}.
\end{equation}
Subsequently, Eq.~\eqref{eq:chargecons} can be written in terms of fluctuations in the chemical potentials,
\begin{subequations}
 \label{eq:M}
\begin{align}
 &\mathbb{M}_{\ell\ell'} \delta \mu_{\ell';\omega} =0, 
 \nonumber \\&
 \frac{1}{T^2} \mathbb{M} = \omega \mathbb{M}_\omega - \kappa_\Omega \mathbb{M}_\Omega + \frac{i}{3\tau_A} \mathbb{I}_A + \frac{i}{3\tau_H} \mathbb{I}_H,
 \label{eq:M_aux2}
\end{align}
where we introduced the parameter
\begin{align}
 \kappa_\Omega = \frac{k \Omega}{T}\,,
 \label{eq_kappa_Omega}
\end{align}
which has a sense of a normalized momentum. We also introduced the following notations:
\begin{align}
 \mathbb{M}^\omega_{\ell\ell'} &= \frac{1}{T^2} \left(\frac{\partial Q_\ell}{\partial \mu_{\ell'}} - \frac{3 Q_\ell Q_{\ell'}}{sT} + \frac{Q_{\ell'} \vec{\mu}}{sT} \cdot \frac{\partial Q_\ell}{\partial \vec{\mu}}\right),\label{eq:Mv}\\
 \mathbb{M}^\Omega_{\ell\ell'} &= \frac{1}{T} \left(\frac{\partial \sigma^\omega_\ell}{\partial \mu_{\ell'}} - \frac{2 \sigma^\omega_\ell Q_{\ell'}}{sT} + \frac{Q_{\ell'} \vec{\mu}}{sT} \cdot \frac{\partial \sigma^\omega_\ell}{\partial \vec{\mu}}\right),
 \label{eq:MOmega}
\end{align}
as well as $\mathbb{I}^A_{\ell\ell'} = \delta_{\ell A} \delta_{\ell' A}$ and $\mathbb{I}^H_{\ell\ell'} = \delta_{\ell H} \delta_{\ell' H}$.
\end{subequations}

\subsection{Unpolarized background}\label{sec:setup:unpol}

The pressure corresponding to an ensemble of non-interacting, massless fermions and anti-fermions with polarizations $\lambda = \pm 1/2$ is given by \cite{Ambrus:2019khr,Morales-Tejera:2024uzg}:
\begin{equation}
 P = -\frac{T^4}{\pi^2} \sum_{\sigma,\lambda} {\rm Li}_4(-e^{\mu_{\sigma,\lambda} / T}),
\end{equation}
where ${\rm Li}_s(z) = \sum_{n = 1}^\infty z^n / n^s$ is the polylogarithm function. In the above, $\mu_{\sigma,\lambda}$ is the effective chemical potential arising from vector, axial and helical imbalance,
\begin{equation}
 \mu_{\sigma,\lambda} = {\vec q}_{\sigma,\lambda} \cdot {\vec \mu} = \sigma \mu_V + 2\lambda \mu_A + 2 \sigma \lambda \mu_H,
\end{equation}
with $\vec{q}_{\sigma,\lambda}$ collecting the vector, axial and helical charges of a fermion ($\sigma = 1$) or antifermion ($\sigma = -1$) with polarization $\lambda = \pm 1/2$, namely
\begin{equation}
 q^V_{\sigma,\lambda} = \sigma, \qquad 
 q^A_{\sigma,\lambda} = 2\lambda, \qquad 
 q^H_{\sigma,\lambda} = 2\lambda \sigma.
\end{equation}
The charge densities and vortical conductivities in the $\beta$ frame read
\begin{align}\label{eq:Charges_Conds}
 Q_\ell &= \frac{\partial P}{\partial \mu_\ell} = -\frac{T^3}{\pi^2} \sum_{\sigma,\lambda} q_{\sigma,\lambda}^\ell {\rm Li}_3(-e^{\mu_{\sigma,\lambda} / T}), \\
 \sigma^\omega_{\ell;\beta} &= \frac{1}{2} \frac{\partial Q_\ell}{\partial \mu_A} = -\frac{T^2}{2\pi^2} \sum_{\sigma,\lambda} 2\lambda q_{\sigma,\lambda}^\ell {\rm Li}_2(-e^{\mu_{\sigma,\lambda} / T}).
\end{align}

Let us now specialize the above expressions to the case of an unpolarized background state characterized by $\mu_A = \mu_H = 0$. Using the following properties of the polylogarithm function,
\begin{align}
 {\rm Li}_4(-e^\alpha) + {\rm Li}_4(-e^{-\alpha}) &= 
 -\frac{7\pi^4}{360} - \frac{\pi^2 \alpha^2}{12} - \frac{\alpha^4}{24},
 \label{eq:Li4sum}\\
 {\rm Li}_3(-e^\alpha) - {\rm Li}_3(-e^{-\alpha}) &= -\frac{\pi^2 \alpha}{6} - \frac{\alpha^3}{6},\label{eq:Li3sum} \\
 {\rm Li}_2(-e^\alpha) + {\rm Li}_2(-e^{-\alpha}) &= -\frac{\pi^2}{6} - \frac{\alpha^2}{2},
\end{align}
we find
\begin{align}
 P &= \frac{7\pi^2 T^4}{180} + \frac{\mu_V^2 T^2}{6}
 + \frac{\mu_V^4}{12\pi^2}, \nonumber \\ 
 Q_V &= \frac{\mu_V T^2}{3}+ \frac{\mu_V^3}{3\pi^2},\ \ \ \sigma_A^\omega = \frac{T^2}{6} + \frac{\mu_V^2}{2\pi^2},\nonumber\\
 \sigma_H^\omega &= \frac{T^2}{\pi^2} \left[{\rm Li}_2(-e^{-\mu_V/T}) - {\rm Li}_2(-e^{\mu_V/T})\right],
 \label{eq:unpol} 
\end{align}
while $Q_A = Q_H = \sigma^\omega_V = 0$. Since $Q_A = 0$, the vortical conductivities in the Landau frame coincide with the corresponding conductivities in the $\beta$ frame:
\begin{equation}
 \sigma^\omega_{\ell} = \sigma^\omega_{\ell;\beta} - \frac{Q_A Q_\ell}{E + P} = \sigma^\omega_{\ell;\beta}.
\end{equation}

Taking into account the following expressions for the 
derivatives of the charge densities and vortical conductivities with respect to the chemical potentials,
\begin{align}
 &\frac{\partial Q_\ell}{ \partial \mu_{\ell'}}{\biggl\rvert}_{\mu_A = \mu_H = 0}  = 2  
 \begin{pmatrix}
  \sigma^\omega_A & 0 & 0 \\
  0 & \sigma^\omega_A & \sigma^\omega_H \\
  0 & \sigma^\omega_H & \sigma^\omega_A
 \end{pmatrix}, &  \nonumber \\& 
 \frac{\partial \sigma^\omega_{\ell;\beta}}{\partial \mu_{\ell'}} = \frac{1}{\pi^2} 
 \begin{pmatrix}
  0 & \mu_V & T L \\
  \mu_V & 0 & 0 \\
  T L & 0 & 0
 \end{pmatrix},
\end{align}
with $L = 2\ln\left(2\cosh \frac{\mu_V}{2T}\right)$,
we find for $\mathbb{M}_\omega$ and $\mathbb{M}_\Omega$ the following expressions \cite{Morales-Tejera:2024uzg}:
\begin{align}
 \mathbb{M}_\omega &= \frac{2}{T^2} 
 \begin{pmatrix}
  \sigma^\omega_A - \frac{T^2}{3} \Delta H & 0 & 0 \\
  0 & \sigma^\omega_A & \sigma^\omega_H \\ 
  0 & \sigma^\omega_H & \sigma^\omega_A
 \end{pmatrix}, &\nonumber\\
 \mathbb{M}_\Omega &= 
 \begin{pmatrix}
  0 & \frac{1}{H} A & \frac{1}{H} B\\
  A & 0 & 0 \\ 
  B & 0 & 0
 \end{pmatrix}.
 \label{eq:unpol_M}
\end{align}
In the above, $\Delta H = H - 1$ with $H = (E + P) / (sT) = 1 + \mu_V Q_V / (sT)$, while $A$ and $B$ are defined as
\begin{align}
 A &= \frac{\alpha_V}{\pi^2} - \frac{Q_V}{3 s}, &
 B &= \frac{HL}{\pi^2} - \frac{2 Q_V}{s T^2}  \sigma^\omega_H.
 \label{eq:unpol_AB}
\end{align}
For future convenience, we list here the large-temperature behaviour of the above functions when $\alpha_V = \mu_V / T$ is small:
\begin{gather}
 \sigma_A^\omega = \frac{T^2}{6} + \frac{\mu_V^2}{2\pi^2}, \qquad 
 \sigma^\omega_H = \frac{2\ln 2}{\pi^2} T^2 \alpha_V + O(\alpha_V^3), \nonumber \\ 
 H = 1 + \frac{15 \alpha_V^2}{7\pi^2} + O(\alpha_V^4), \nonumber\\
 L = 2 \ln 2 + O(\alpha_V^2), \quad 
 A = \frac{2\alpha_V}{7\pi^2} + O(\alpha_V^3), \nonumber \\ 
 B = \frac{2\ln 2}{\pi^2} + \frac{7\pi^2 - 120 \ln 2}{28\pi^4} \alpha_V^2 + O(\alpha_V^4).
 \label{eq:unpol_largeT}
\end{gather}
At large chemical potential, when $\lvert \alpha_V \rvert \rightarrow \infty$, we have
\begin{subequations}\label{eq:unpol_largemu}
\begin{gather}
 H = \frac{\alpha_V^2}{\pi^2} \left(1 + \frac{23\pi^2}{15 \alpha_V^2} + O(\alpha_V^{-4})\right), \nonumber \\ 
 A = \frac{2\alpha_V}{3\pi^2} \left(1 - \frac{4\pi^2}{15\alpha_V^2} + O(\alpha_V^{-4})\right), \nonumber\\
 \frac{4s_V Q_v}{\pi^2 s} = \frac{4 s_V}{\pi^2 \alpha_V} (H - 1) \simeq \nonumber \\ \frac{4 s_V \alpha_V}{\pi^4} \left(1 + \frac{8\pi^2}{15 \alpha_V^2} + O(\alpha_V^{-4})\right),
\end{gather} 
where $s_V ={\rm sgn}(\mu_V)$ of $\mu_V$ is the sign of the vector chemical potential.
In what concerns the quantities $L$, $B$ and $\sigma_H$, their large-$\alpha_V$ behaviour receives exponentially-damped corrections of the form $e^{-\lvert \alpha_V \rvert}$, as shown below:
\begin{gather}
 L \simeq \lvert \alpha_V \rvert + 2 e^{-\lvert \alpha_V \rvert}, \nonumber \\  
  B - s_V A \simeq \frac{2}{\pi^2} \left(H + \frac{2 s_V Q_V}{s}\right) e^{-\lvert \alpha_V \rvert}, \nonumber \\ 
 \sigma^\omega_A - s_V \sigma^\omega_H \simeq \frac{2T^2}{\pi^2} e^{-\lvert \alpha_V \rvert}.
\end{gather}
\end{subequations}

We are now ready to solve ${\rm det} \mathbb{M} = 0$. Starting from Eq.~\eqref{eq:M_aux2}, we can compute the determinant of $\mathbb{M}$ by taking the following steps:
\begin{multline}
 {\rm det}\left(\frac{1}{T^2} \mathbb{M}\right)  = {\rm det}(\omega \mathbb{M}_\omega - \kappa_\Omega \mathbb{M}_\Omega) \\ 
 + \frac{i}{3\tau_A} 
 \left\lvert\begin{array}{cc}
    \frac{2\omega}{T^2} \left(\sigma_A^\omega - \frac{T^2}{3} \Delta H\right)  &  -\frac{\kappa_\Omega}{H} B \\
    -\kappa_\Omega B & \frac{2\omega}{T^2} \sigma_A^\omega
 \end{array}\right\rvert\\
 + \frac{i}{3\tau_H} 
 \left\lvert\begin{array}{cc}
    \frac{2\omega}{T^2} \left(\sigma_A^\omega - \frac{T^2}{3} \Delta H\right)  &  -\frac{\kappa_\Omega}{H} A \\
    -\kappa_\Omega A & \frac{2\omega}{T^2} \sigma_A^\omega
 \end{array}\right\rvert \\ 
 - \frac{2\omega}{9 T^2 \tau_{A}\tau_H} \left(\sigma^\omega_A
 - \frac{T^2}{3} \Delta H\right)=0\,.
\end{multline}
The first term evaluates to
\begin{multline}
 {\rm det}(\omega \mathbb{M}_\omega - \kappa_\Omega \mathbb{M}_\Omega) = \\  
 \left(\frac{2 \omega}{T^2}\right)^3 \left(\sigma^\omega_A - \frac{T^2}{3} \Delta H \right) [(\sigma^\omega_A)^2 - (\sigma^\omega_H)^2] \\
 - \frac{2\omega \kappa_\Omega^2}{H T^2}[(A^2 + B^2) \sigma_A^\omega - 2 A B \sigma^\omega_H].
\end{multline}
Solving the other $2 \times 2$ determinants finally gives
\begin{multline}
 {\rm det}\left(\omega\mathbb{M}_v - \kappa_\Omega \mathbb{M}_\Omega + \frac{i}{3\tau_H} \mathbb{I}_H + \frac{i}{3\tau_{A}} \mathbb{I}_A\right) \\
 = \frac{2\omega}{T^2} \left\{\left(\frac{2 \omega}{T^2}\right)^2 \left(\sigma^\omega_A - \frac{T^2}{3} \Delta H \right) [(\sigma^\omega_A)^2 - (\sigma^\omega_H)^2] \right.  \\ \left. - 
 \frac{\kappa_\Omega^2}{H} [(A^2 + B^2) \sigma_A^\omega - 2 A B \sigma^\omega_H]\right\}\\
 + \frac{i}{3} \left[\left(\frac{2\omega}{T^2}\right)^2 \sigma_A^\omega \left(\sigma_A^\omega - \frac{T^2}{3} \Delta H\right) \left(\frac{1}{\tau_H} + \frac{1}{\tau_{A}}\right) \right.  \\ \left. - \frac{\kappa_\Omega^2}{H}\left(\frac{A^2}{\tau_H}+ \frac{B^2}{\tau_{A}}\right) \right] \\
 - \frac{2\omega}{9 T^2 \tau_{A}\tau_H} \left(\sigma^\omega_A
 - \frac{T^2}{3} \Delta H\right) = 0.
\label{eq:QCD_detM}
\end{multline}
Given a particular solution $\omega_*$ to the previous equation, i.e. a collective mode of the system, the perturbations of the chemical potential are related through $\mathbb{M}_{\ell\ell'}\delta \mu_{\ell'}=0$. Specifically, solving the second and third rows ($\ell = A, H$) gives:
\begin{align}
 &\delta\mu^*_A = \nonumber \\ & \dfrac{\left[\frac{2\omega_*}{T^2}(A \sigma^\omega_A - B \sigma^\omega_H) + \frac{i}{3\tau_H} A\right] \kappa_\Omega \delta \mu^*_V}{\left(\frac{2\omega_*}{T^2}\right)^2 [(\sigma^\omega_A)^2 - (\sigma^\omega_H)^2] + \frac{2 i \omega_*}{3T^2} \sigma^\omega_A\left(\frac{1}{\tau_A} + \frac{1}{\tau_H}\right) - \frac{1}{9\tau_A \tau_H}},\nonumber\\
 &\delta\mu^*_H = \nonumber \\ & \dfrac{\left[\frac{2\omega_*}{T^2}(B \sigma^\omega_A - A \sigma^\omega_H) + \frac{i}{3\tau_A} B\right] \kappa_\Omega \delta \mu^*_V}{\left(\frac{2\omega_*}{T^2}\right)^2 [(\sigma^\omega_A)^2 - (\sigma^\omega_H)^2] + \frac{2 i \omega_*}{3T^2} \sigma^\omega_A\left(\frac{1}{\tau_A} + \frac{1}{\tau_H}\right) - \frac{1}{9\tau_A \tau_H}}.
    \label{eq:QCD_mu}
\end{align}
The previous solutions are valid so long as the denominator does not vanish for the particular mode $\omega_*$. Looking ahead, in Subsec.~\ref{sec:QED:cvw} we will encounter the situation when the denominator does vanish. Then, it is more convenient to solve $\mathbb{M}_{\ell\ell'}\delta \mu_{\ell'}=0$ for the first and second rows ($\ell=V,A$) and take $\delta\mu_H$ as the reference amplitude. The relations between the amplitudes are given by: 
\begin{align}
    \delta\mu_V^* &= \dfrac{\left[\frac{2\omega_*}{T^2} (A\sigma_H^\omega-B\sigma_A^\omega)-\frac{i}{3\tau_A} B \right]\kappa_\Omega  \delta\mu_H^*}{A^2 \kappa_\Omega^2 - \frac{2  \omega_*}{T^2}H\left(\sigma_A-\frac{1}{3}T^2 \Delta H\right)\left( \frac{2\omega_*}{T^2}\sigma_A^\omega + \frac{i}{3\tau_A} \right)}\,,\nonumber \\
    \delta\mu_A^* &= \dfrac{\left[\left(\frac{2\omega_*}{T^2}\right)^2 H \left(\sigma_A^\omega-\frac{1}{3}T^2 \Delta H\right)\sigma_H^\omega-AB\kappa_\Omega^2 \right]\delta\mu_H^*}{A^2 \kappa_\Omega^2 - \frac{2  \omega_*}{T^2}H\left(\sigma_A-\frac{1}{3}T^2 \Delta H\right)\left( \frac{2\omega_*}{T^2}\sigma_A^\omega + \frac{i}{3\tau_A} \right)}\,.
    \label{eq:QCD_mu2}
\end{align}
Note that the above relations contain an implicit dependence on the helicity relaxation time, $\tau_H$, through the mode angular frequency, $\omega_*$.
We stress that Eqs. \eqref{eq:QCD_mu} and \eqref{eq:QCD_mu2} are equivalent for a given mode $\omega_*$ provided the denominators do not vanish. 

Before ending this section, we quote from Ref.~\cite{Morales-Tejera:2024uzg} the solutions for the angular velocities satisfying Eq.~\eqref{eq:QCD_detM} in the case of conserved axial and helical charges, corresponding to the $\tau_A, \tau_H \rightarrow \infty$ limit:
\begin{align}
 &\omega_{\rm im.} = 0, \nonumber \\ &
 \omega_h^\pm = \pm \frac{\kappa_\Omega T^2}{2} \sqrt{\frac{\sigma^\omega_A(A^2 + B^2) - 2 A B \sigma^\omega_H}
 {H(\sigma^\omega_A - \frac{T^2}{3} \Delta H)[(\sigma^\omega_A)^2 - (\sigma^\omega_H)^2]}}.
 \label{eq:cons_omega}
\end{align}
The mode $\omega_{\rm im.} = 0$ corresponds to a non-propagating mode for which the vector perturbation does not fluctuate, $\delta \mu_V^{\rm im.} = 0$. We added the ``${\rm im.}$'' subscript to refer to this mode, as in the presence of a finite helicity relaxation time, this mode develops a non-vanishing imaginary part. Instead, the axial and helical perturbations are linked through
\begin{equation}
 A \delta \mu^{\rm im.}_A + B \delta \mu^{\rm im.}_H = 0.
 \label{eq:cons_mu0}
\end{equation}
The other two modes make up the helical vortical wave. Taking $\delta \mu_V^\pm$ as the reference fluctuation amplitude, the axial and helical amplitudes become
\begin{align}
    \delta\mu^{h;\pm}_A &= \dfrac{T^2 (A\sigma_A^\omega - B\sigma_H^\omega) \kappa_\Omega}{2\omega_h^\pm [(\sigma^\omega_A)^2-(\sigma^\omega_H)^2]} \delta \mu^{h;\pm}_V\,, \nonumber \\ 
    \delta\mu_H &= \dfrac{T^2 (B\sigma_A^\omega - A\sigma_H^\omega) \kappa_\Omega}{2\omega_h^\pm [(\sigma^\omega_A)^2-(\sigma^\omega_H)^2]} \delta \mu^{h;\pm}_V\,.
    \label{eq:cons_muh}
\end{align}
In Sec.~\ref{sec:QED}, we shall study Eqs.~\eqref{eq:QCD_detM} and \eqref{eq:QCD_mu} for the cases when $0 < \tau_H < \infty$ (the helical charge has a finite relaxation time and thus it is not conserved) and $\tau_A \rightarrow \infty$ (the axial charge is conserved). The case with finite $\tau_A$ will be studied in Sec.~\ref{sec:QCD}. In both cases, we will take various limits, such as the high temperature or the degenerate gas limit, discussing the effect of charge non-conservation on the wave spectrum uncovered in the case of the plasma with conserved V/A/H charges, discussed in Ref.~\cite{Morales-Tejera:2024uzg}.

\section{Non-conserved helical charge}
\label{sec:QED}

In this section, we take into account the non-conservation of the helical charge due to local scattering processes. At the same time, we keep not just the vector but also the axial charge conserved, which amounts to letting $\tau_A \rightarrow \infty$. This situation can be encountered, e.g., in a QED (electron-positron) plasma in the absence of external electromagnetic fields, when the axial anomaly vanishes, and the axial current is conserved both at the classical and quantum levels. 

As in the previous section, we will consider a plasma with vanishing helical and axial background (mean) chemical potentials. The assumption of vanishing helical chemical potential is justified by the fact that at finite values of $\tau_H$, the violation of the conservation of helical charge prevents introducing the helical chemical potential as a Lagrange multiplier in the grand canonical ensemble. The assumption of the vanishing $\mu_A$ background is not imposed physically but is technically convenient to simplify the analysis. The generalization of our discussion in this section to the case of finite $\mu_A$ is relatively straightforward. All in all, Eq.~\eqref{eq:dis2} becomes 
\begin{equation}\label{eq:QED:diss}
    \partial_{\mu}J^{\mu}_V =0\,,
    \ \ 
    \partial_{\mu}J^{\mu}_A =0\,,
    \ \ 
    \partial_{\mu}J^{\mu}_H = -\frac{T^2}{3 \tau_H} \delta \mu_H\,.
\end{equation}
Physically, the non-conservation of helicity in QED arises from helicity-violating pair-annihilation processes of the form $e_R^+e_L^-\to e_L^+ e_R^-$. Each such process violates helicity by two units while preserving chirality. Analogous processes are mediated by gluon interchange in QCD. For small helical imbalance, the helical relaxation time arising from such processes was estimated for QCD in the weakly-interacting regime as \cite{Ambrus:2019khr,Helicity:2024err}
\begin{multline}\label{eq:Hell_relax}
 \tau_H^{\rm QCD} \simeq \left(\dfrac{250 \ {\rm MeV}}{k_B T}\right)\left(\dfrac{1}{\alpha_{\rm QCD}}\right)^2\left(\dfrac{2}{N_f}\right)\\\times 4.8\ {\rm fm}/c.
\end{multline}
An extension of the above expression to the case of finite background vector chemical potential is provided in Appendix~\ref{app:tauH}.

In the case $\tau_A \rightarrow \infty$, the matrix $\mathbb{M} / T^2$ in Eq.~\eqref{eq:M_aux2} reduces to 
\begin{equation}\label{eq:Mat_Cons_Axial}
 \frac{1}{T^2} \mathbb{M} = \omega \mathbb{M}_\omega - \kappa_\Omega \mathbb{M}_\Omega + \frac{i}{3\tau_H} \mathbb{I}_H,
\end{equation}
while the Eq.~\eqref{eq:QCD_detM} showing ${\rm det}(\mathbb{M} / T^2) = 0$ becomes 
\begin{multline}
 \frac{2\omega}{T^2} \left\{\left(\frac{2 \omega}{T^2}\right)^2 \left(\sigma^\omega_A - \frac{T^2}{3} \Delta H \right) [(\sigma^\omega_A)^2 - (\sigma^\omega_H)^2] \right. \\ \left.-
 \frac{\kappa_\Omega^2}{H} [(A^2 + B^2) \sigma_A^\omega - 2 A B \sigma^\omega_H]\right\}\\
 + \frac{i}{3\tau_H} \left[\left(\frac{2\omega}{T^2}\right)^2 \sigma_A^\omega \left(\sigma_A^\omega - \frac{T^2}{3} \Delta H\right) - \frac{\kappa_\Omega^2 A^2}{H}\right] = 0.
 \label{eq:QED_detM}
\end{multline}
The mode amplitudes in Eq.~\eqref{eq:QCD_mu} reduce to
\begin{align}
    \delta\mu^*_A &= \dfrac{\left(A\sigma^\omega_A-B\sigma^\omega_H + \frac{i T^2 A}{6\omega_* \tau_H}\right) \kappa_\Omega \delta \mu^*_V}{\frac{2\omega_*}{T^2} [(\sigma^\omega_A)^2-(\sigma^\omega_H)^2] + \frac{i}{3\tau_H} \sigma^\omega_A},\nonumber \\ 
    \delta\mu^*_H &= \dfrac{\left(B\sigma^\omega_A-A\sigma^\omega_H\right)  \kappa_\Omega \delta \mu_V^*\,}{\frac{2\omega_*}{T^2} [(\sigma^\omega_A)^2-(\sigma^\omega_H)^2] + \frac{i}{3\tau_H} \sigma^\omega_A}\,.
    \label{eq:QED_mu}
\end{align}

As opposed to the idealised case when all charges are conserved, corresponding to the limit $\tau_A, \tau_H \rightarrow \infty$ and summarized in Eqs.~\eqref{eq:cons_omega}--\eqref{eq:cons_muh}, the determinant 
\begin{align}
{\rm det}\left(-\kappa_\Omega \mathbb{M}_\Omega + \frac{i}{3\tau_H} \mathbb{I}_H\right) = -\frac{i}{3H\tau_H} \kappa^2_\Omega A^2
\end{align}
no longer vanishes and the trivial mode $\omega_{\rm im} = 0$ is no longer a solution of the equation ${\rm det}(\mathbb{M}) = 0$. Hence, all three modes are nontrivial. In the following subsections, we will analyze the properties of the angular frequencies $\omega$ in the following special cases: the small chemical potential limit (Subsec.~\ref{sec:QED:smallmu}), the small $\tau_H$ limit (when the helicity degree of freedom is frozen, see Subsec.~\ref{sec:QED:cvw}) and the large chemical potential limit (Subsec.~\ref{sec:QED:largemu}). Subsec.~\ref{sec:QED:spectra} presents our general conclusions on the modes spectrum for the considered system.

\subsection{Small vector chemical potential limit}\label{sec:QED:smallmu}

Considering the normalized chemical potential $\alpha_V = \mu_V / T$ a small quantity, we seek a power series solution of the energy dispersion $\omega$ given as
\begin{equation}
   \omega = \omega_0 + \omega_1 \alpha_V + \omega_2 \alpha_V^2 + O(\alpha_V^3)\,.
   \label{eq:QED_w_expansion}
\end{equation}
We now expand $\det(\mathbb{M})$ in powers of $\alpha_V$ and equate it recursively to zero, repeating the steps order by order with respect to $\alpha_V$. To find $\omega_0$, we set $\mu_V = 0$ in Eq.~\eqref{eq:QED_detM} and take the leading-order term in the $\alpha_V$ expansion shown in Eq.~\eqref{eq:unpol_largeT}, arriving at
\begin{equation}
 \frac{\omega_0}{27} \left[\omega_0\left(\frac{i}{\tau_H} + \omega_0\right) - k^2 c_h^2\right] = 0\,,
 \label{eq:QED_w0_eq}
\end{equation}
where $c_h$ represents the velocity of the helical vortical wave in a neutral, non-dissipating plasma,
\begin{equation}
    c_h = \frac{6 \ln 2}{\pi^2} \frac{ \Omega}{T}\,.
    \label{eq:ch}
\end{equation}

We identify the mode corresponding to $\omega_0 = 0$ as the axial vortical mode, $\omega_\AVW$, while the helical vortical modes $\omega_h^\pm$ have
\begin{align}
 &\omega_{h;0}^\pm = -\frac{i}{2\tau_H} \pm k c_h \sqrt{1 - \frac{k_{\rm th}^2}{k^2}}, \nonumber \\ &
 k_{\rm th} = \frac{1}{2\tau_H c_h} = \frac{\pi^2}{12 \ln 2} \frac{T}{\Omega \tau_H}.
 \label{eq:QED_kth}
\end{align}
The above relation provides the dispersion relation for the HVW in a neutral plasma, which is fully consistent with the previously reported result for the HVW in the same limit [c.f. Eq. (118) in Ref.\cite{Ambrus:2019khr}].
Notice that we work with a dissipative (lossy) medium so that the apparent (in fact, infinite) superluminal group velocity $v_{\rm gr.} = \partial {\rm Re}\, [\omega(k)]/\partial k$ at the merging point $k = k_{\rm th}$ does not violate causality as long as the asymptotic causality condition at large momentum, $k \to \infty$, is fulfilled~\cite{Pu:2009fj}.

The above solution indicates that the HVW propagates only when the wavenumber $k$ exceeds the threshold value $k_{\rm th}$ given by the second relation in Eq.~\eqref{eq:QED_kth}. At wavenumbers lower than $k_{\rm th}$, the modes $\omega^\pm_h$ are purely dissipative.

At first order with respect to $\alpha_V$, Eq.~\eqref{eq:QED_detM} gives
\begin{equation}
 \frac{\omega_1}{27}\left[\omega_0\left(\frac{2i}{\tau_H} + 3\omega_0\right) - k^2 c_h^2\right] = 0\,.
\end{equation}
This condition is valid for both the axial and the vector-helical modes. For the axial mode, $\omega_{\AVW;0} = 0$ and the square brackets evaluate to $-k^2 c_h^2 \neq 0$. For the vector-helical modes, $\omega^\pm_{h;0}(i\tau_H^{-1} + \omega^\pm_{h;0}) =k^2 c_h^2$ and the square brackets evaluate to $2k^2 c_h^2 -i\tau_H^{-1} \omega^\pm_{h;0}$, being non-vanishing unless $k = k_{\rm th}$. We thus conclude that the first-order contributions to the velocities vanish, $\omega_{\AVW;1} = \omega^\pm_{h;1} = 0$, for both the axial and the helical-vortical modes. Thus, the axial mode is of second order with respect to $\alpha_V$. To find its expression, we expand Eq.~\eqref{eq:QED_detM} to second order, while considering $\omega \rightarrow \omega_\AVW = \omega_{\AVW;2} \alpha_V^2\,$:
\begin{align}
 &-\frac{4}{3} \left(\frac{\kappa_\Omega \ln 2}{\pi^2 }\right)^2 \left[\omega_{\AVW;2} + \frac{i}{\tau_H (7 \ln 2)^2}\right] = 0 \nonumber \\ & \Rightarrow \quad 
 \omega_{\AVW;2} = -\frac{i}{\tau_H (7 \ln 2)^2}.
 \label{eq:QED_w2a}
\end{align}
Thus, the axial mode is purely dissipative. In the limit $\tau_H \rightarrow \infty$ when helicity is conserved, it is clear that $\omega_{\AVW; 2} \rightarrow 0$ and indeed, $\omega_{\AVW} = 0$ at any order. This observation allows us to identify the mode of Eq.~\eqref{eq:QED_w2a} with the mode $\omega_{\rm im.}$ obtained in the case when all charges are conserved as it was summarized in Eq.~\eqref{eq:cons_omega}.

The relation between the fluctuations amplitudes can be found by taking the large-temperature expansion of the expressions in Eq.~\eqref{eq:QED_mu}, using Eqs.~\eqref{eq:unpol_largeT}. In the case of the axial mode, we take $\delta \mu_A^\AVW$ as the reference amplitude and obtain the following relations between the fluctuations of the chemical potentials in this mode
\begin{subequations}
\begin{align}
 \delta \mu_V^\AVW & \simeq -\frac{i}{k \tau_H c_h} \frac{\alpha_V}{7 \ln 2} \delta \mu_{A}^\AVW\,, \\
 \delta \mu^\AVW_H & \simeq -\frac{\alpha_V}{7 \ln 2} \delta \mu^\AVW_{A}\,.
\end{align}
 \label{eq:QED_a_dmu}
\end{subequations}
Hereafter, the approximate relation ``$\simeq$'' implies that the relation is valid in the leading order in $\alpha_V$ with higher-order corrections omitted.

For the vector-helical modes, we obtain
\begin{align}
 \delta \mu^{h;\pm}_H &\simeq \frac{\omega^\pm_{h;0}}{k c_h} \delta \mu_V^{h;\pm} \nonumber \\ &= 
 \left(-i k_{\rm th} \pm \sqrt{k^2 - k_{\rm th}^2}\right) \frac{\delta \mu^{h;\pm}_V}{k}, \nonumber\\
 \delta \mu^{h;\pm}_A &\simeq \left(\frac{6 \kappa_\Omega}{7\pi^2 \omega^\pm_{h;0}} - \frac{2 \omega^\pm_{h;0}}{\kappa_\Omega}\right) \alpha_V \delta \mu_{V}^\pm \nonumber\\
 &= \left[\frac{i}{\tau_H}\left(\frac{\pi^2}{84(\ln 2)^2} + 1\right) \right. \nonumber \\ &\left. \pm 2 c_h \sqrt{k^2 - k_{\rm th}^2} \left(\frac{\pi^2}{84 (\ln 2)^2} - 1\right) \right] \frac{\alpha_V}{\kappa_\Omega} \delta \mu^\pm_V. 
 \label{eq:QED_vh_dmu}
\end{align}
The inverse proportionality between $\delta \mu^\AVW_V$ and $k / k_{\rm th} = 2 c_h \tau_H k$ implied by Eq.~\eqref{eq:QED_a_dmu} reveals the breakdown of the small chemical potential expansion, considered in this section, at large wavelengths or small $\tau_H$. This effect appears due to the finite helical relaxation time $\tau_H$, which changes the low-$k$ wave spectrum significantly compared to the case of conserved charges. We will address the $k \rightarrow 0$ limit (or, equivalently, the small $\tau_H$ limit) in the following subsection. 

Let us now consider as the initial state a harmonic perturbation in the vector charge:
\begin{align}
 &\bar{\mu}_V(0, z) = \mu_V + \delta \bar{\mu}_{V;0} \cos(k z), \nonumber \\ &
 \bar{\mu}_A(0,z) = \bar{\mu}_H(0,z) = 0.
 \label{eq:harmV_init}
\end{align}
To leading order in $\alpha_V$, this setup assumes that the amplitudes in the axial mode vanish, $\delta \mu_\ell^\AVW \simeq 0$. The amplitudes $\delta \mu_V^{h;\pm}$ and $\delta \mu_H^{h;\pm}$ corresponding to the vector-helical modes satisfy 
\begin{align}
 \delta \mu_V^{h;\pm} &\simeq \frac{\delta \bar{\mu}_{V;0}}{2} \left(1 \pm \frac{i k_{\rm th}}{\sqrt{k^2 - k_{\rm th}^2}}\right), \nonumber \\ 
 \delta \mu_H^{h;\pm} &\simeq \pm \frac{k\delta \bar{\mu}_{V;0}}{2 \sqrt{k^2 - k_{\rm th}^2}},
 \label{eq:harmV_amplitudes}
\end{align}
while $\delta \bar{\mu}_{A;0}^{h;\pm}$ is proportional to $\alpha_V$, as shown in Eq.~\eqref{eq:QED_vh_dmu}. The full solution reads:
\begin{align}
 \delta\bar{\mu}_V(t,z) &\simeq \delta \bar{\mu}_{V;0} e^{-t/2\tau_H} \cos(k z) \nonumber\\
 & \hspace{-40pt} \times \left[\cos\left(c_h t \sqrt{k^2 - k_{\rm th}^2}\right)+ \frac{\sin\left(c_h t \sqrt{k^2 - k^2_{\rm th}}\right)}{2 \tau_H c_h \sqrt{k^2 - k_{\rm th}^2}} 
 \right],\nonumber\\
 \delta\bar{\mu}_H(t,z) &\simeq \frac{k \delta \bar{\mu}_{V;0} e^{-t / 2\tau_H}}{\sqrt{k^2 - k_{\rm th}^2}}  \nonumber\\
 &\times \sin(k z) \sin\left(c_h t \sqrt{k^2 - k_{\rm th}^2}\right).
 \label{eq:harmV_QED}
\end{align}

In the infrared, when $k < k_{\rm th} = 1 / 2 \tau_H c_h$, the square root becomes imaginary, $\sqrt{k^2 - k_{\rm th}^2} = i \sqrt{k_{\rm th}^2 - k^2}$, and the trigonometric functions become hyperbolic ones. In the late-time limit, we have
\begin{align}
 \delta \bar{\mu}_V {\biggl\rvert}_{k < k_{\rm th}} &\rightarrow \frac{\delta \bar{\mu}_{V;0}}{2} \left(1 + \frac{k_{\rm th}}{\sqrt{k_{\rm th}^2 - k^2}}\right)\times \nonumber \\ & \times e^{-\frac{t}{2\tau_H} (1 - \sqrt{1 - k^2 / k_{\rm th}^2})} \cos(kz), \nonumber\\
 \delta \bar{\mu}_H {\biggl\rvert}_{k < k_{\rm th}} &\rightarrow \frac{\delta \bar{\mu}_{V;0} k^2 \tau_H c_h}{2 \sqrt{k_{\rm th}^2 - k^2}} \nonumber \\ & \times e^{-\frac{t}{2\tau_H} (1 - \sqrt{1 - k^2 / k_{\rm th}^2})} \sin(kz).
 \label{eq:harmV_IR}
\end{align}
It can be seen that the modes with $k \rightarrow 0$ suffer negligible damping. When $k > k_{\rm th} = 1 / 2 \tau_H c_h$, normal propagation resumes, and all modes are damped according to the factor $e^{-t / 2\tau_H}$. 

\begin{figure}[t]
\centering
\begin{tabular}{c}
 \includegraphics[width=.89\linewidth]{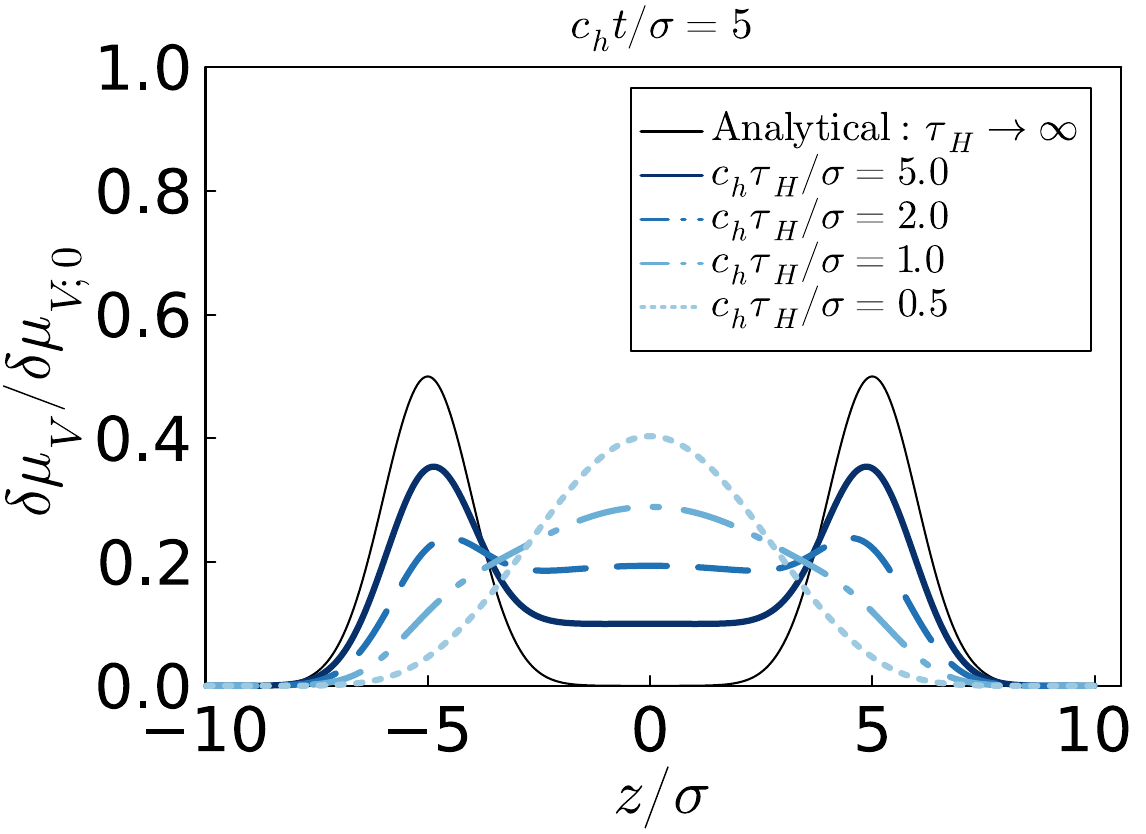}\\
 \includegraphics[width=.89\linewidth]{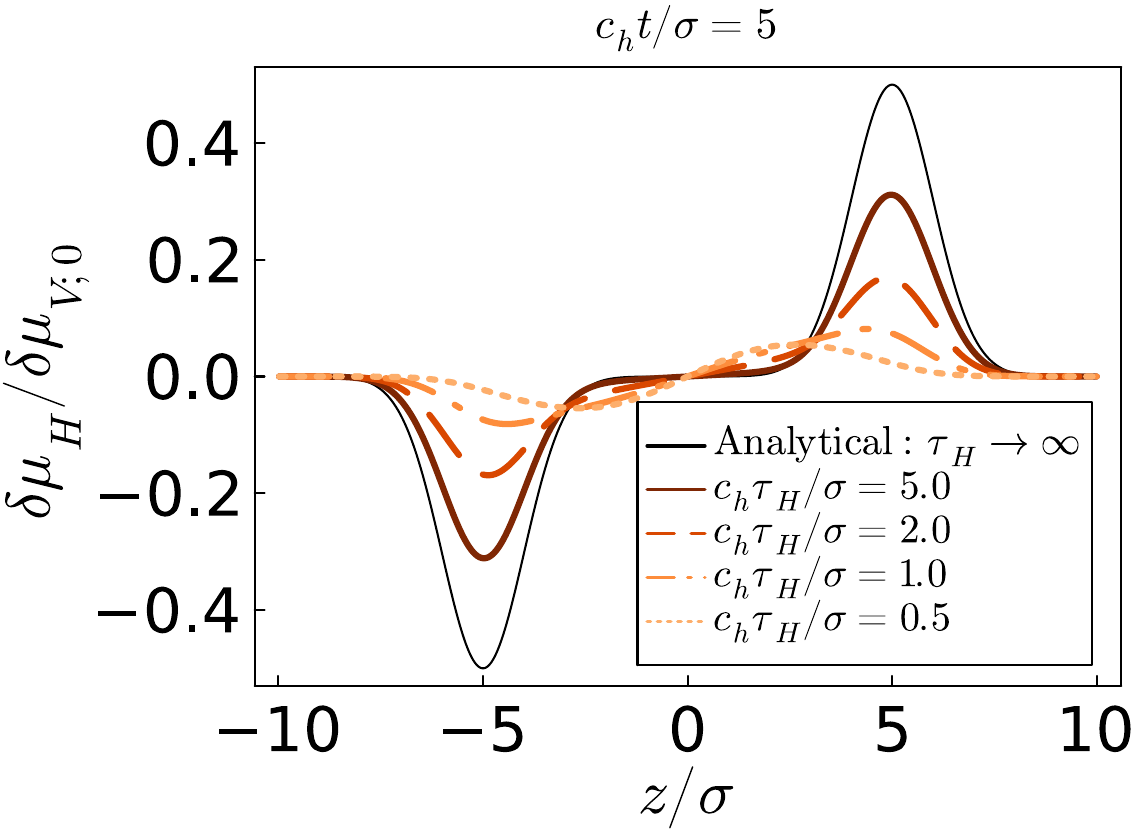}
\end{tabular}
\caption{Effect of the helicity relaxation time $\tau_H$ on the propagation of the initial Gaussian perturbation of the vector charge, shown in Eq.~\eqref{eq:gaussV_init}, in the case of conserved axial charge ($\tau_A \rightarrow \infty$). The profiles correspond to various values of $\tau_H$ and are represented with respect to $z / \sigma$, where $\sigma$ is the Gaussian width. The background fluid is neutral and unpolarized ($\mu_V = \mu_A = \mu_H = 0$).
 \label{fig:HVW_tauH}
}
\end{figure}

Let us now consider an initial Gaussian perturbation of the vector charge:
\begin{align}
 &\bar{\mu}_V(0, z) = \mu_V + \delta \bar{\mu}_{V;0} e^{-z^2 /2\sigma^2}, \nonumber \\ & 
 \bar{\mu}_A(0,z) = \bar{\mu}_H(0,z) = 0.
 \label{eq:gaussV_init}
\end{align}
Considering a neutral background state with $\mu_V = 0$, the time evolution of this configuration can be expressed in integral form as
\begin{multline}
 \bar{\mu}_V(t, z) \simeq \delta \bar{\mu}_{V;0}k e^{-t/2\tau_H} \sigma \sqrt{2\pi} \int_{-\infty}^\infty dk\, e^{-\frac{1}{2} \sigma^2 k^2 + i k z} \\\times
 \left[\cos\left(c_h t \sqrt{k^2 - k_{\rm th}^2}\right) + 
 \frac{\sin\left(c_h t \sqrt{k^2 - k_{\rm th}^2}\right)}{2 \tau_H c_h \sqrt{k^2 - k_{\rm th}^2}} \right].
 \label{eq:gaussV_tauH_sol}
\end{multline}
In the limit when the helicity charge is conserved, $\tau_H \rightarrow \infty$, the square brackets on the second line of Eq.~\eqref{eq:gaussV_tauH_sol} evaluate to $\cos(c_h k t)$ and the integral with respect to $k$ can be performed analytically (see also Eq.~(3.36) of Ref.~\cite{Morales-Tejera:2024uzg}):
\begin{align}
 &\lim_{\tau_H \rightarrow \infty} \delta \bar{\mu}_V(t, z) =\frac{\delta \bar{\mu}_{V;0}}{2} \nonumber \\ & \qquad\qquad \left[e^{-(z - c_h t)^2 / 2\sigma^2} + e^{-(z + c_h t)^2 / 2\sigma^2}\right],\nonumber\\
 &\lim_{\tau_H \rightarrow \infty} \delta \bar{\mu}_H(t, z) =\frac{\delta \bar{\mu}_{V;0}}{2} \nonumber \\ & \qquad\qquad \left[e^{-(z - c_h t)^2 / 2\sigma^2} - e^{-(z + c_h t)^2 / 2\sigma^2}\right].
 \label{eq:HVW_gauss}
\end{align}

In Fig.~\ref{fig:HVW_tauH}, we represented $\delta \bar{\mu}_V(t,z)$ and $\delta \bar{\mu}_H(t,z)$ with respect to $z / \sigma$, when $c_h t / \sigma = 5$, for the case of a neutral background ($\mu_V = \mu_A = \mu_H = 0$). We considered various values of $c_h \tau_H / \sigma$. It can be seen that, compared to the case when the helicity charge is conserved and $\tau_H \rightarrow \infty$ (shown with the black line), the propagation features are damped as $\tau_H$ is decreased. At the lowest value of $c_h \tau_H / \sigma =0.5$, $\delta \mu_V(t,z)$ simply decays on the time scale given by $\tau_H$.

\subsection{Small \texorpdfstring{$\tau_H$}{\texttau H} limit: Recovering the CVW}\label{sec:QED:cvw}

Equations~\eqref{eq:QED_a_dmu} and \eqref {eq:QED_vh_dmu} of the previous subsection show that it is problematic to take the $k,\tau_H \rightarrow 0$ limit from the small chemical potential expansion. In order to access the small $\tau_H \kappa_\Omega$ regime, we expand the quantity $\text{w} = \tau_H \omega$ and $\delta \mu_{\ell}$ in powers of $\tau_H$:
\begin{align}
    &\text{w} = \tau_H \omega = \text{w}_0 + \text{w}_1 \tau_H + \dots, \nonumber \\ &
    \delta \mu_{\ell} = \delta \mu_{\ell;0} + \delta \mu_{\ell;1} \tau_H + \dots.
\end{align}
It is convenient to work with $\text{w}$ as in the $\tau_H \rightarrow 0$ limit, we can expect $\omega$ to have a $\tau_H^{-1}$ leading-order contribution, see e.g. Eq.~\eqref{eq:QED_kth}. Additionally, from Eq.~\eqref{eq:Mat_Cons_Axial}, we observe that $\text{w}$ is a function of the combination $\kappa_\Omega \tau_H = k \Omega \tau_H / T$ [see Eq.~\eqref{eq_kappa_Omega}] and consequently, the small $\tau_H$ expansion naturally encapsulates the results for small wavenumbers.

Multiplying Eq.~\eqref{eq:QED_detM} by $\tau_H^3$, we get to zeroth order in $\tau_H$:
\begin{align}
 &\left(\frac{2 \text{w}_0}{T^2}\right)^2 \left(\sigma_A^\omega - \frac{T^2}{3} \Delta H\right)\nonumber \\ & \times \left(\frac{i}{3} \sigma^\omega_A + \frac{2 \text{w}_0}{T^2} [(\sigma^\omega_A)^2 - (\sigma^\omega_H)^2]\right) = 0,
\end{align}
which gives as solutions $\text{w}^\pm_{\AVW;0} = 0$ as a double root (we will see shortly that these are axial modes, as they involve the axial chemical potential fluctuations), as well as the solution
\begin{equation}
 \text{w}_{h;0} = -\frac{i T^2 \sigma_A^\omega}{6[(\sigma^\omega_A)^2 - (\sigma^\omega_H)^2]}.
 \label{eq:QED_smallk_w0h}
\end{equation}
The above corresponds to a purely dissipative helical mode, with $\text{w}_{h;0} = -i + O(\alpha_V^2)$ at large temperature and $\text{w}_{h;0} \simeq -\frac{i \pi^2}{24} e^{\lvert \alpha_V \rvert}$ in the degenerate limit. 

Going now to the next-to-leading order, the axial modes, characterized by $\tau_H \omega_\AVW = \text{w}_{\AVW;1} \tau_H + \dots$, have their leading term given by
\begin{align}
 &\frac{i}{3} \left[\left(\frac{2 \text{w}^\pm_{\AVW;1}}{T^2}\right)^2 \sigma_A^\omega\left(\sigma^\omega_A - \frac{T^2}{3} \Delta H \right) - \frac{A^2}{H} \kappa_\Omega^2 \right] = 0 \nonumber \\ &\Rightarrow 
 \text{w}^\pm_{\AVW;1} = \pm \frac{A T^2 \kappa_\Omega}{2 \sigma^\omega_A \sqrt{H}} \left(1 - \frac{T^2 \Delta H}{3 \sigma^\omega_A}\right)^{-1/2}.
\end{align}
Using Eqs.~\eqref{eq:unpol_largeT} and \eqref{eq:unpol_largemu},
it can be seen that $\text{w}^\pm_{\AVW;1} = \pm 6 \alpha_V \kappa_\Omega / (7 \pi^2)$ in the large-$T$ limit and $\pm 2\pi \kappa_\Omega / (\alpha_V^2 \sqrt{3})$ in the degenerate limit. We remark that the two propagating modes $\text{w}^\pm_{\AVW;1}$ are necessarily different from the ones found in the small chemical potential expansion since those were not propagating at small wavenumber $k$. Therefore, we can anticipate that at finite but small chemical potential, two modes are propagating for small $k$. As the wavenumber $k$ increases, these modes will stop propagating, and we will have three purely dissipating modes until we reach yet another critical value of $k$, where the modes found in the small chemical potential limit start propagating again -- see Sec.~\ref{sec:QED:spectra} for more details.

For the helical mode with $\tau_H\omega_h = \text{w}_{h;0} + \text{w}_{h;1} \tau_H + \dots$, the first-order correction satisfies
\begin{align}
 &\frac{4 \text{w}_{h;0} \text{w}_{h;1}}{T^4}\left(\sigma^\omega_A - \frac{T^2}{3} \Delta H\right) \nonumber \\ &\times
 \left(\frac{6 \text{w}_{h;0}}{T^2} [(\sigma_A^\omega)^2 - (\sigma_H^\omega)^2] + \frac{2i}{3} \sigma^\omega_A\right) = 0.
 \label{eq:QED_smallk_w1h}
\end{align}
The term inside the second pair of parentheses evaluates to $-i \sigma^\omega_A / 3 \neq 0$, by virtue of Eq.~\eqref{eq:QED_smallk_w0h}. Thus, Eq.~\eqref{eq:QED_smallk_w1h} implies that $\text{w}^h_1 = 0$.

The relation between the fluctuation amplitudes $\delta\mu_\ell$ can be extracted by expanding Eq.~\eqref{eq:QED_mu} in powers of $\tau_H$. For the axial modes, we find, to leading order in $\tau_H$,
\begin{align}
 \delta\mu_A^{\AVW;\pm} &\simeq \dfrac{AT^2 \kappa_\Omega}{2 \text{w}_{\AVW;1}^\pm\sigma_A^\omega} \delta\mu_V^{\AVW;\pm} 
 \nonumber \\ &= \pm \sqrt{H} \left(1 - \frac{T^2 \Delta H}{3 \sigma^\omega_A}\right)^{1/2} \delta \mu_V^{\AVW;\pm}\,, \nonumber\\
 \delta\mu_H^{\AVW;\pm} &\simeq -\frac{3 i}{\sigma^\omega_A} \left( B \sigma^\omega_A -A\sigma_H^\omega\right) \kappa_\Omega \tau_H \delta\mu_V^{\AVW;\pm}\,.\label{eq:CVW_delta_mu}
\end{align}
As for the helical mode, the denominator in Eq.~\eqref{eq:QED_mu} vanishes at $\omega_*=\omega_{h;0}$ and the above relations are not applicable. Therefore we resort to Eq. \eqref{eq:QCD_mu2} in the limit $\tau_A\to \infty$. Expanding the latter for small $\tau_H$ gives
\begin{align}
    &\delta\mu_V^h \simeq \frac{T^2 }{2 \text{w}_{h;0} H\sigma^\omega_A} \frac{B \sigma^\omega_A - A \sigma^\omega_H}{\sigma_A^\omega - \frac{T^2}{3} \Delta H} \kappa_\Omega \tau_H \delta \mu^h_{H}\,, \nonumber \\ & 
    \delta\mu_A^h \simeq -\dfrac{\sigma_H^\omega}{\sigma_A^\omega} \delta\mu_H^h.
\end{align}

The Chiral Vortical Wave (CVW) can be recovered as follows. The traditional derivation of the CVW \cite{Jiang:2015cva} did not include fluctuations of the helical chemical potential. The freezing of the helical degree of freedom is naturally implemented if one requires instantaneous dissipation of helicity fluctuations, i.e. $\tau_H\to 0\,$.  
The angular frequencies $\omega_h = \text{w}_h / \tau_H$ and $\omega^\pm_\AVW = \text{w}^\pm_\AVW / \tau_H$ of the three modes found at small $\tau_H$ can be written in the leading order as:\footnote{The angular frequencies $\omega_h$ and $\omega^\pm_\AVW$ in Eq.~\eqref{eq:omega:AVW} of this article should not be confused with their non-dissipative counterparts, $\omega^\pm_h$ and $\omega_\AVW$, obtained in, respectively, Eqs.~(77) and (90)
of the companion paper~\cite{Morales-Tejera:2024uzg}.}
\begin{align}
  & \omega_h = -\frac{i T^2 \sigma_A^\omega}{6 \tau_H[(\sigma^\omega_A)^2 - (\sigma^\omega_H)^2]}\,,\nonumber \\ & 
   \omega^\pm_\AVW = \pm \frac{A T^2}{2 \sigma^\omega_A \sqrt{H}} \left(1 - \frac{T^2 \Delta H}{3 \sigma^\omega_A}\right)^{-1/2} \frac{k \Omega}{T}\,.
   \label{eq:omega:AVW}
\end{align}
As $\tau_H\to0$, we find that the helical mode dissipates very quickly, removing any fluctuation of helical chemical potential. In contrast, the axial modes give a propagating wave mixing only axial and vector degrees of freedom, with $\delta \mu_H = O(\tau_H)$ [see Eq.~\eqref{eq:CVW_delta_mu}], i.e. the Chiral Vortical Wave 
\cite{Jiang:2015cva, Gorbar:2017toh}. In fact, the pair of modes in Eq.~\eqref{eq:omega:AVW} coincides exactly with the result obtained in Ref.~\cite{Gorbar:2017toh} for the Chiral Vortical Wave. Using the definitions for $A$, $H$, $\Delta H$ and $\sigma_A$ given below Eq. \eqref{eq:unpol_M}, the pair of modes in Eq. \eqref{eq:omega:AVW} become
\begin{align}
    &\omega_a^\pm =\dfrac{6 \pi T \mu_V (\pi^2 T^2 + 5\mu_V^2)k \Omega}{\sqrt{7\pi^4 T^4 +6\pi^2 T^2\mu_V^2 +15\mu_V^4}}\nonumber\\& \times \dfrac{1}{\sqrt{(\pi^2 T^2+3\mu_V^2)(7\pi^4T^4 + 30\pi^2 T^2\mu_V^2 +15\mu_V^4)}}\,,
    \label{eq_CWV}
\end{align}
which is the same as the modes obtained in Eq.~(63) of Ref. \cite{Gorbar:2017toh} upon substitution of Eq.~(26) in the same paper. It is important to note that there is a discrepancy between Eq.~\eqref{eq_CWV} and the result of Ref.~\cite{Jiang:2015cva}, where the fluctuations in the energy-momentum sector were not taken into account (see Ref.~\cite{Gorbar:2017toh} for a detailed discussion).

Note that the CVW is only present at finite (vector) chemical potential $\mu_V$. In agreement with this property, the angular frequency of our axial modes, given by the second relation in Eq.~\eqref{eq:omega:AVW}, does not vanish provided $\mu_V \neq 0$ as it follows from the definition~\eqref{eq:unpol_AB} of the coefficient $A$ and the absence of the leading $\mu_V$ behaviour in other terms in this expression.

We remark that at finite (non-zero) $\tau_H$, the Chiral Vortical Wave (CVW) is propagating for small wavenumbers, but it can stop propagating when the wavenumber exceeds certain finite value. These aspects will be further clarified in Sec.~\ref{sec:QED:spectra} (see also Fig.~\ref{fig:k*}).

\begin{figure}[!ht]
    \centering
    \begin{tabular}{c}
    \includegraphics[width=.92\linewidth]{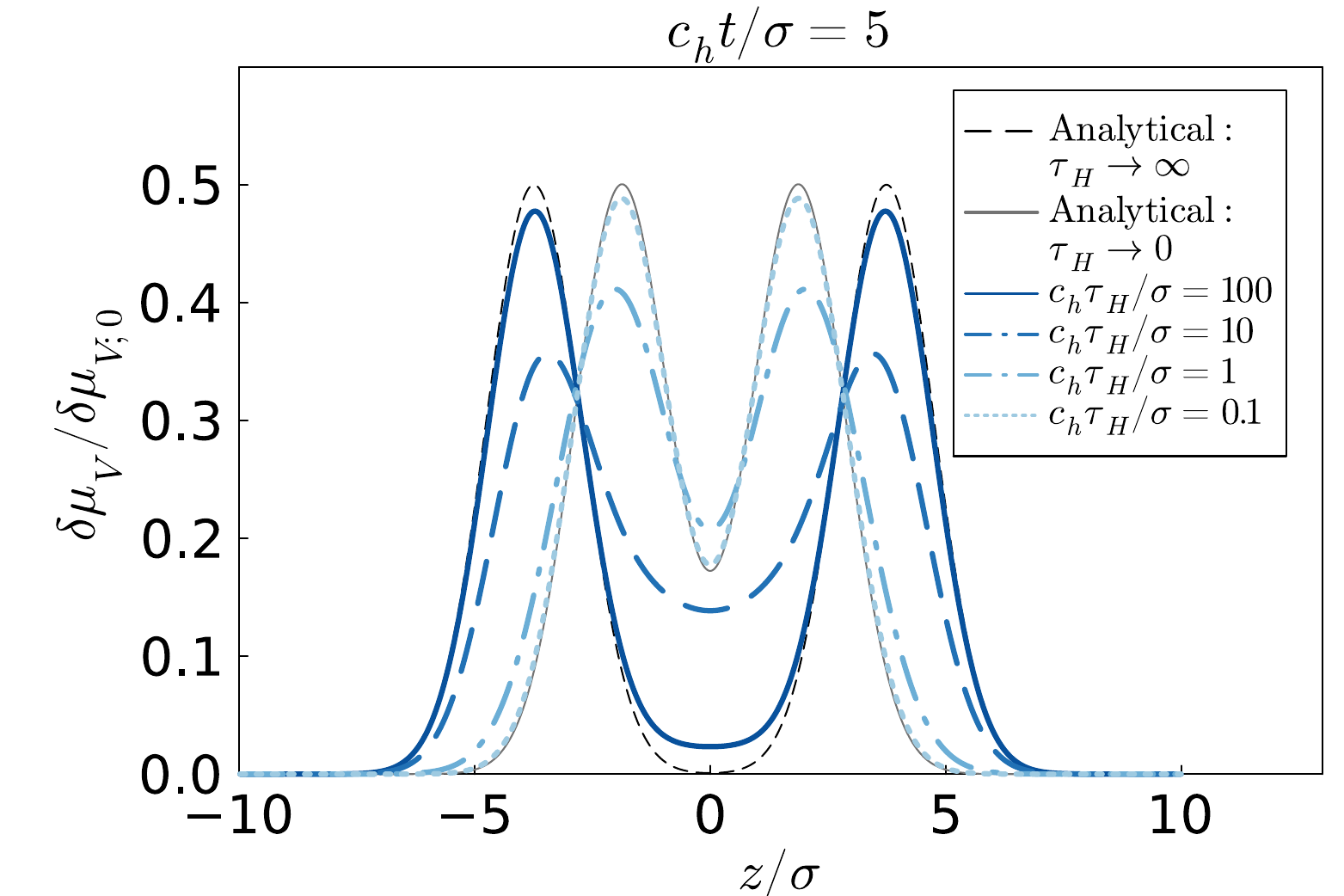}\\
    \includegraphics[width=.92\linewidth]{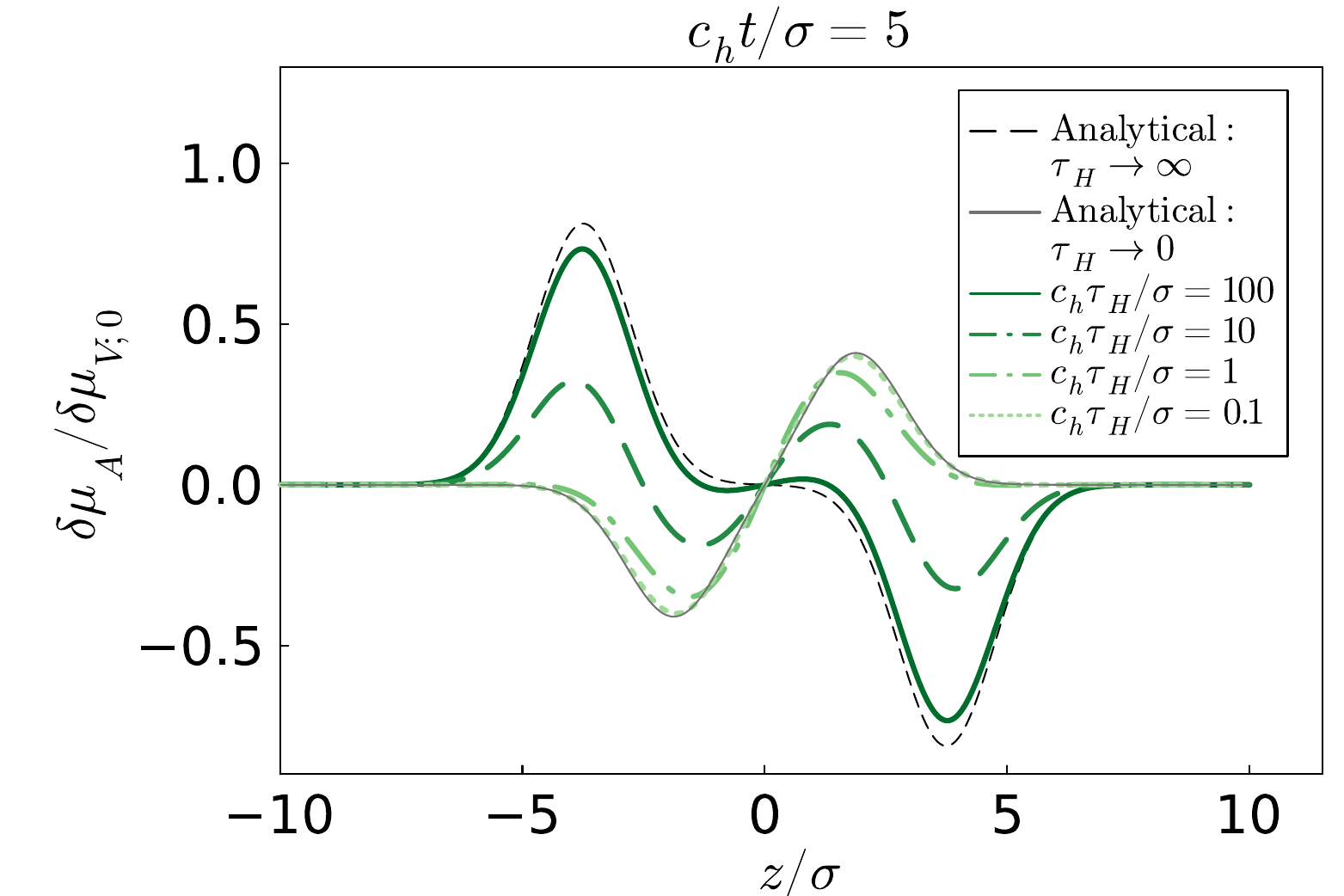}\\
    \multicolumn{1}{c}{\includegraphics[width=.92\linewidth]{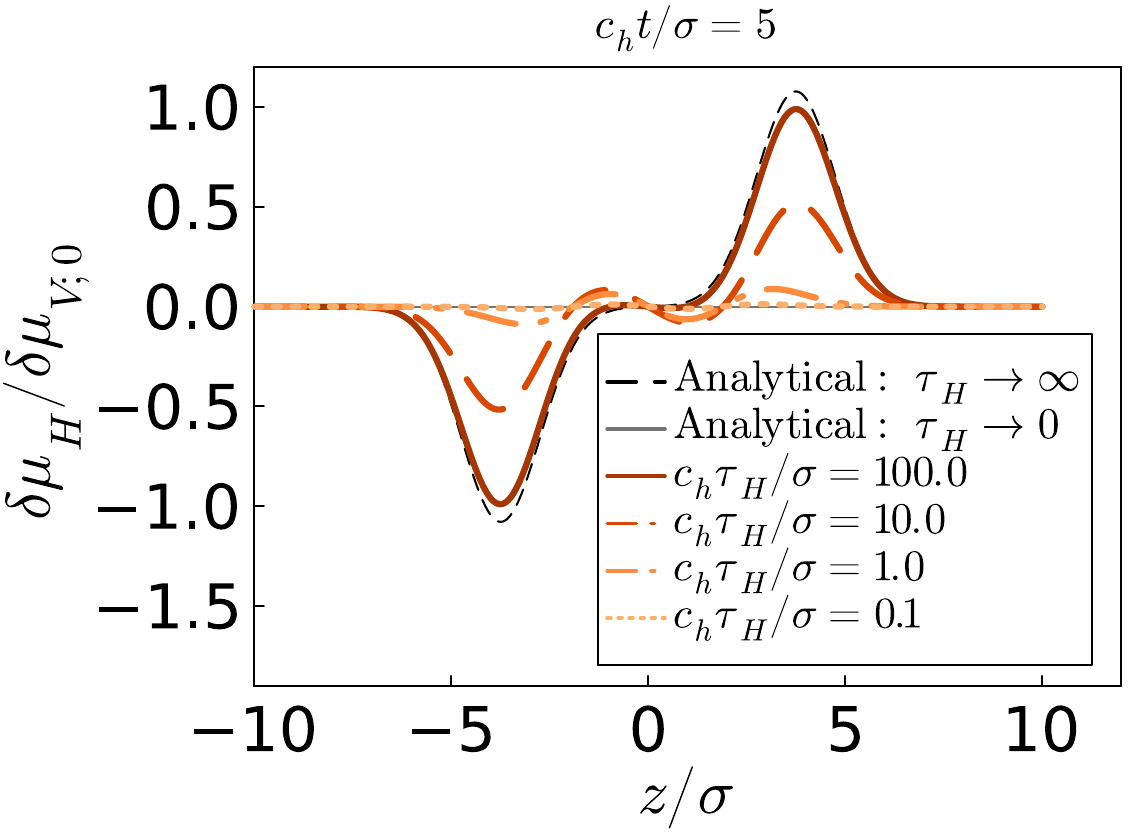}}
    \end{tabular}
    \caption{Profiles of the perturbations in the vector, axial and helical chemical potentials at time $c_h t / \sigma = 5$ for the Gaussian initial conditions in Eq.~\eqref{eq:gaussV_init}, set by the initial Gaussian width $\sigma$ and the velocity $c_h$ of the helical vortical wave in a neutral plasma with conserved axial charge~\eqref{eq:ch}, for various values of the helical relaxation time $\tau_H$. The helicity-preserving limit $\tau_H \rightarrow \infty$ corresponds to the helical vortical wave solutions shown in Eq.~\eqref{eq:HVW_gauss}, while the helicity-frozen limit $\tau_H \rightarrow 0$ corresponds to the axial vortical wave described in Eq.~\eqref{eq:AVW_gauss}.We took $\alpha_V=13/6$ for the background state.
    \label{fig:CVW}
    }
\end{figure}

Let us illustrate the propagation properties of the CVW by considering the initial Gaussian profile for the vector chemical potential perturbation in Eq.~\eqref{eq:gaussV_init}. To leading order, only the axial modes contribute, such that 
\begin{equation}
 \sum_{\omega = \omega^\pm_\AVW} \delta \mu^{\AVW;\pm}_V(k) = \frac{\sigma}{\sqrt{2\pi}} \delta \mu_V e^{-\sigma^2 k^2 / 2}.
\end{equation}
Given the relation \eqref{eq:CVW_delta_mu} between $\delta \mu^{\AVW;\pm}_V$ and $\delta \mu^{\AVW;\pm}_A$, as well as the constraint $\sum_{\omega = \omega^\pm_\AVW} \delta \mu^{\AVW;\pm}_A = 0$, we find
\begin{equation}
 \delta \mu^{\AVW;\pm}_A(k) = \pm \frac{\sigma}{2} \sqrt{\frac{H}{2\pi} \left(1 - \frac{T^2 \Delta H}{3\sigma_A^\omega}\right)} \delta \mu_V e^{-\sigma^2 k^2 / 2}.
\end{equation}
This leads to the following propagating Gaussian pulse solutions:
\begin{align}
 \bar{\mu}_V(t,z) &= \mu_V + \frac{\delta \mu_V}{2} \left[e^{-(z - v_\AVW t)^2 / 2\sigma^2} + e^{-(z + v_\AVW t)^2 / 2\sigma^2}\right],\nonumber\\
 \bar{\mu}_A(t,z) &= \frac{\delta\mu_V}{2} \sqrt{H\left(1 - \frac{T^2 \Delta H}{3 \sigma^\omega_A}\right)} \nonumber\\
 &\times \left[e^{-(z - v_\AVW t)^2 / 2\sigma^2} - e^{-(z + v_\AVW t)^2 / 2\sigma^2}\right].
 \label{eq:AVW_gauss}
\end{align}
The above solution agrees with the $\tau_H \rightarrow 0$ limit shown in Fig.~\ref{fig:CVW}, where we took 
an unpolarized background state ($\mu_A = \mu_H = 0$) with $\alpha_V = 13/6 \simeq 2.16$, corresponding to a ratio $c_h(\alpha_V) / c_a(\alpha_V) \simeq 2$ between the velocities of the pure helical-vortical and chiral-vortical waves. 
This figure also demonstrates the smooth transition from the HVW in the helicity-conserving limit to the CVW as $\tau_H$ is decreased from $\infty$ to $0$.

\subsection{Large vector chemical potential limit}\label{sec:QED:largemu}

The characteristic Eq.~\eqref{eq:QED_detM} for generic vector chemical potential is a cubic equation in the following form:
\begin{multline}
    a(\alpha_V) \omega^3 + \dfrac{i}{\tau_H}  b(\alpha_V) \omega^2 - \kappa^2_\Omega c(\alpha_V) \omega- \dfrac{i}{\tau_H} \kappa_\Omega^2 d(\alpha_V)\\
     =0\,,
    \label{eq:QED_detM_largemu}
\end{multline}
where the coefficients $a$, $b$, $c$ and $d$ are non-negative and can be read off from Eq.~\eqref{eq:QED_detM}:
\begin{align}
 a &= \frac{8}{T^6} \left(\sigma^\omega_A - \frac{T^2}{3} \Delta H \right) \left[(\sigma^\omega_A)^2 - (\sigma^\omega_H)^2\right], \nonumber\\
 b &= \frac{4}{3T^4} \sigma_A^\omega \left(\sigma_A^\omega - \frac{T^2}{3} \Delta H\right),\nonumber\\
 c &= \frac{2}{HT^2} \left[(A^2 + B^2) \sigma_A^\omega - 2 A B \sigma^\omega_H\right], 
\ \ \ d = \frac{A^2}{3H}.
\label{eq:abcd}
\end{align}
We first notice that Eq.~\eqref{eq:QED_detM_largemu} always supports a purely imaginary solution, due to the fact that under $\omega \to i \omega_I$, the characteristic equation turns into a polynomial equation of third order with real coefficients, with at least one real solution. The remaining two modes come in pairs with opposite signs for the real part. In the large chemical potential limit, the coefficients satisfy
\begin{align}
    \lim_{\lvert \alpha_V \rvert \to \infty} a(\alpha_V) &\to \dfrac{8\alpha_V^4}{3 \pi^6}e^{-\lvert \alpha_V \rvert}\,, 
    \lim_{\lvert \alpha_V \rvert\to \infty} b(\alpha_V) \to \dfrac{\alpha_V^4}{9\pi^4}\,,\\
    \lim_{\lvert \alpha_V \rvert\to \infty} c(\alpha_V) &\to \dfrac{32}{9\pi^4}e^{-\lvert \alpha_V \rvert}\,,
    \lim_{\lvert \alpha_V \rvert\to \infty} d(\alpha_V) = \dfrac{4}{27\pi^2}\,.
\end{align}
Consequently, as $\lvert \alpha_V \rvert \to \infty$, the system hosts one purely imaginary mode and two purely propagating modes with real-valued energy dispersion relations, respectively:
\begin{equation}
 \omega_{\rm im.} = -\dfrac{i\pi^2}{24\tau_H}e^{\lvert \alpha_V \rvert}\,, \qquad 
 \omega_\pm = \pm \dfrac{2\pi \Omega}{\alpha_V^2 T \sqrt{3}} k\,,
 \label{eq:QED_deg_omega}
\end{equation}
where we used the definition of $\kappa_\Omega$ in Eq.~\eqref{eq_kappa_Omega}.
From the above results, we find that as we increase the vector chemical potential, the non-propagating mode $\omega_{\rm im.}$ dissipates exponentially fast, whereas the phase speed $\omega_{\pm} / k = \pm 2\pi \Omega / \alpha_V^2 T \sqrt{3}$ of the other two modes vanishes quadratically, as $\alpha_V^{-2}$. This result is consistent with Fig.~\ref{fig:im}, which will be discussed in Sec.~\ref{sec:QED:spectra}.

\begin{figure*}[!ht]
    \centering
    \begin{tabular}{cc}
    \includegraphics[width=.46\linewidth]{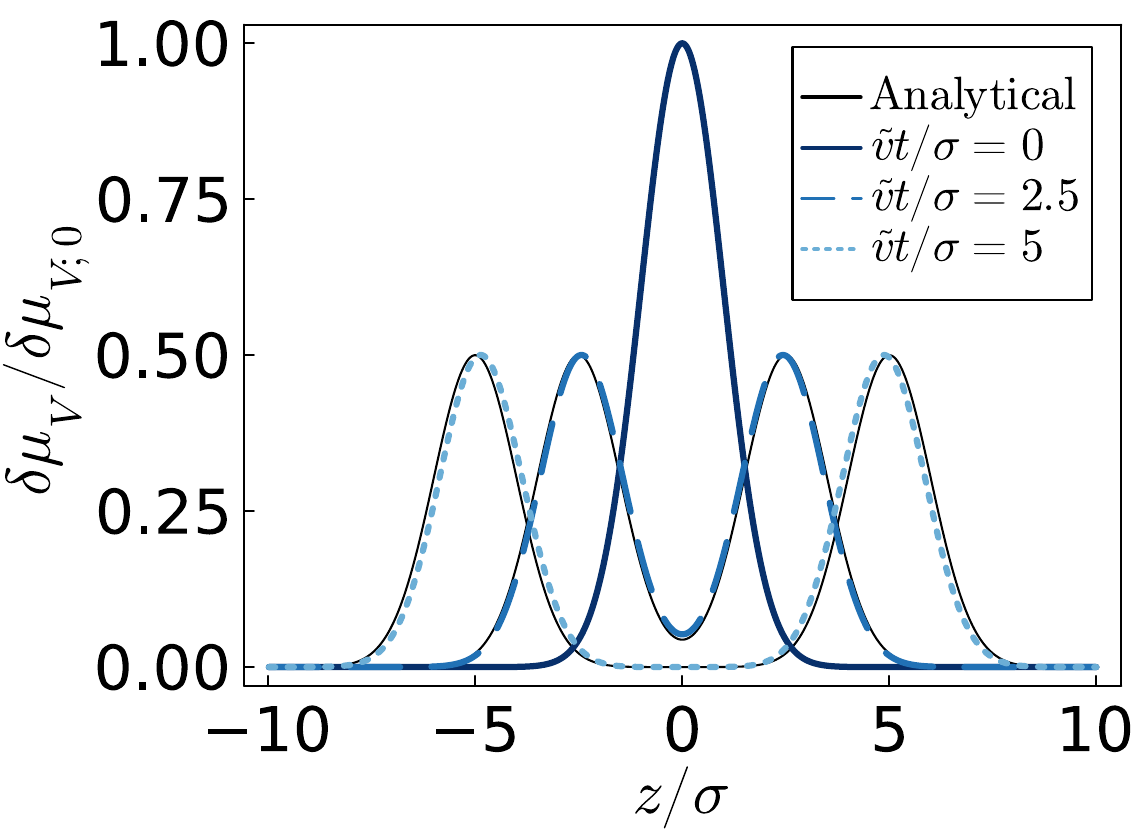} &
    \includegraphics[width=.46\linewidth]{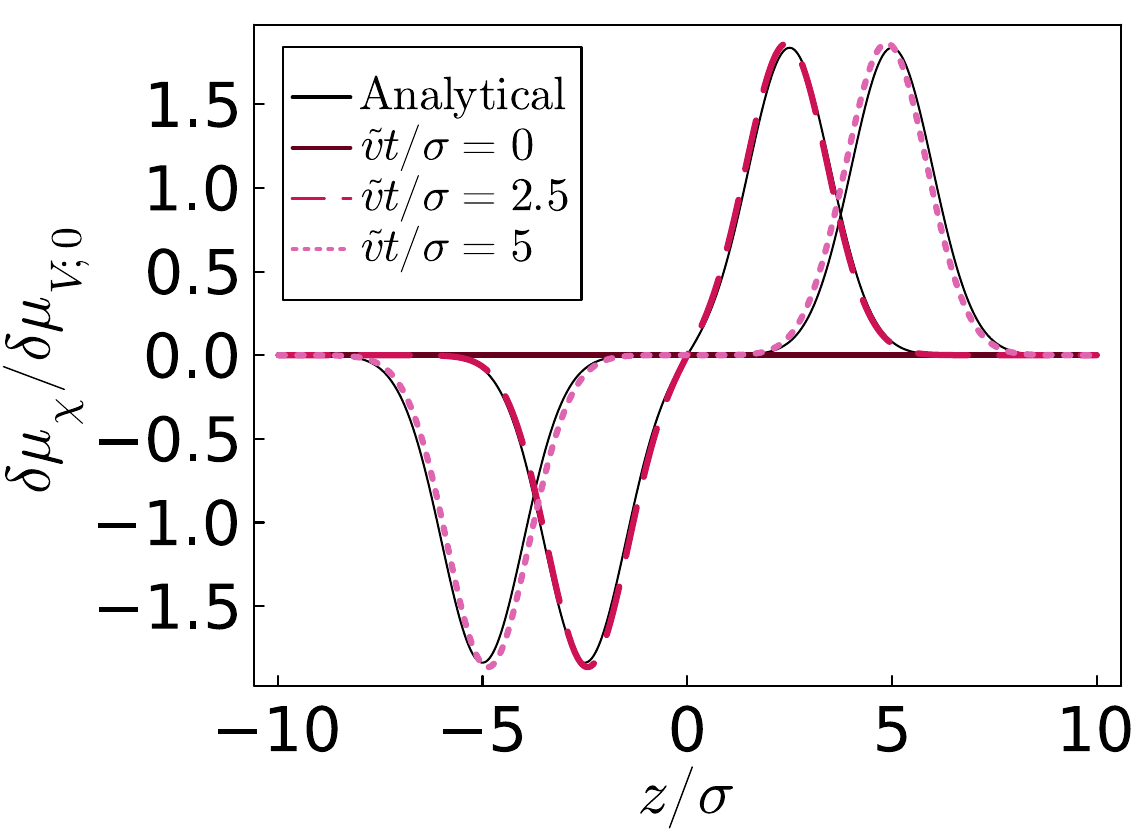}\\
   \includegraphics[width=.46\linewidth]{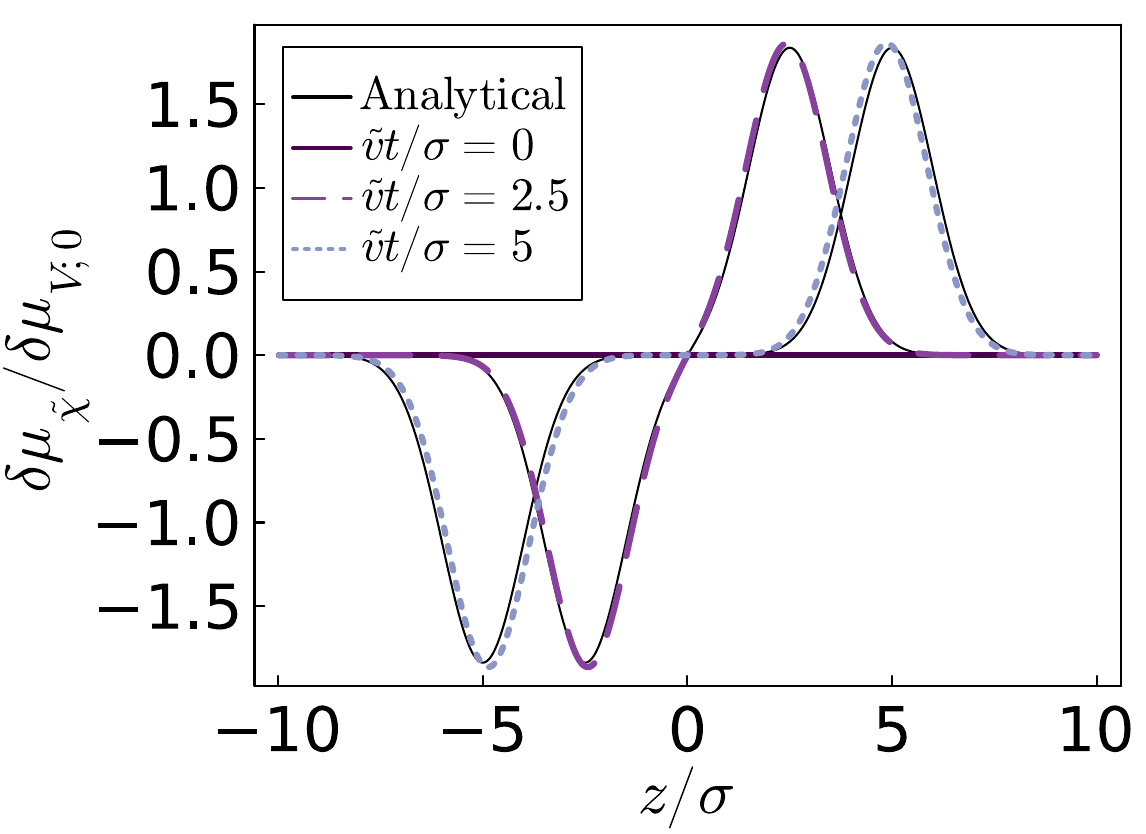} &
    \includegraphics[width=.46\linewidth]{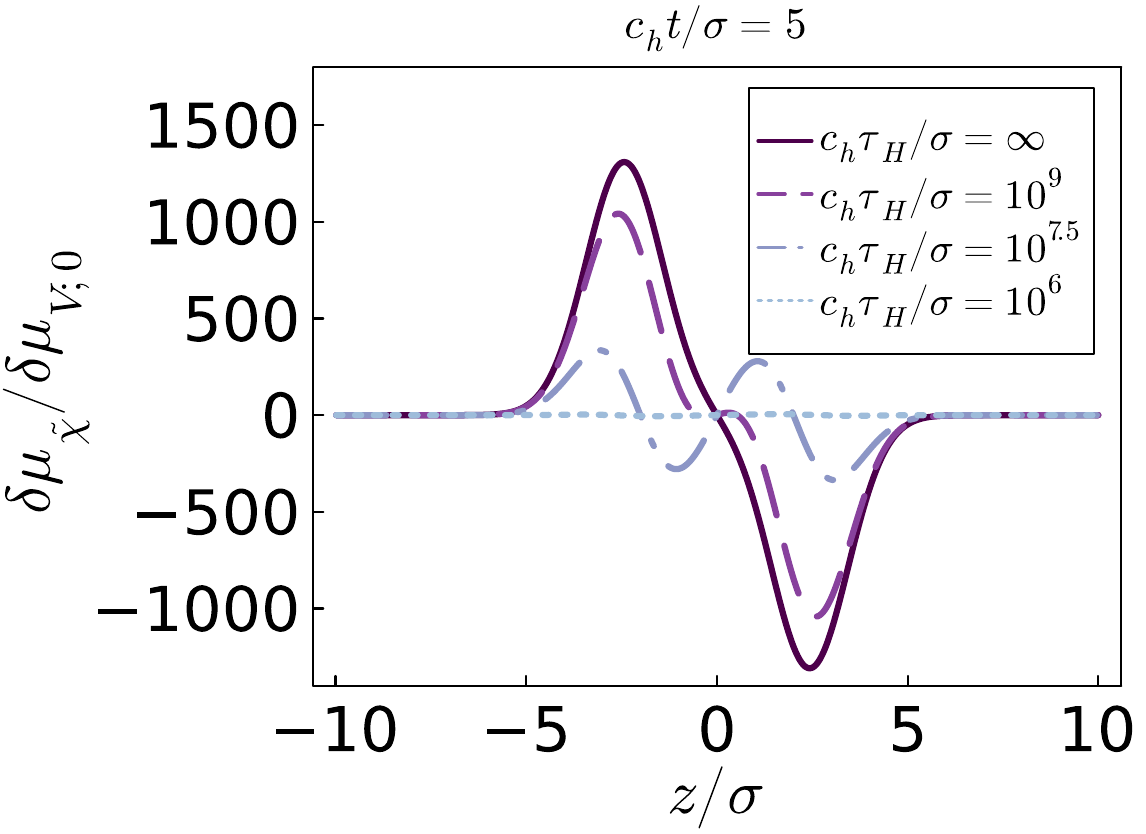}
    \end{tabular}
    \caption{Propagation of an initial Gaussian perturbation in the vector chemical potential, shown in Eq.~\eqref{eq:gaussV_init}, for the degenerate limit discussed in Sec.~\ref{sec:QED:largemu}. The profiles are represented at three time instances, corresponding to $\tilde{v} t / \sigma = 0$, $2.5$ and $5$, where the phase velocity of the propagating solution in the degenerate limit $\tilde{v}$ is given in Eq.~\eqref{eq_tilde_v_deg}. The background state has $\alpha_V=20$ and $\tau_H = 2 \sigma / \tilde{v}$. We also represented the analytical solution for the limit $\alpha_V \rightarrow \infty$, shown in Eq.~\eqref{eq:QED_deg_gauss}.
    The bottom right panel shows the dependence of $\delta \mu_{\tilde{\chi}}$ on $\tau_H$.
    \label{fig:QED_deg}
    }
\end{figure*}

The general solution for $\delta\mu_\chi = \delta \mu_A + s_V \delta\mu_H$ and $\delta\mu_{\tilde{\chi}} = \delta\mu_A - s_V \delta\mu_H$ for arbitrary vector chemical potential $\mu_V$, where $s_V = {\rm sgn}(\mu_V)$, and arbitrary helical relaxation time, is given by
\begin{align}
    &\delta\mu_{\chi}= \nonumber\\&  \frac{ \kappa_\Omega T^2  \left(A  T^2 \tau_H^{-1}-6 i (A+s_VB) \omega_*  (\sigma_A^\omega-s_V\sigma_H^\omega)\right)}{2 \omega_*  \left( \sigma_A^\omega T^2 \tau_H^{-1}-6 i \omega_*
   \left[(\sigma_A^\omega)^2-(\sigma_H^\omega)^2\right]\right)}\delta\mu_V\,,\nonumber\\
    &\delta\mu_{\tilde{\chi}}= \nonumber\\& \frac{ \kappa_\Omega T^2 \left(A  T^2 \tau_H^{-1}-6 i (A-s_VB) \omega_*  (\sigma_A^\omega+s_V\sigma_H^\omega)\right)}{2 \omega_* \left( \sigma_A^\omega T^2 \tau_H^{-1}-6 i \omega_* 
   \left[(\sigma_A^\omega)^2-(\sigma_H^\omega)^2\right]\right)}\delta\mu_V\,.
   \label{eq:QED_muchi}
\end{align}
From the previous expressions, it appears that the degenerate limit $\mu_V\to\infty$ and the helicity-conservation limit $\tau_H^{-1}\to 0$ do not commute. Naively, we could take $\lvert \mu_V \rvert/T\to \infty$ for some finite $\tau_H^{-1}$. In such case, the amplitudes for the propagating modes $\omega_\pm$ read
\begin{align}
    \delta\mu_{\chi}^{\pm} &\simeq \delta\mu_{\tilde{\chi}}^{\pm} \simeq \frac{ \kappa_\Omega T^2 }{2 \omega^\pm  }\dfrac{A}{\sigma^\omega_A}\delta\mu_V^\pm  \simeq \pm \dfrac{\lvert \alpha_V \rvert}{\pi \sqrt{3}} \delta\mu_V^\pm\,, \label{eq:QED_muchi_largemu}
%
\end{align}
where the factors of $\tau_H^{-1}$ have cancelled out and the result is well defined as $\tau_H^{-1}\to 0$. 

Regarding the exponentially dissipating modes, the denominator of the amplitudes \eqref{eq:QED_muchi} vanishes and we resort to the $\tau_A\to\infty$ limit of \eqref{eq:QCD_mu2}. Taking $\delta\mu_{\tilde{\chi}}$ as the reference perturbation we find:
\begin{align}\label{eq:dmu_im}
      \delta\mu_{\chi}^{\textrm{im.}} &\simeq -\dfrac{2}{\alpha_V^2}e^{-\lvert \alpha_V \rvert}\delta\mu_{\tilde{\chi}}^{\textrm{im.}}\,,\nonumber\\
      \delta\mu_V^{\textrm{im.}} &\simeq -\dfrac{72 i \kappa_{\Omega}\tau_H}{\pi^2 \alpha_V^2}e^{-2\lvert \alpha_V \rvert}\delta\mu_{\tilde{\chi}}^{\textrm{im.}}.
\end{align}


In Fig.~\ref{fig:QED_deg}, we show the evolution of the initial Gaussian profile of the vector chemical potential of Eq.~\eqref{eq:gaussV_init} in the nearly-degenerate case when $\alpha_V = 20$. The plots show the profiles of $\delta \bar{\mu}_V(t,z)$, $\delta \bar{\mu}_A(t,z)$ and $\delta \bar{\mu}_H(t, z)$ at initial time and at times satisfying $\tilde{v} t / \sigma = 2.5$ and $5$, where 
\begin{align}
    \tilde{v} = \frac{\lvert {\rm Re}(\omega_\pm)\rvert}{k} = \frac{2\pi \Omega}{\alpha_V^2 T \sqrt{3}}\,,
\label{eq_tilde_v_deg}
\end{align}
is the phase velocity of the propagating solution in the degenerate limit [see Eq.~\eqref{eq:QED_deg_omega}]. We also represent the analytical solution, which reads 
\begin{align}
 \delta \bar{\mu}_V(t,z) &= \frac{\delta \mu_V}{2} \left(e^{-(z - \tilde{v} t) / 2\sigma^2} + e^{-(z + \tilde{v} t) / 2\sigma^2}\right),\nonumber\\
 \delta \bar{\mu}_\chi(t,z) &= \frac{\lvert \alpha_V \rvert}{\sqrt{3\pi}} \frac{\delta \mu_V}{2} \left(e^{-(z - \tilde{v} t) / 2\sigma^2} - e^{-(z + \tilde{v} t) / 2\sigma^2}\right),
 \label{eq:QED_deg_gauss}
\end{align}
while $\delta\bar{\mu}_{\tilde{\chi}}(t,z) \simeq \delta \bar{\mu}_\chi(t,z)$. As shown in Fig.~\ref{fig:QED_deg}, the above solution provides a good match to the numerical solution in the considered limit at the normalized vector chemical potential $\alpha_V = \mu_V / T = 20$. The amplitude of $\delta \bar{\mu}_{\tilde{\chi}} = \delta \bar{\mu}_A - s_V \delta \bar{\mu}_H$ is finite, contrary to the degenerate limit of the case when the charge is conserved when the amplitude of this mode increases like $\alpha_V^4$ (see Fig.~5 and Eq.~(149) from Ref.~\cite{Morales-Tejera:2024uzg}). To demonstrate the regularization effect of $\tau_H$, panel (d) shows the decrease of the ratio $\delta \bar{\mu}_{\tilde{\chi}} / \delta \mu_V$ from $O(10^3)$ at $\tau_H \rightarrow \infty$ to $O(1)$ at $c_h\tau_H/\sigma = 10^6 $. This effect becomes apparent when $\tau_H$ is decreased below a value of the order of $e^{\lvert \alpha_V \rvert}$.

Prior to concluding this section, we address the non-commutativity of the large vector chemical potential ($\mu_V \to \infty$) and the conserved helicity ($\tau_H\to 0$) limits. To this end, we need to know the dependence of the helicity relaxation time on the vector chemical potential. This problem is addressed in Appendix \ref{app:tauH}, and a fit to numerical data suggests that the helicity relaxation time diverges in the degenerate limit according to  

\begin{align}
    \tau_H^{-1} (\mu_V\gg T) &= \frac{\tau^{-1}_{\textrm{deg.}}}{\lvert \alpha_V \rvert} e^{-\lvert \alpha_V \rvert}\,,
\end{align}
with $\tau_{\textrm{deg.}} \simeq 8.432 \times 10^{-3}\,{\rm fm}/c$. Taking this dependence into account, we can take the degenerate limit and the helicity conservation limit in a correlated manner. While the propagating modes $\omega_{\pm}$ remain unaffected, the non-propagating mode $\omega_{\textrm{im.}}$ becomes, to leading order, 

\begin{equation}
    \omega_{\textrm{im.}} = -\dfrac{i\pi^2}{24 \tau_{\textrm{deg.}}\lvert \alpha_V \rvert}\,.
\end{equation}
Therefore, the dissipation is inhibited as we increase the chemical potential, contrary to the naive $\alpha_V\to\infty$ limit in Eq. \eqref{eq:QED_deg_omega}. Finally, taking the limit for the amplitudes \eqref{eq:dmu_im} gives
\begin{align}
 \delta\mu_{\chi}^{\textrm{im.}} &\simeq -\dfrac{2}{\alpha_V^2}e^{-\lvert \alpha_V \rvert}\delta\mu_{\tilde{\chi}}^{\textrm{im.}}\,,\nonumber\\
 \delta\mu_V^{\textrm{im.}} &\simeq -\dfrac{72 i \kappa_{\Omega}\tau_{\textrm{deg}}}{\pi^2 \lvert \alpha_V \rvert}e^{-\lvert \alpha_V \rvert}\delta\mu_{\tilde{\chi}}^{\textrm{im.}}.
\end{align}
We conclude that the fluctuation amplitudes $\delta \mu_\chi^{\rm im}$ and $\delta \mu_V^{\rm im}$ are exponentially suppressed compared to the amplitude $\delta \mu^{\rm im}_{\tilde{\chi}}$. In the above limit, we obtain $\delta \mu_V^\pm \simeq \delta \mu_V/2$ and $\delta \mu_{\tilde{\chi}}^{\rm im} \simeq \delta \mu_{\tilde{\chi}} - \delta \mu_\chi$ and thus none of the modes have exponentially-large amplitude. The amplitude of the propagating modes $\delta \mu^\pm_\chi$ and $\delta \mu^\pm_{\tilde{\chi}}$ still grows linearly with 
$\lvert \alpha_V \rvert$ [see Eq.~\eqref{eq:QED_muchi_largemu}], substantially milder than the quartic dependence $\delta \mu^{h;\pm}_{\tilde{\chi}} \sim \alpha_V^4$ found when the helicity charge is strictly conserved [see Eq.~(149) in Ref.~\cite{Morales-Tejera:2024uzg}].

\subsection{Generic properties of spectra}\label{sec:QED:spectra}

In Subsec. \ref{sec:QED:smallmu} we have seen that, at small vector chemical potential, there is a critical wavenumber above which two of the three modes start to propagate. At the same time, for small wavenumbers at finite chemical potential, we showed in Subsec. \ref{sec:QED:cvw} that again, there are two modes that always propagate. These two results are compatible with each other only if, for small chemical potential, the modes that propagate at small $k$ turn into non-propagating modes at some finite $k$. We will shortly see that this is indeed the case. 

\begin{figure}[!ht]
    \centering
    \includegraphics[width=0.92\linewidth]{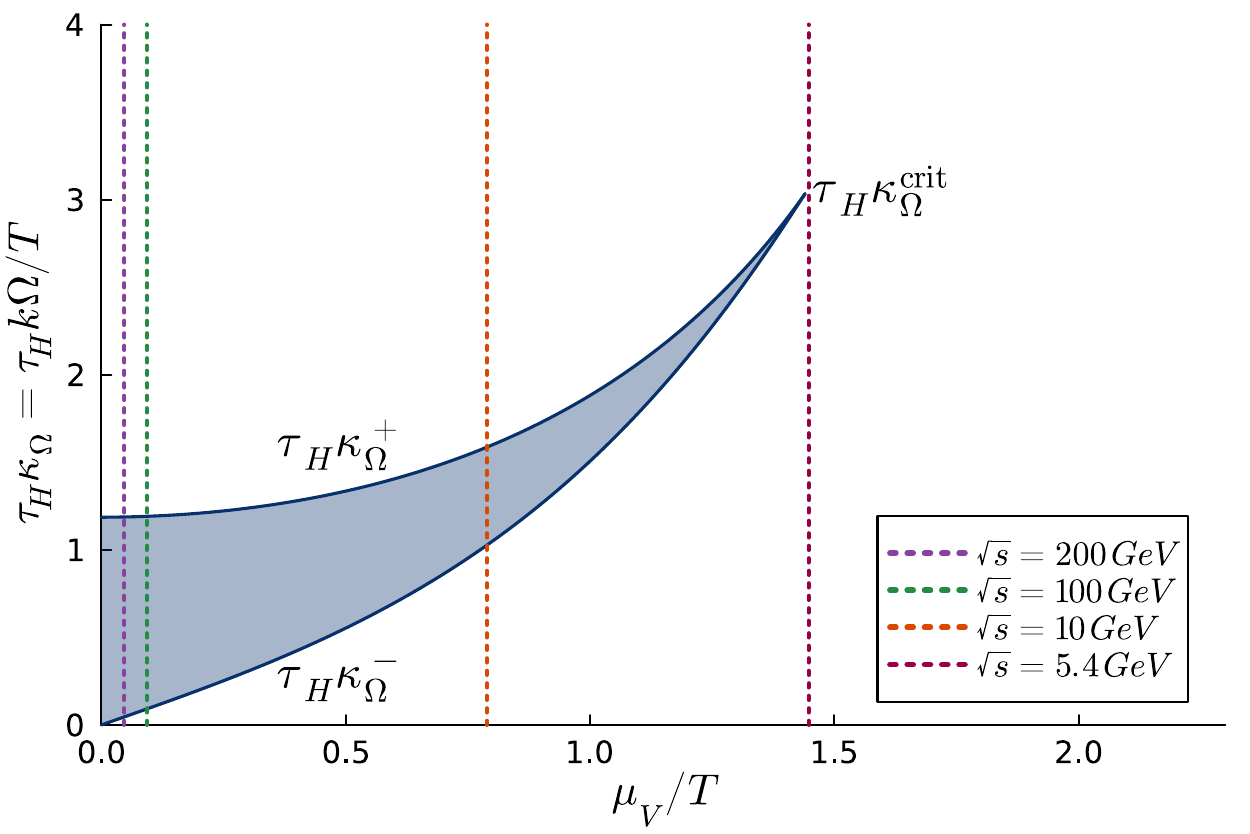}
    \caption{Propagating points $\kappa_\Omega^+$ (upper curve) and $\kappa_\Omega^-$ (lower curve) as functions of the normalized vector chemical potential $\alpha_V = \mu_V/T$ for the unpolarized plasma. In the shaded region, there are no propagating modes. The boundaries $\kappa_\Omega = \kappa_\Omega^\pm$ of the shaded region, defined in Eq.~\eqref{eq:QED_kappapm}, join each other at the critical value $\kappa_\Omega^{\rm crit.}$ given in Eq.~\eqref{eq:QED:kappa_crit}. The vertical lines mark the energy of the collision $\sqrt{s}$ corresponding to the given value of the normalized chemical potential $\alpha_V$ [see a discussion around Eq.~\eqref{eq:T_muB}].
    }
    \label{fig:k*}
\end{figure}

Propagating modes come in pairs with the same imaginary part, whereas non-propagating modes have a vanishing real part. Thus, one is able to compute the critical values $k_*$ at which modes start/stop propagating by looking at purely imaginary degenerate solutions of the characteristic equation, Eq.~\eqref{eq:QED_detM_largemu}. These can be obtained by setting to $0$ the derivative with respect to $\omega$ of the left-hand side of Eq.~\eqref{eq:QED_detM_largemu}:
\begin{equation}
 3a \omega^2 + \frac{2i b}{\tau_H} \omega - \kappa_\Omega^2 c = 0\,.
 \label{eq:QED_detM_largemu_lim}
\end{equation}
The solution of the above equation reads:
\begin{equation}
 \omega_\pm = -\frac{i b}{3a\tau_H} \left(1 \pm \sqrt{1 - \frac{3 a c}{b^2} (\tau_H\kappa_\Omega)^2}\right)\,.
 \label{eq:QED_omegapm}
\end{equation}
To find the critical wavenumbers $k_*$, we impose that both Eq.~\eqref{eq:QED_detM_largemu} and \eqref{eq:QED_detM_largemu_lim} hold simultaneously. Multiplying Eq.~\eqref{eq:QED_detM_largemu_lim} by $\omega/3$ and subtracting the result from Eq.~\eqref{eq:QED_detM_largemu} leads to
\begin{equation}
 \frac{ib}{3\tau_H} \omega_\pm^2 - \frac{2 c}{3} \kappa_\Omega^2 \omega_\pm - \frac{i}{\tau_H} \kappa_\Omega^2 d = 0\,.
\end{equation}
Multiplying again Eq.~\eqref{eq:QED_detM_largemu_lim} by $i b / 9 a \tau_H$ and subtracting the result from the above equation leads to the relation 
\begin{equation}
 \left(\frac{2b^2}{9a \tau_H^2} -\frac{2c}{3} \kappa_\Omega^2 \right) \omega_\pm = \frac{i}{\tau_H}\left(d - \frac{bc}{9a}\right) \kappa_\Omega^2\,.
\end{equation}
Substituting now the solution in Eq.~\eqref{eq:QED_omegapm} gives the following equation for $\tau_H \kappa_\Omega$:
\begin{align}
& 4a c^3 (\kappa_\Omega \tau_H)^6 - (b^2 c^2 + 18 abcd - 27 a^2 d^2) (\kappa_\Omega \tau_H)^4 \nonumber\\&+ 4 b^3 d (\tau_H \kappa_\Omega)^2 = 0.
\end{align}
The solution $\kappa_\Omega = 0$ corresponds to $\omega = 0$ and is thus spurious. The other two solutions for $\kappa_\Omega^2$ read 
\begin{align}
 (\kappa_\Omega^\pm)^2& = \frac{1}{8 a c^3 \tau_H^2} \left[b^2 c^2 + 18abcd - 27a^2 d^2\right. \nonumber\\&\left. \pm \sqrt{bc - ad} (bc - 9ad)^{3/2}\right].
 \label{eq:QED_kappapm}
\end{align}
In the case of a neutral background, when $\alpha_V = 0$, the coefficients $a$, $b$, $c$ and $d$ evaluate to
\begin{equation}
 a = b = \frac{1}{27}, \qquad 
 c = \frac{1}{3} \left(\frac{2 \ln 2}{\pi^2}\right)^2, \qquad 
 d = 0.
\end{equation}
In this case, Eq.~\eqref{eq:QED_kappapm} gives $(\kappa_\Omega^-)^2 = 0$, meaning no infrared modes can propagate. The second solution satisfies
\begin{equation}
 \tau_H \kappa_\Omega^+ = \frac{\pi^2}{12 \ln 2}.
\end{equation}
From here, we can recover the threshold wavenumber required for wave propagation, $k_{\rm th} = \kappa^+_\Omega T / \Omega = 1 / (2 c_h \tau_H)$, with $c_h = 6 \Omega \ln 2 / (\pi^2 T)$, as found in Eq.~\eqref{eq:QED_kth}. 

The trajectories of $\kappa_\Omega^\pm \equiv k^\pm \Omega/T$ as functions of $\alpha_V = \mu_V / T$ are shown in Fig.~\ref{fig:k*}. The shaded region has no propagating modes. At zero chemical potential, we recover the known result that the modes do not propagate until the critical wavenumber in Eq.~\eqref{eq:QED_kth} is reached, as demonstrated above.
As $\alpha_V$ is increased, both $\kappa_\Omega^+$ and $\kappa_\Omega^-$ increase, the latter at a faster rate. 
Close to $\alpha_V = 0$, one can estimate the leading-order increase as:
\begin{align}
 &\kappa_\Omega^- = \dfrac{\pi^2}{21 \tau_H \ln^2 2 } \alpha_V + O(\alpha_V^3), \nonumber\\&
 \kappa_\Omega^+ = \frac{\pi^2}{12\tau_H \ln 2} + \dfrac{1.135}{\tau_H} \alpha_V^2 + O(\alpha_V^4).
\end{align}
The region of no propagation shrinks down to a point when $\alpha_V$ reaches a threshold value, given by the equation $bc = 9ad$, namely
\begin{equation}
 (B^2 - 8A^2) (\sigma_A^\omega)^2 - 2 A B \sigma^\omega_A \sigma^\omega_H + 9A^2 (\sigma^\omega_H)^2 = 0.
\end{equation}
The above equation is solved when $\alpha_V \simeq 1.4471$, in which case $\kappa_\Omega = k \Omega / T$ evaluates to 
\begin{equation}
 \kappa_\Omega^{\rm crit} \simeq \frac{3.0642}{\tau_H}\,.
 \label{eq:QED:kappa_crit}
\end{equation}
All three modes merge at this critical value of $\kappa_\Omega$ for the given chemical potential.

\begin{figure*}[!ht]
\centering
\includegraphics[width=0.92\linewidth]{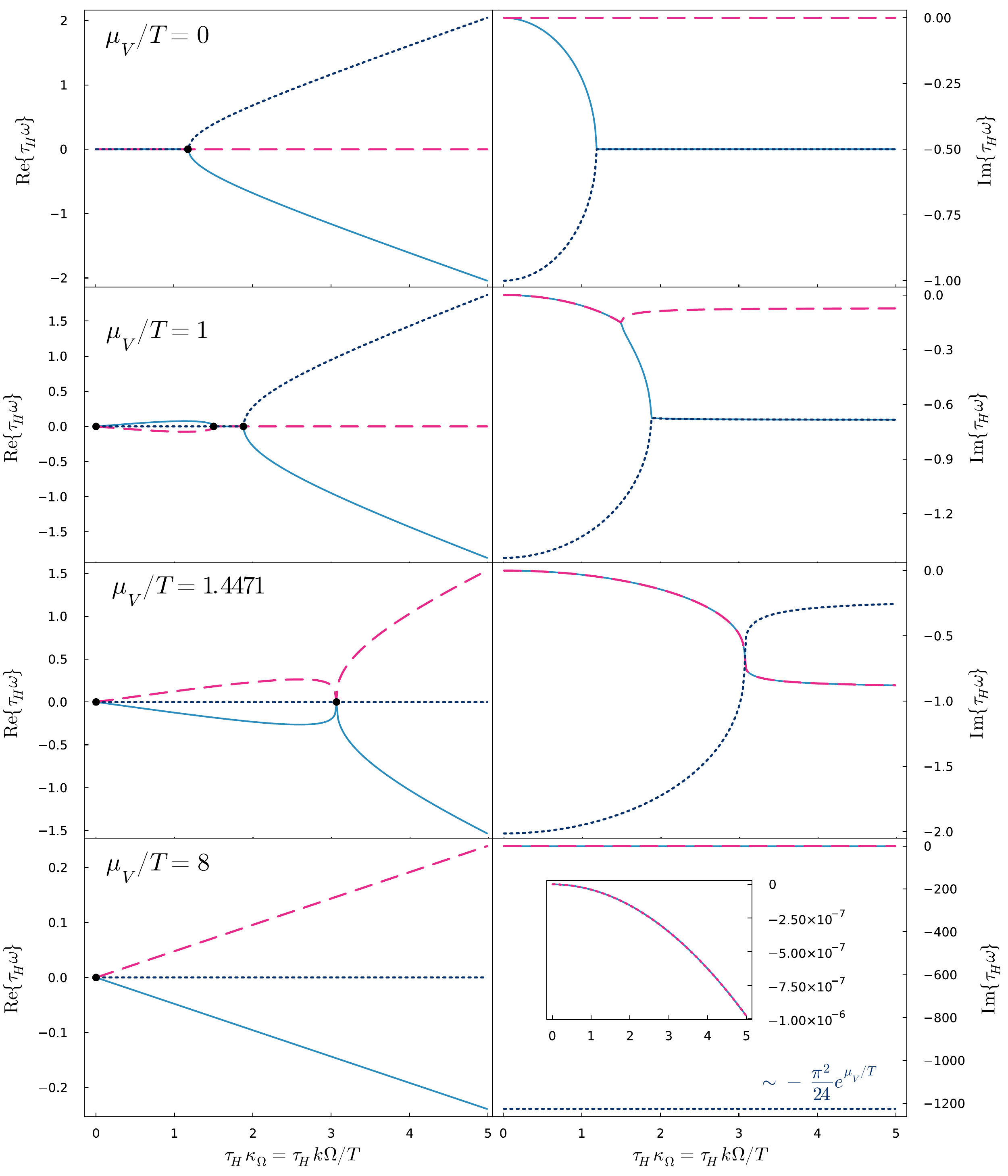}
\caption{\small Collective excitations of the charge sector in the unpolarized plasma with dissipating helical charge. Each row shows the mode spectrum for a different value of the dimensionless vector chemical potential, $\alpha_V = \mu_V / T$. The left (right) plots correspond to the real (imaginary) part of the dimensionless angular frequency, $ \tau_H \omega$. The black dots mark the points where the wave propagation starts or stops, i.e. when ${\rm Re}(\tau_H \omega)$ vanishes identically. Colors are correlated between the left and right plots. }
\label{fig:im}
\end{figure*}

The wave spectrum is shown as the real and imaginary parts of $ \tau_H \omega$ on the left and right columns of Fig.~\ref{fig:im}. The dimensionless chemical potential $\alpha_V = \mu_V / T$ increases from $0$ on the first line to $8$ on the last line. At $\alpha_V = 0$, ${\rm Re}(\tau_H \omega) = 0$ for $\tau_H \kappa_\Omega < \tau_H k_{\rm th} \Omega / T = \pi^2 / 12\ln 2 \simeq 1.18657$. At $\alpha_V = 1$, ${\rm Re}(\tau_H \omega)$ is nonvanishing for $0 < \kappa_\Omega < \kappa_\Omega^- \simeq 1.51 \tau_H^{-1}$ and when $\kappa_\Omega > \kappa_\Omega^+ \simeq 1.88 \tau_H^{-1}$, with an intermediate range of finite extent where there is no propagation. At the critical value $\alpha_V = 1.4471$, this range is reduced to a single point, $\tau_H \kappa_\Omega^+ = \tau_H \kappa_\Omega^- \equiv \tau_H \kappa_\Omega^{\rm crit} = 3.0642$. Otherwise, for $\alpha_V > 1.4471$, all modes are propagating.  

As we pointed out in section \ref{sec:QED:cvw}, the branch of propagating modes at a smaller wavenumber in Fig.~\ref{fig:im} can be thought of as the generalization of the Chiral Vortical Wave previously studied in \cite{Jiang:2015cva,Gorbar:2017toh} to fluids possessing an independent helical degree of freedom. On the other hand, the modes propagating at a larger wavenumber in Fig.~\ref{fig:im} correspond to the generalization of the Helical Vortical Wave found in \cite{Ambrus:2019khr} for finite vector chemical potential. As the chemical potential is increased, the modes propagate for any wavelength, and distinguishing between CVW and HVW modes becomes problematic.

Figure~\ref{fig:k*} deserves a further comment. In heavy-ion collisions, the ratio $\mu_V/T$ of the plasma is correlated with the energy of the collision $\sqrt{s}$. We can take the parametrization from~\cite{Cleymans:2005xv} for the following freeze-out values:
\begin{equation}
    T(\mu_B) = a-b\mu_B^2-c\mu_B^4\,, 
    \qquad
    \mu_B(\sqrt{s})=\dfrac{d}{1+f\sqrt{s}}\,,
    \label{eq:T_muB}
\end{equation}
where $\mu_B = 3 \mu_V$ and the fitted parameters are 
\begin{gather}
    a=0.166(2)\ \text{GeV}\,, \qquad b=0.139(16)\ \text{GeV}\,, \nonumber\\ c=0.053(21)\ \text{GeV}\,, \qquad
    d=1.308(28)\ \text{GeV}\,, \nonumber\\f=0.273(8)\ \text{GeV}^{-1}\,.
\end{gather}
We use this data set in Fig.~\ref{fig:k*} to show different collision energies, which are marked in the figure by vertical lines.
It can be seen that at $\sqrt{s} = 5.4$ GeV (or lower energies), the chemical potential is high enough that all wavelengths propagate. As the collision energy is increased, the chemical potential decreases, leading to the development of a non-propagation gap between the infrared ($\kappa_\Omega < \kappa_\Omega^-$) and ultraviolet ($\kappa_\Omega > \kappa_\Omega^+$) propagating modes.
The same considerations will apply to Fig.~\ref{fig:tricrit}.

\section{Non-conserved axial and helical charges}\label{sec:QCD}

In this section, we study the spectra of collective excitations when both the helical and the axial charges are not conserved. Following the lines of Eq.~\eqref{eq:dis2}, we include the dissipation of axial charge through its chemical potential:
\begin{equation}\label{eq:QCD_diss}
 \partial_{\mu}J^{\mu}_V =0\,, \
 \partial_{\mu}J^{\mu}_A =- \dfrac{\mu_A T^2}{3 \tau_A}\,,
    \ 
 \partial_{\mu}J^{\mu}_H = -\dfrac{\mu_H T^2}{3 \tau_H}\,.
\end{equation}

The modes supported by Eqs.~\eqref{eq:QCD_diss} can be obtained by solving the characteristic equation, ${\rm det}(T^{-2} \mathbb{M}) = 0$, which reduces to Eq.~\eqref{eq:QCD_detM}. In what follows, we solve this equation in various limits: the small chemical potential in Sec.~\ref{sec:QCD:smallmu}; small axial relaxation times $\tau_A, \tau_H \rightarrow 0$ in Sections~\ref{sec:QCD:tauA} and \ref{sec:QCD:tauH}; the large wavenumber limit $k \rightarrow \infty$ in Sec.~\ref{sec:QCD:largek}; and the large chemical potential limit $\lvert \mu_V \rvert \gg T$ in Sec.~\ref{sec:QCD:largemu}. We summarize our findings in Sec.~\ref{sec:QCD:spectra}

\subsection{Small chemical potential}\label{sec:QCD:smallmu}

Writing $\omega = \omega_0 + \omega_1 \alpha_V + \omega_2 \alpha_V^2 + \dots$ as in Eq.~\eqref{eq:QED_w_expansion},
the small chemical potential limit of Eq.~\eqref{eq:QCD_detM} extends Eq.~\eqref{eq:QED_w0_eq} to
\begin{equation}
 \frac{1}{27} \left(\frac{i}{\tau_{A}} + \omega_0\right) \left[\omega_0\left(\frac{i}{\tau_H} +\omega_0\right) - k^2 c_h^2 \right] = 0.
\end{equation}
While the pair of helical modes, $\omega^\pm_{h;0}$, remain unchanged with respect to Eq.~\eqref{eq:QED_kth}, the axial mode in Eq.~\eqref{eq:QED_w2a} now picks up a zeroth-order dissipative contribution:
\begin{equation}
 \omega_{\AVW;0} = -\frac{i}{\tau_{A}}, \qquad 
 \omega^\pm_{h;0} = -\frac{i}{2\tau_H} \pm k c_h \sqrt{1 - \frac{k_{\rm th}^2}{k^2}},
 \label{eq:QCD_w0h}
\end{equation}
with $k_{\rm th} = 1 /(2 \tau_H c_h)$ given in Eq.~\eqref{eq:QED_kth}. Compared to the situation encountered in Sec.~\ref{sec:QED}, we see that the axial mode is now damped on a time scale given by $\tau_A$. The propagation properties of the helical vortical wave, uncovered in Eq.~\eqref{eq:QED_kth}, remain identical as long as $\mu_V / T = 0$.

At first order with respect to $\alpha_V = \mu_V / T$, all three modes receive vanishing contributions,
\begin{equation}
 \omega_{\AVW;1} =\omega^\pm_{h;1} = 0,
\end{equation}
while at second order, Eq.~\eqref{eq:QED_w2a} gets modified to
\begin{multline}
 \omega_{\AVW;2} = \frac{3i}{\pi^2 \tau_{A}(\tau_{A} - \tau_H - \tau_H\tau_{A}^2 k^2 c_h^2)} 
 \\\times \left[(\tau_{A} - \tau_H)\left(1 + \frac{12}{49 \pi^2} \kappa_\Omega^2 \tau_{A}^2\right) \right.\\
 \left. + 
 \tau_H\frac{48}{\pi^2} (\ln 2)^2 \left(1 + \frac{3}{28 \pi^2} \kappa_\Omega^2 \tau_{A}^2\right)\right].
 \label{eq:QCD_w2a}
\end{multline}
Recall that the small chemical potential expansion in the conserved axial charge ($\tau_A \rightarrow \infty$) case does not work for small wavenumbers~$k$. Now, the situation is slightly different: the expansion is unreliable whenever the denominator appearing in the expression for $\omega_{\AVW;2}$ vanishes, and the small $\alpha_V$ expansion breaks down. This happens at the ``breakdown'' wavenumber,
\begin{equation}
 k_{\rm br.} = \dfrac{1}{c_h \sqrt{\tau_A}}\sqrt{\dfrac{1}{\tau_H}-\dfrac{1}{\tau_A}}
 \,, \qquad \tau_H < \tau_A \,.
 \label{eq:QCD_kstar}
\end{equation}
Note that in the limit where the axial charge is conserved ($\tau_A \to \infty$), the breakdown wavenumber vanishes, $k_{\rm br.}\to 0\,$. The same happens when $\tau_A = \tau_H$. 
On the other hand, the value of $k_{\rm br.}$ reaches a maximum for $\tau_A=2 \tau_H$, giving $k_{\rm br.} = 1/(2c_h \tau_H)$, which curiously coincides with the threshold wavenumber $k_{\rm th}$ at which the helical wave starts to propagate, derived in Eq.~\eqref{eq:QED_kth}. 

The breakdown of the small chemical potential expansion was already encountered in Sec.~\ref{sec:QED:smallmu} for conserved axial charge at the level of the fluctuation amplitudes (c.f.~Eq. \eqref{eq:QED_a_dmu}), giving $k_{\rm br.}=0$. 
A closer look at the small wavenumber limit in Sec. \ref{sec:QED:cvw} revealed that the axial modes were propagating in the presence of a finite vector chemical potential $\mu_V$ (see Eq. \eqref{eq:omega:AVW}). The propagating nature of the axial modes was not captured in the small $\mu_V$ expansion of Sec. \ref{sec:QED:cvw}. Analo\-gously, at finite $\tau_A$ and $\mu_V$, we anticipate that the axial modes may be propagating in the vicinity of $k_{\rm br.}$. This intuition will be confirmed with the explicit computation of the spectrum in Sec. \ref{sec:QCD:spectra} (see Fig. \ref{fig:imqcd}).

\subsection{Small \texorpdfstring{$\tau_{A}$}{\texttau A}  limit}\label{sec:QCD:tauA}

We can study the limit when the axial charge dissipates much faster than the helical charge. This is equivalent to freezing the axial degree of freedom. In this limit, the axial mode dissipates very quickly, according to the dispersion relation:
\begin{equation}
 \omega_{\AVW} \simeq -\frac{i T^2 \sigma^\omega_A}{6 \tau_{A} [(\sigma^\omega_A)^2 - (\sigma^\omega_H)^2]}\,,
\end{equation}
whereas the dispersion relation for the vector and helical fluctuations is
\begin{align}\label{eq:QCD_HVW_noA}
   &\omega^\pm_{h;0} \simeq\nonumber \\ & \frac{T^2}{12\tau_H\sigma^\omega_A} \left(-i \pm \sqrt{\frac{36 B^2 \sigma^\omega_A}{H(\sigma^\omega_A - \frac{T^2}{3} \Delta H)} \tau_H^2 \kappa_\Omega^2 - 1}\right).
\end{align}
Therefore, the vector and helical fluctuations propagate provided that the normalized wavenumber $\kappa_\Omega = k \Omega / T$ [see Eq.~\eqref{eq_kappa_Omega}] exceeds some critical value satisfying
\begin{equation}
    \tau_H^2 \kappa_{\Omega;*}^2 = \frac{H}{36 B^2 \sigma^\omega_A} \left(\sigma^\omega_A - \frac{T^2}{3} \Delta H\right).
    \label{eq:QCD_tauA0_kth}
\end{equation}
The corresponding threshold wavenumber ranges from $k_* = T \kappa_{\Omega;*} / (\tau_H \Omega) = 1 / (2 \tau_H c_h)$ at vanishing chemical potential [recovering Eq.~\eqref{eq:QED_kth}] to 
a value $k_* = T \kappa_{\Omega;*} / \Omega$ satisfying
$\tau_H \kappa_{\Omega;*} = \pi / 4\sqrt{3} \simeq 0.45$ at infinite chemical potential. Note that propagation is inhibited for small wavenumbers at any value of the vector chemical potential, contrary to the situation found for conserved axial charge ($\tau_A\to\infty$) in Sec. \ref{sec:QED:spectra} (see Eq.~\eqref{eq:QED_kappapm} and Fig.~\ref{fig:k*}) where any non-zero chemical potential gave rise to propagating modes in the infrared region with large wavelengths.

\begin{figure}[ht]
    \centering
    \includegraphics[width=.92\linewidth]{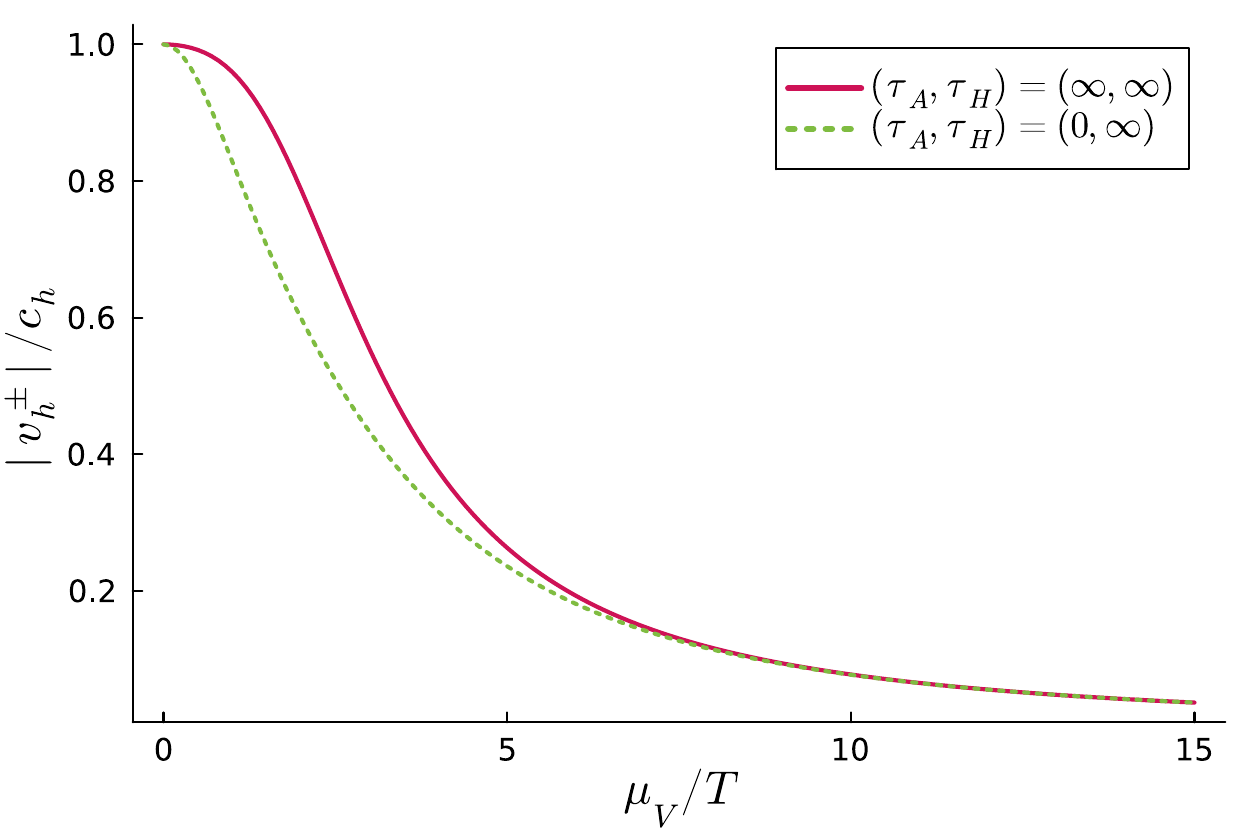}
    \caption{Propagation velocity for the helical vortical wave in the limit of conserved helicity for the cases when chirality is either conserved or frozen.}
    \label{fig:Comparison_velocities}
\end{figure}

In the case of conserved helicity, when $\tau_H\to\infty$, the helical vortical modes are propagating for any wavelength, and their dispersion relation Eq.~\eqref{eq:QCD_HVW_noA} reduces to 
\begin{equation}
    \lim_{(\tau_A,\tau_H)\to(0,\infty)}\omega_{h:0}^\pm = \pm \frac{\kappa_\Omega B T^2}{2\sqrt{H\sigma^\omega_A(\sigma^\omega_A - \frac{T^2}{3} \Delta H)} } \,.
\end{equation}
In this limit, the modes are qualitatively similar to the helical vortical modes found for the unpolarised plasma with conserved charges ($\tau_A,\tau_H\to \infty$) given in \eqref{eq:cons_omega}. The small vector chemical potential limit of both expressions agree quantitatively in the leading order but differ thereafter:
\begin{subequations}
\begin{align}
    \lim_{\substack{\tau_A \rightarrow 0 \\ \tau_H \rightarrow \infty}} \omega_{h:0}^\pm (\mu_V\ll T) &=\pm c_h k \mp 0.0978\dfrac{k \Omega}{T}\alpha_V^2 \nonumber\\
    & \hspace{.2\linewidth} +O(\alpha_V^4)\,,\\ 
    \lim_{\substack{\tau_A \rightarrow \infty \\ \tau_H \rightarrow \infty}} \omega_{h:0}^\pm (\mu_V\ll T) &=\pm c_h k \mp 0.0124\dfrac{k \Omega}{T}\alpha_V^2 \nonumber\\
    & \hspace{.2\linewidth} +O(\alpha_V^4)\,,
\end{align}
\end{subequations}
where we have used the expansions given in Eq.~\eqref{eq:unpol_largeT} and the definitions for $\kappa_\Omega$ and $c_h$ given in Eqs.~\eqref{eq_kappa_Omega} and \eqref{eq:ch}. In the large chemical potential limit, we can employ the relations \eqref{eq:unpol_largemu} to show that both dispersion relations agree quantitatively up to exponentially small corrections. In Fig.~\ref{fig:Comparison_velocities}, we show the propagation velocity $v^{\pm}_h \equiv \omega_{h;0}^\pm /k$ for conserved helicity in both cases, i.e. conserved axial charge $\tau_A\to\infty$ and frozen axial degree of freedom $\tau_A\to0$. Notably, the freezing of the axial degree of freedom results in a diminishing of the propagation velocity at finite vector chemical potential as opposed to the case when chirality is present and conserved.

\subsection{\texorpdfstring{Small $\tau_H$}{\texttau H} limit}
\label{sec:QCD:tauH}

Similarly to the previous subsection, we can study the spectra when the fluctuations of the helical degree of freedom dissipate very quickly, implying, in other words, that this degree of freedom is frozen.
The analysis in this section extends that of Sec.~\ref{sec:QED:cvw} to the case of a finite (non-infinite) axial relaxation time $\tau_A$. As discussed in Sec.~\ref{sec:QED:cvw}, freezing the helicity degree of freedom opens up the propagation channel for the chiral vortical wave. On the other hand, a finite helical relaxation time inhibits the propagation of the helical vortical wave at large wavelengths. We therefore anticipate that, in a similar manner, a finite axial relaxation time $\tau_A$ will inhibit the propagation of the axial vortical wave at large wavelengths.

Multiplying Eq.~\eqref{eq:QCD_detM} by $\tau_H^3$ and considering that $\tau_H \omega$ is finite when $\tau_H \rightarrow 0$, we find the purely dissipative mode:
\begin{equation}
 \omega_h \simeq -\frac{i T^2 \sigma^\omega_A}{6 \tau_H [(\sigma^\omega_A)^2 - (\sigma^\omega_H)^2]},
\end{equation}
which agrees with Eq.~\eqref{eq:QED_smallk_w0h}.
The angular frequencies for the other two(axial) modes, $\omega^\pm_{\AVW}$, are finite when the helical mode freezes, $\tau_H \rightarrow 0$, satisfying
\begin{multline}
    \omega^\pm_{\AVW} \simeq 
    \frac{T^2}{12 \sigma^\omega_A \tau_A} \\\times
    \left(-i \pm \sqrt{\frac{36 A^2 \sigma^\omega_A}{H(\sigma^\omega_A - \frac{T^2}{3} \Delta H)} \tau_A^2\kappa_\Omega^2 - 1}\right)\,.
    \label{eq:QCD_tauH0_kth}
\end{multline}
The above expression generalizes Eq.~\eqref{eq:omega:AVW}, giving the energy dispersion relation for the CVW to the case when the axial charge is not conserved. As already noted in Sec.~\ref{sec:QED:cvw}, the vector and axial fluctuations do not propagate at vanishing chemical potential. At finite chemical potential 
and finite (i.e., non-infinite) $\tau_A$, they propagate provided that $k$ exceeds some critical value. Using the large temperature expansion \eqref{eq:unpol_largeT}, we find that the critical value behaves as $\tau_A \kappa_* = 7\pi^2 / (12\alpha_V)$ for small chemical potential. In contrast, it finds a minimum in the limit of infinite chemical potential: $\tau_{A} \kappa_*^{\rm min} = k_*^{\rm min} \tau_A \Omega / T = \pi / 4\sqrt{3} \simeq 0.45$, which coincides with the large chemical potential limit of Eq.~\eqref{eq:QCD_tauA0_kth} giving the propagation threshold in the case when $\tau_A \rightarrow 0$. Since the helical and axial degrees of freedom correspond to the same physical quantity at infinite vector chemical potential, it is reasonable that freezing of either the helical or the axial chemical potentials gives the same result in this degenerate limit.

\subsection{Large wavenumber limit}\label{sec:QCD:largek}

As seen in Sec.~\ref{sec:QCD:smallmu}, a finite (non-infinite) relaxation time brings significant modifications in the wave spectrum for small wavenumbers $k$ [cf.~\eqref{eq:QCD_w0h}]. We now consider the opposite limit of large wavenumbers (small wavelengths) in order to demonstrate that the relaxation time of the axial and/or helical charges does not affect the wave spectrum in the UV limit. Specifically, we find
\begin{align}
 \omega^\pm_h &= \pm k  c_h(\alpha_V) + O(k^0)\,, \nonumber\\ 
 c_h(\alpha_V) &= \frac{\Omega T}{2} \sqrt{\frac{\sigma^\omega_A(A^2 + B^2) - 2 A B \sigma^\omega_H}
 {H(\sigma^\omega_A - \frac{T^2}{3} \Delta H)[(\sigma^\omega_A)^2 - (\sigma^\omega_H)^2]}}\,,
\end{align}
where $c_h(\alpha_V)$ is the velocity of the helical vortical wave in an unpolarized fluid with vector imbalance $\mu_V = \alpha_V T$, in the absence of helicity- or chirality-violating interactions, as obtained in Eq.~(123) of Ref.~\cite{Morales-Tejera:2024uzg}. The axial mode turns out to be purely dissipative:
\begin{equation}
 \omega_\AVW = -\frac{i T^2}{6 \tau_{A}\tau_H} \frac{B^2\tau_H + A^2 \tau_{A}}{(A^2 + B^2) \sigma^\omega_A - 2 A B \sigma^\omega_H} + O(k^{-1}).
\end{equation}
At large wavenumbers, the phase velocity of the axial mode is imaginary and its imaginary part will settle to a constant negative value, which depends on the dimensionless vector chemical potential $\alpha_V = \mu_V / T$ and on the axial and helicity relaxation times. These results will be verified in Figure~\ref{fig:imqcd}.

\subsection{Large chemical potential limit}\label{sec:QCD:largemu}

In order to study the large chemical potential limit, we solve the eigenvalue equation Eq.  \eqref{eq:QCD_detM} perturbatively in $T/\lvert \mu_V \rvert$. Using the expansion in Eq.~\eqref{eq:unpol_largemu}, we can compute the energy dispersion relation as a Taylor series in $e^{-\lvert \mu_V \rvert/T}$. Expanding
\begin{equation}
    \omega = \omega_{-1} e^{\lvert \mu_V \rvert/T} + \omega_0 + O(e^{-\lvert \mu_V \rvert/T})\,,
\end{equation}
the characteristic equation Eq. \eqref{eq:QCD_detM} becomes, to leading order,
\begin{multline}
 \dfrac{4i \sigma_A}{3 T^4}  \left(\sigma_A-\frac{1}{3}T^2 \Delta H\right)\left(\frac{1}{\tau_A} + \frac{1}{\tau_H}-\frac{24 i}{\pi^2} \omega_{-1}\right)\\ 
 \times e^{2\lvert \mu_V \rvert/T} \omega_{-1}^2 + O(e^{\lvert \mu_V \rvert/T}) = 0\,.
\end{multline}
Therefore, we find
\begin{align}
    &\omega_{-1}^\pm = 0 \hspace{0.5cm} \textrm{(double root)}\,, \nonumber\\& \omega_{-1}^{\textrm{im.}} = -\dfrac{i \pi^2}{24}\left(\frac{1}{\tau_H}+\frac{1}{\tau_A}\right)\,.
\end{align}
The contributions $\omega_0^\pm$ to modes with $\omega_{-1}^\pm = 0$ come from the next-to-leading order contribution to Eq. \eqref{eq:QCD_detM}, namely
\begin{multline}
 -\frac{i A^2 \kappa_\Omega^2}{3H} \left(\frac{1}{\tau_A} + \frac{1}{\tau_H}\right) + 
 \frac{2i \omega_0^\pm}{9 T^2} \left(\sigma_A^\omega - \frac{T^2}{3} \Delta H\right) \\\times 
 \left[\frac{6 \sigma^\omega_A}{T^2}\left(\frac{1}{\tau_A} + \frac{1}{\tau_H}\right) \omega_0^\pm + \frac{i}{\tau_A \tau_H}\right] = 0.
\end{multline}
Expanding $\omega_0$ in powers of $T/\lvert \mu_V \rvert$ gives, to leading order in $T / \lvert \mu_V \rvert$:
\begin{multline}\label{eq:QCD_w:largemu}
   \omega_\pm \simeq \dfrac{-i\pi^2 T^2}{6(\tau_A+\tau_H) \mu_V^2} \\\times
    \left(1\pm \sqrt{1-\frac{48}{\pi^2}(\tau_A+\tau_H)^2\kappa_\Omega^2}\right).
\end{multline}
Note that when both relaxation times are finite, there is no propagation for small wavenumbers. This is in contrast to the result obtained in Subsec.~\ref{sec:QED:largemu}, where we considered $\tau_A=\infty$, and the analogous modes propagate for all wavelengths. Notably, the $\pm$ modes are sensitive to an emergent effective relaxation time $\tau_{\rm eff}=\tau_A + \tau_H$ which can be traced to the interdependence of the axial and helical degrees of freedom in the degenerate limit. From Eq.~\eqref{eq:QCD_w:largemu}, we extract the critical wavenumber $k_*$ above which these two modes start propagating:
\begin{equation}
 k_* = \dfrac{\kappa_\Omega^* T}{\Omega} = \frac{\pi}{4\sqrt{3}}\frac{T}{(\tau_A+\tau_H)\Omega}\,.
 \label{eq:k_yet_another_star}
\end{equation}

\begin{figure}[t]
    \centering
    \begin{tabular}{c}
    \includegraphics[width=.89\linewidth]{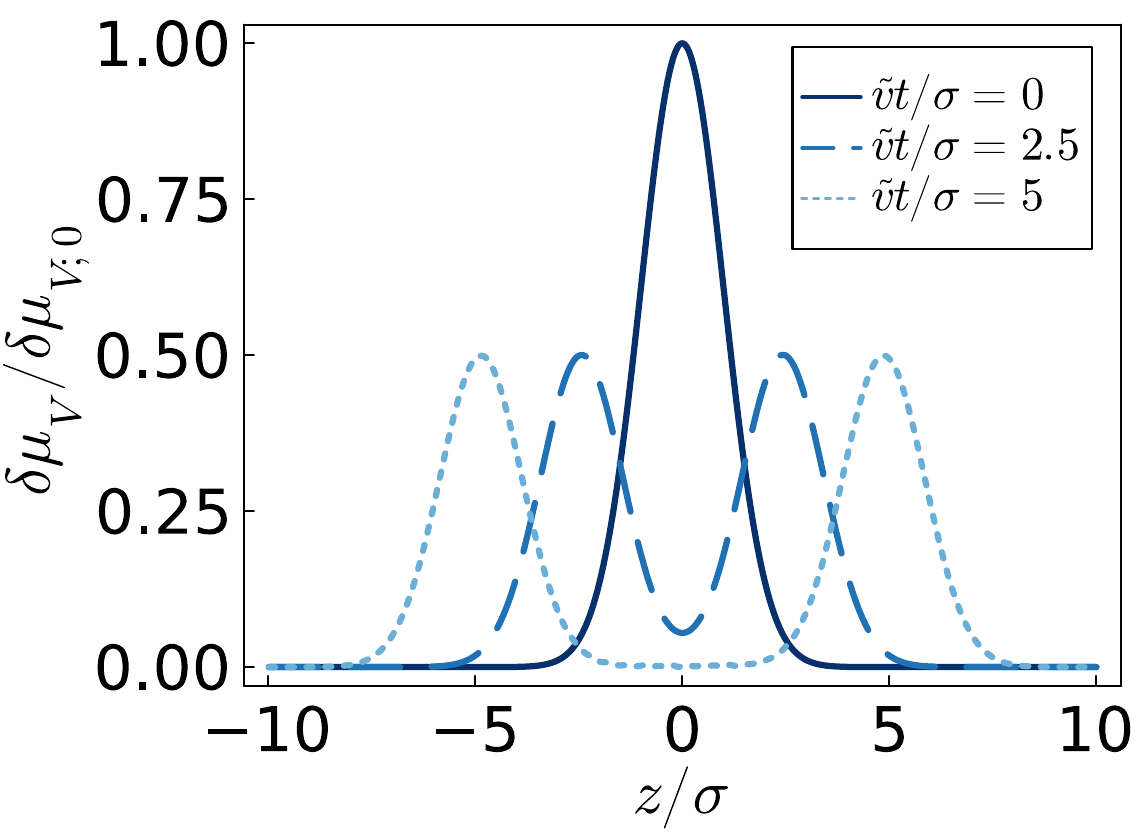}\\
    \includegraphics[width=.89\linewidth]{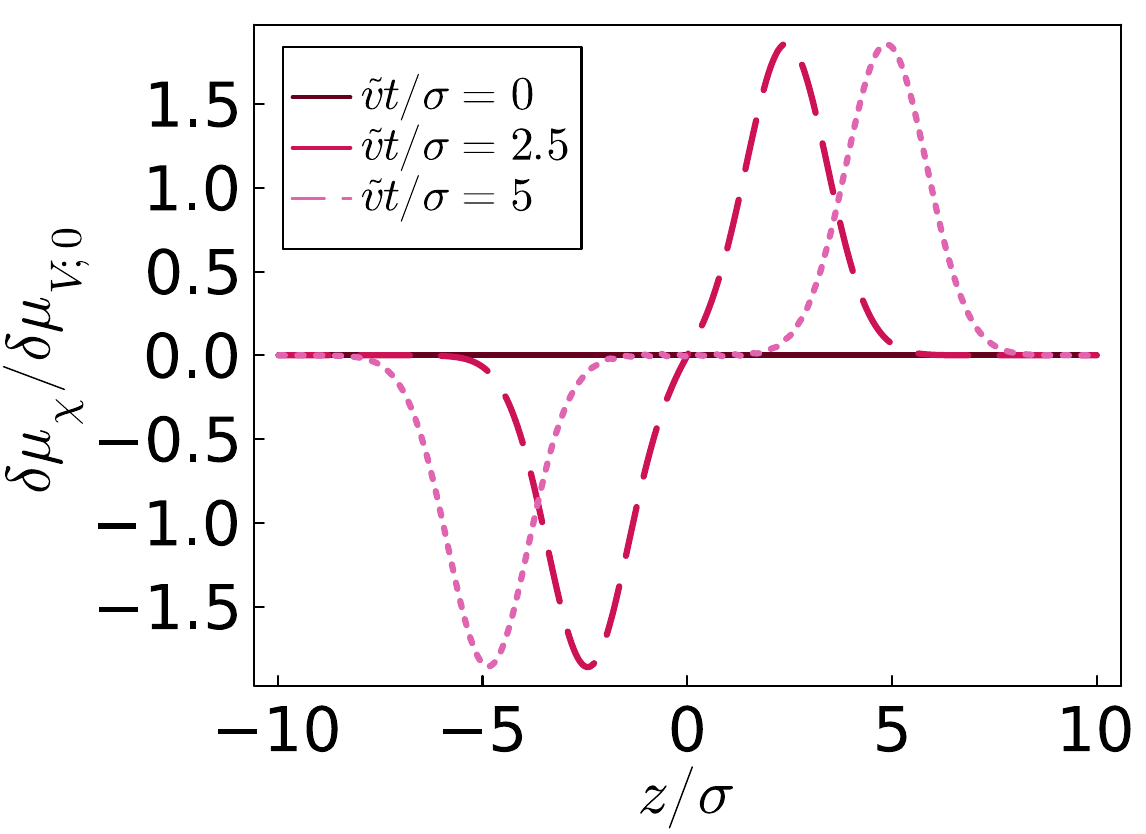}\\
    \multicolumn{1}{c}{\includegraphics[width=.89\linewidth]{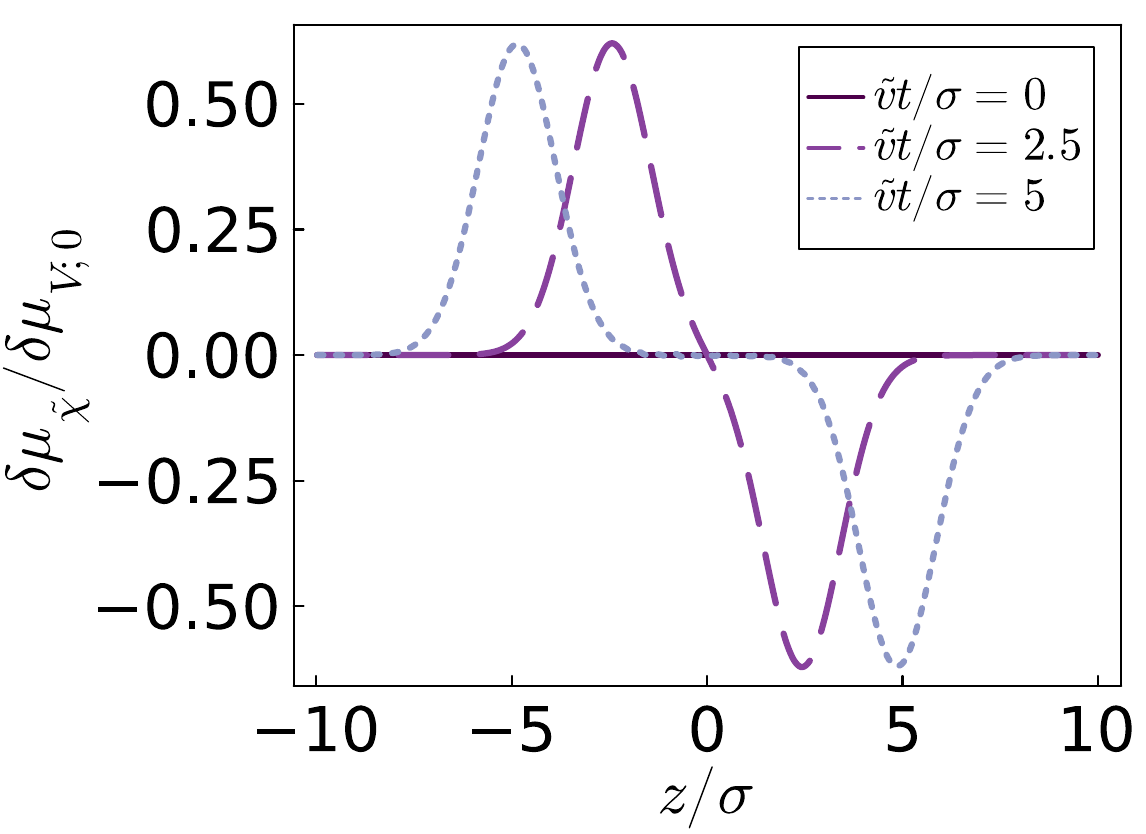}}
    \end{tabular}
    \caption{Same as Fig.~\ref{fig:QED_deg}, for the case when the axial charge is not conserved. For definiteness, we took $\tilde{v} \tau_A/\sigma = 1$, such that $\tau_A = \tau_H / 2$, where $\tilde{v}$ is defined in Eq.~\eqref{eq_tilde_v_deg}, with the angular velocities $\omega_\pm$ given by Eq.~\eqref{eq:QCD_w:largemu}.
    \label{fig:QCD_deg}
    }
\end{figure}

We now consider the amplitudes of the perturbations in the vector, axial, and helical chemical potentials. As already discussed in Sec.~\ref{sec:QED:largemu}, it is convenient to discuss the combinations $\delta\mu_{\chi} = \delta\mu_A + s_V \delta\mu_H$ and $\delta\mu_{\tilde\chi} = \delta\mu_A - s_V \delta\mu_H$, where $s_V = {\mathrm{sgn}\, \mu_V}$ is the sign of the vector chemical potential, that  follow from Eq.~\eqref{eq:QCD_mu}:
\begin{align}
 &\delta\mu^*_\chi = \kappa_\Omega \delta \mu^*_V \times \nonumber\\& \dfrac{\frac{2\omega_*}{T^2} (A + s_V B)(\sigma^\omega_A - s_V \sigma^\omega_H) + \frac{i}{3} \left(\frac{A}{\tau_H} + \frac{s_V B}{\tau_A}\right)}{\left(\frac{2\omega_*}{T^2}\right)^2 [(\sigma^\omega_A)^2 - (\sigma^\omega_H)^2] + \frac{2 i \omega_*}{3T^2} \sigma^\omega_A\left(\frac{1}{\tau_A} + \frac{1}{\tau_H}\right) - \frac{1}{9\tau_A \tau_H}} ,\nonumber\\
 &\delta\mu^*_{\tilde\chi} =\kappa_\Omega \delta \mu^*_V\times \nonumber\\&\dfrac{\frac{2\omega_*}{T^2}(A - s_V B)(\sigma^\omega_A + s_V\sigma^\omega_H) + \frac{i}{3} \left(\frac{A}{\tau_H} - \frac{s_V B}{\tau_A} \right)}{\left(\frac{2\omega_*}{T^2}\right)^2 [(\sigma^\omega_A)^2 - (\sigma^\omega_H)^2] + \frac{2 i \omega_*}{3T^2} \sigma^\omega_A\left(\frac{1}{\tau_A} + \frac{1}{\tau_H}\right) - \frac{1}{9\tau_A \tau_H}} .
 \label{eq:QCD_muchi}
\end{align}
The previous result is exact. As we approach the degenerate limit, Eq.~\eqref{eq:unpol_largemu} shows that $A\to s_V B$ and $\sigma_A\to s_V \sigma_H$ up to exponentially small corrections, and the amplitudes for the two propagating modes \eqref{eq:QCD_w:largemu} simplify to 
\begin{align}
 \delta\mu_{\chi}^\pm &\simeq  \frac{A T^2}{4\sigma_A^\omega}\frac{\kappa_\Omega(\tau_A + \tau_H)}{\omega_\pm(\tau_A + \tau_H) + \frac{iT^2}{6\sigma^\omega_A}} \delta\mu_V^\pm \nonumber\\
 &\simeq -\frac{\kappa_\Omega(\tau_A + \tau_H)}{3\alpha_V^2 \omega_\mp(\tau_A + \tau_H)} \dfrac{\lvert \mu_V \rvert}{T} \delta\mu_V^\pm, \nonumber\\
 \delta\mu_{\tilde{\chi}}^\pm &\simeq \frac{AT^2}{4\sigma_A^\omega}\frac{\kappa_\Omega(\tau_A-\tau_H)}{  \omega_\pm(\tau_A+\tau_H) + \frac{iT^2}{6\sigma_A} }\delta\mu_V^\pm \nonumber\\
 &\simeq -\dfrac{\kappa_\Omega(\tau_A-\tau_H)}{3\alpha_V^2 \omega_\mp(\tau_A+\tau_H)} \dfrac{\lvert \mu_V \rvert}{T} \delta\mu_V^\pm.
\end{align}
As in the case with $\tau_A \rightarrow \infty$ considered in Eq.~\eqref{eq:QED_muchi_largemu}, the amplitudes of $\delta \mu^\pm_\chi$ and $\delta \mu^\pm_{\tilde{\chi}}$ are $\lvert \alpha_V \rvert = \lvert \mu_V \rvert / T$ times larger than $\delta \mu_V^\pm$, due to the behaviour $\omega_\mp \sim \alpha_V^{-2}$. Contrary to Eq.~\eqref{eq:QED_muchi_largemu}, the degeneracy between $\delta \mu^\pm_\chi$ and $\delta \mu^\pm_{\tilde{\chi}}$ is lifted whenever both $\tau_A$ and $\tau_H$  are finite (i.e., non-zero and non-infinite).

Regarding the exponentially-dissipating modes, the denominators of the amplitudes \eqref{eq:QCD_muchi} vanish and we take the degenerate limit in Eqs.~\eqref{eq:QCD_mu2}. Taking $\delta \mu_{\tilde{\chi}}$ as the reference amplitude we find 

\begin{align}
    &\delta\mu_{\chi}^{\textrm{im.}} \simeq -\dfrac{2(\tau_A-\tau_H)}{(\tau_A+\tau_H)\alpha_V^2}e^{-\lvert \alpha_V \rvert} \delta  \mu_{\tilde{\chi}}^{\textrm{im.}}\,, \nonumber\\&  \delta\mu_{V}^{\textrm{im.}} \simeq -\dfrac{72 i \kappa_\Omega \tau_A\tau_H}{\pi^2(\tau_A+\tau_H)\alpha_V^2}e^{-2\lvert \alpha_V \rvert} \delta \mu_{\tilde{\chi}}^{\textrm{im.}}
\end{align}


We present in Fig.~\ref{fig:QCD_deg} the propagation properties of an initial Gaussian fluctuation in the vector chemical potential, given in Eq.~\eqref{eq:gaussV_init}, similar to the one studied in Fig.~\ref{fig:QED_deg} ($\alpha_V = 20$, $\tau_H = 2 \sigma /\tilde{v}$; see the end of Sec.~\ref{sec:QED:largemu} for details). As opposed to Fig.~\ref{fig:QED_deg}, here we take a finite axial relaxation time, which we set to $\tau_A = \tau_H / 2$. While the propagation of $\delta \bar{\mu}_V(t,z)$ and of $\delta \bar{\mu}_\chi(t,z)$ remains unchanged, the amplitude of $\delta \bar{\mu}_{\tilde{\chi}}(t, z)$ is reduced by a factor of about $3$ compared to the case when the axial charge is conserved. 

\subsection{Generic properties of spectra}\label{sec:QCD:spectra}

\begin{figure}[!ht]
    \centering
    \includegraphics[scale=0.35]{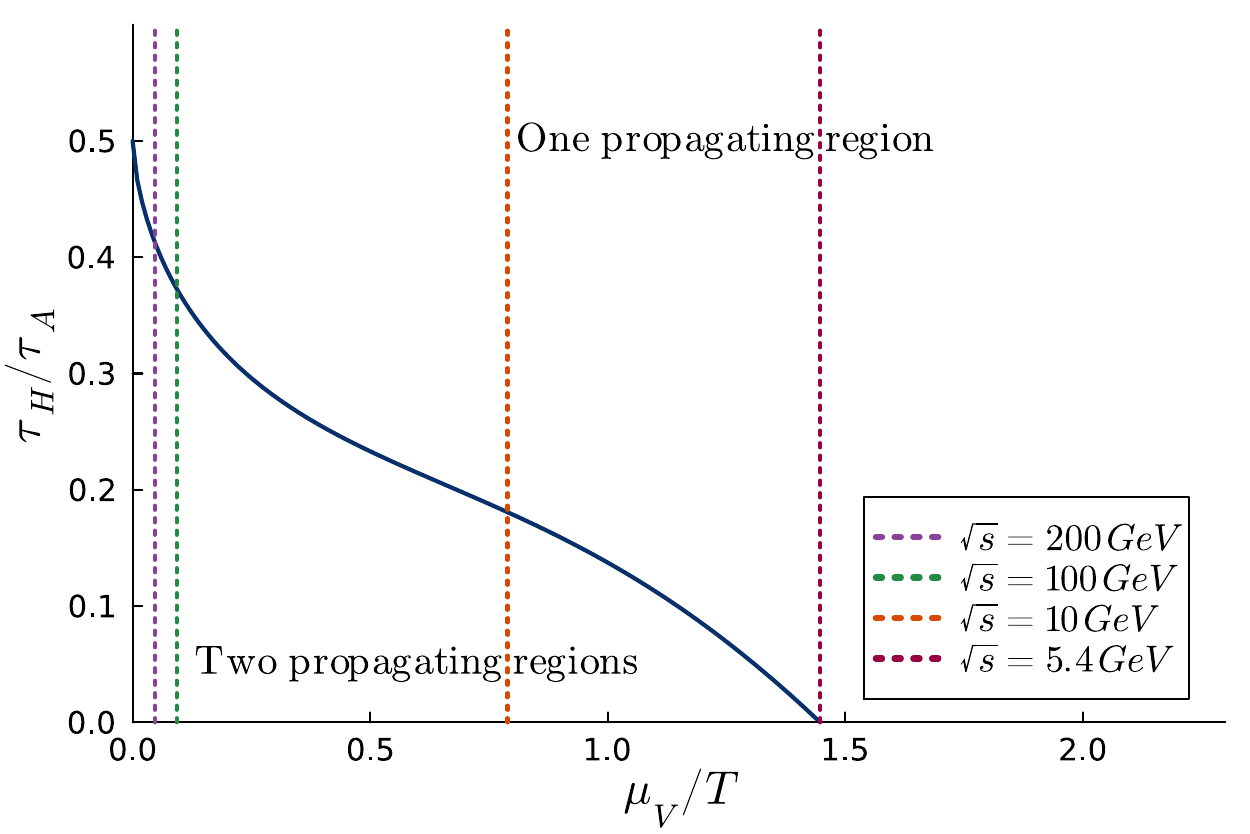}
    \caption{Solution of Eq.~\eqref{eq:merg} showing the values of the parameters (blue curve) for which the spectrum of excitations contains a merging point. The curve divides the parameter space into two regions: Above the curve, the spectrum has only one branch of propagating modes; Below the curve, the spectrum displays two disconnected branches of propagating modes.  
    \label{fig:tricrit}
    }
\end{figure}

\begin{figure*}[!ht]
\centering
\includegraphics[width=0.92\linewidth]{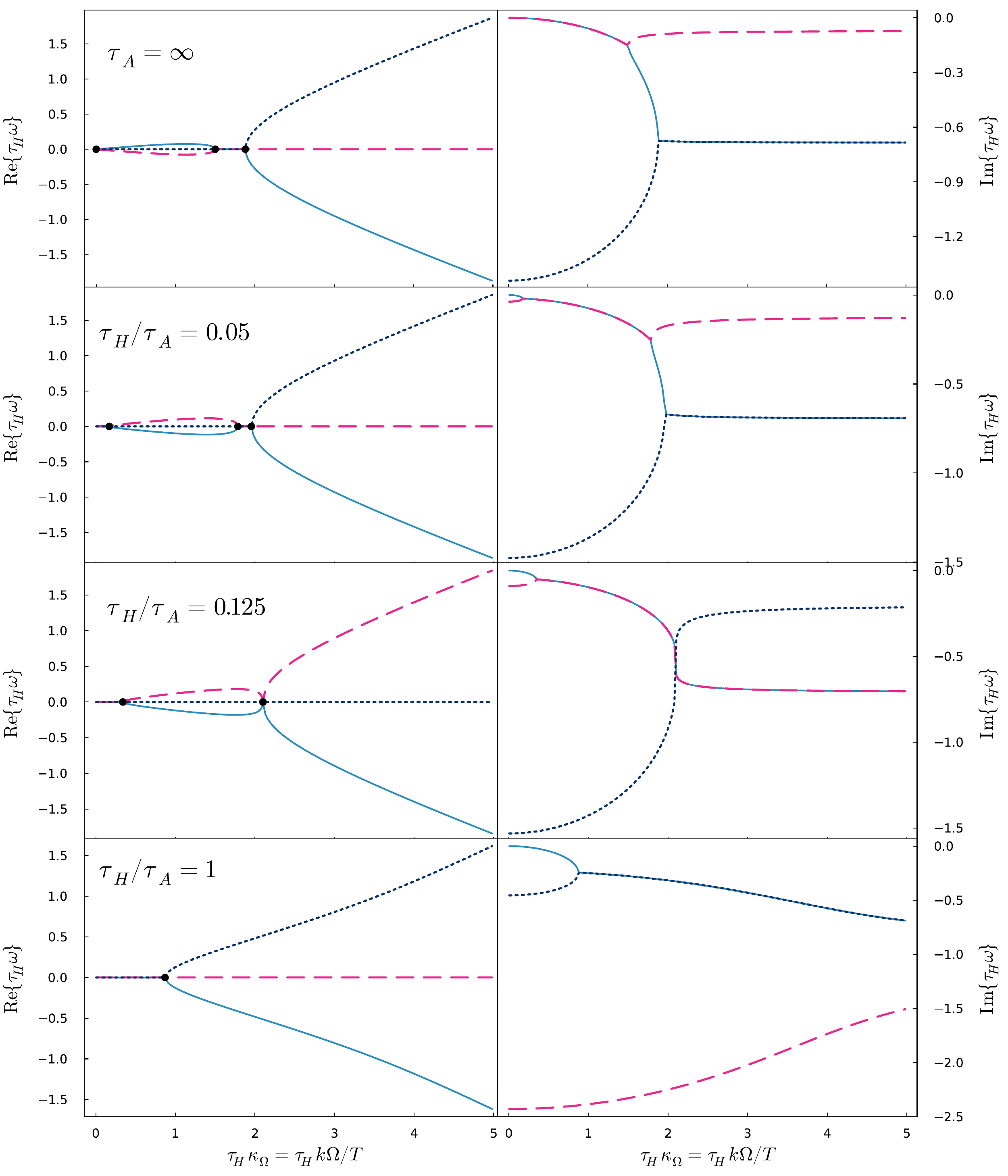}
\caption{Collective excitations of the charge sector in the unpolarized plasma with dissipating axial and helical charges for $\mu_V/T=1\,$. Each row shows the spectrum for a different value of the ratio $\tau_H / \tau_A$. The left (right) plots correspond to the real (imaginary) part of the dimensionless angular frequency, $\tau_H \omega$. The lines with identical styles at the left and right plots correspond to the same solutions. The black dots mark the points where propagation starts or stops.}
\label{fig:imqcd}
\end{figure*}

In Sec.~\ref{sec:QED:spectra}, we have seen that there exists a value of the vector chemical potential which divides the spectrum of excitations into two regions, depending on whether there exist one or two branches of propagating modes (c.f. Fig. \ref{fig:im}). The chemical potential separating both regions was found to be $\mu_V = 1.4471\, T$, and the wavenumber at which the two propagating branches merge satisfies $\tau_H \kappa_\Omega = 3.0642$. In this section, the presence of a finite axial relaxation time $\tau_A$ provides an extra parameter, and we expect to find a one-parameter family of such points. We will impose the existence of the merging point in order to find the curve separating both regions in terms of $\mu_V$ and $\tau_H/\tau_A\,$. 

Similarly to Eq.~\eqref{eq:QED_detM_largemu}, we write the characteristic Eq.~\eqref{eq:QCD_detM} as 
\begin{align}
    &a(\alpha_V)\omega^3 + i \left(\frac{1}{\tau_A}+ \frac{1}{\tau_H}\right) b(\alpha_V) \omega^2 -\nonumber\\& \left(c(\alpha_V)\kappa_\Omega^2 + \dfrac{\varsigma(\alpha_V)}{\tau_A\tau_H}\right)\omega - i \kappa_\Omega^2 \delta(\alpha_V,\tau_A,\tau_H) = 0\,, 
    \label{eq:detM_abcd}
\end{align}
where $a\,,b$ and $c$ are already given in \eqref{eq:abcd}, while $\varsigma$ and $\delta$ are defined as 

\begin{equation}
    \varsigma = \dfrac{2}{9 T^2}\left(\sigma_A^\omega - \frac{1}{3} T^2\Delta H\right)\,, \ \ \  \delta = \dfrac{A^2 \tau_A + B^2 \tau_H}{3 H \tau_A \tau_H}\,.
    \label{eq:vars_delta}
\end{equation}
At the merging point $\omega_{\textrm{merg.}}$, all three solutions of Eq. \eqref{eq:detM_abcd} coincide. Therefore, the first and second derivatives of Eq. \eqref{eq:detM_abcd} with respect to $\omega$ also vanish:
\begin{align}
    3 a \omega_{\textrm{merg.}}^2 + 2 i b\left(\dfrac{1}{\tau_A} + \dfrac{1}{\tau_H}\right)\omega_{\textrm{merg.}} &\nonumber\\ -\left(c\kappa_{\textrm{merg.}}^2 + \dfrac{\varsigma}{\tau_A\tau_H}\right) &=0\,, \nonumber\\
    6 a \omega_{\textrm{merg.}} + 2 i b\left(\dfrac{1}{\tau_A} + \dfrac{1}{\tau_H}\right) &= 0\,.
\end{align}
Solving the above equations for $\omega_{\textrm{merg.}}$ and $\kappa_{\textrm{merg.}}^2$ gives
\begin{align}
    \omega_{\textrm{merg.}} &= -\dfrac{ib}{3a} \left(\dfrac{1}{\tau_A} + \dfrac{1}{\tau_H}\right)\, \nonumber\\ \kappa_{\textrm{merg.}}^2 &= \dfrac{b^2}{3 a c}\left(\dfrac{1}{\tau_A} + \dfrac{1}{\tau_H}\right)^2 - \dfrac{\varsigma}{c}\frac{1}{\tau_A\tau_H}\,.
\end{align}
Replacing $\omega_{\textrm{merg.}}$ and $\kappa_{\textrm{merg.}}^2$ into Eq. \eqref{eq:detM_abcd} gives the equation in parameter space that determines the existence of the merging point:
\begin{multline}
 \dfrac{b^3}{27 a^2}\left(\dfrac{1}{\tau_A} + \dfrac{1}{\tau_H}\right)^3 - \dfrac{b^2 \delta}{3 a c} \left(\dfrac{1}{\tau_A} + \dfrac{1}{\tau_H}\right)^2 \\
 + \dfrac{\delta \varsigma}{c}\dfrac{1}{\tau_A\tau_H} = 0\,.
\end{multline}
From the definitions in Eqs. \eqref{eq:abcd} and \eqref{eq:vars_delta}, we can rewrite the previous equation as
\begin{multline}
    \dfrac{(\sigma_A^\omega)^3\left(\frac{1}{\tau_A} + \frac{1}{\tau_H}\right)^3}{\left[(\sigma_A^\omega)^2-(\sigma_H^\omega)^2\right]} \\
    +  \dfrac{27\left[(\sigma_A^\omega)^2  -(\sigma_H^\omega)^2\right]\left(\frac{A^2}{\tau_H}+\frac{B^2}{\tau_A}\right)}{\tau_A\tau_H\left[(A^2 + B^2)\sigma_A^\omega -2 A B \sigma_H^\omega\right]} \\  -  \dfrac{9(\sigma_A^\omega)^2 \left(\frac{A^2}{\tau_H}+\frac{B^2}{\tau_A}\right)\left(\frac{1}{\tau_A} + \frac{1}{\tau_H}\right)^2}{ \left[(A^2 + B^2)\sigma_A^\omega -2 A B \sigma_H^\omega\right]}
    =0\,,
    \label{eq:merg}
\end{multline}
where we multiplied by $729\left[(\sigma_A^\omega)^2-(\sigma_H^\omega)^2\right]$ and divided by $\left(\sigma_A^\omega - \frac{1}{3} T^2\Delta H\right)$ to have a more compact expression. Finally, multiplying Eq. \eqref{eq:merg} by $\tau_H^3$, it is manifest that it depends on the ratio of lifetimes $\tau_H/\tau_A\,$. 

The solutions of Eq. \eqref{eq:merg} are shown in Figure~\ref{fig:tricrit}. The spectra along the blue line feature a point where all three modes coincide, which appear when two branches of propagating modes merge together (see, e.g., Figure~\ref{fig:im} in the previous section). We recover the result of the previous section that for a conserved axial charge, i.e., $\tau_H / \tau_A\to 0$, the merging point exists for $\mu_V\simeq 1.4471 \,T$. The curve divides the parameter space into two regions: below the curve, there will be two branches of propagating modes, as it happens in the second line of Figure~\ref{fig:imqcd}; the complementary region contains only one branch of propagating modes, similarly to the last line in Fig.~\ref{fig:imqcd}. Furthermore, the curve in Fig.~\ref{fig:tricrit} has a global maximum at $\mu_V/T=0$: $\tau_A=2\tau_H$. Thus, we can assert that for $\tau_A < 2\tau_H$, there will only be one branch of propagating modes, regardless of the value of the vector chemical potential. 

In Fig.~\ref{fig:imqcd}, we show the spectrum of excitations for fixed $\mu_V/T=1$ and varying the ratio of axial and helical lifetimes, which can be understood as varying $\tau_A$ for fixed $\tau_H$. As anticipated from Fig.~\ref{fig:tricrit}, two branches of propagating modes exist for sufficiently small $\tau_H/\tau_A$. As we increase the ratio, these branches collide, and above a threshold value, there is only one branch of propagating modes. It should also be noted that for a non-conserved axial charge, i.e., finite $\tau_A$, there are no propagating modes for small wavenumber. 

\begin{figure}[!ht]
    \centering
    \begin{tabular}{c}
    \includegraphics[width=.89\linewidth]{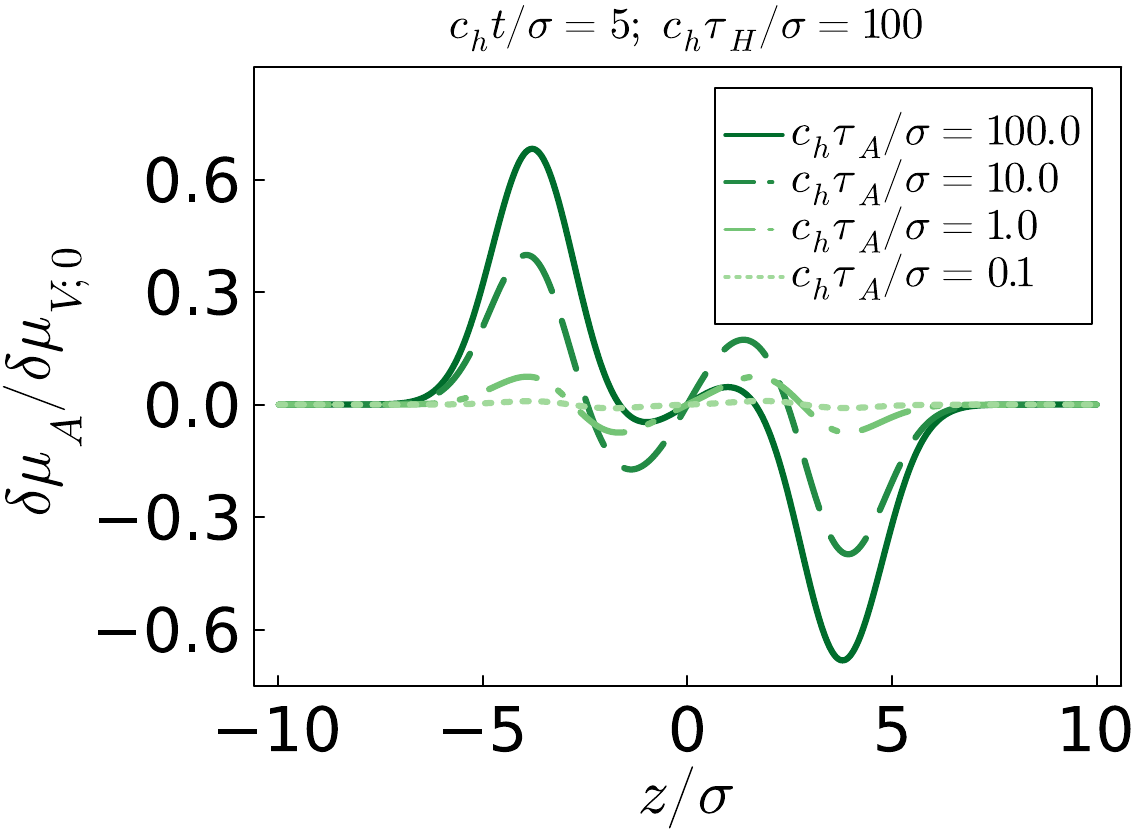}\\
    \includegraphics[width=.89\linewidth]{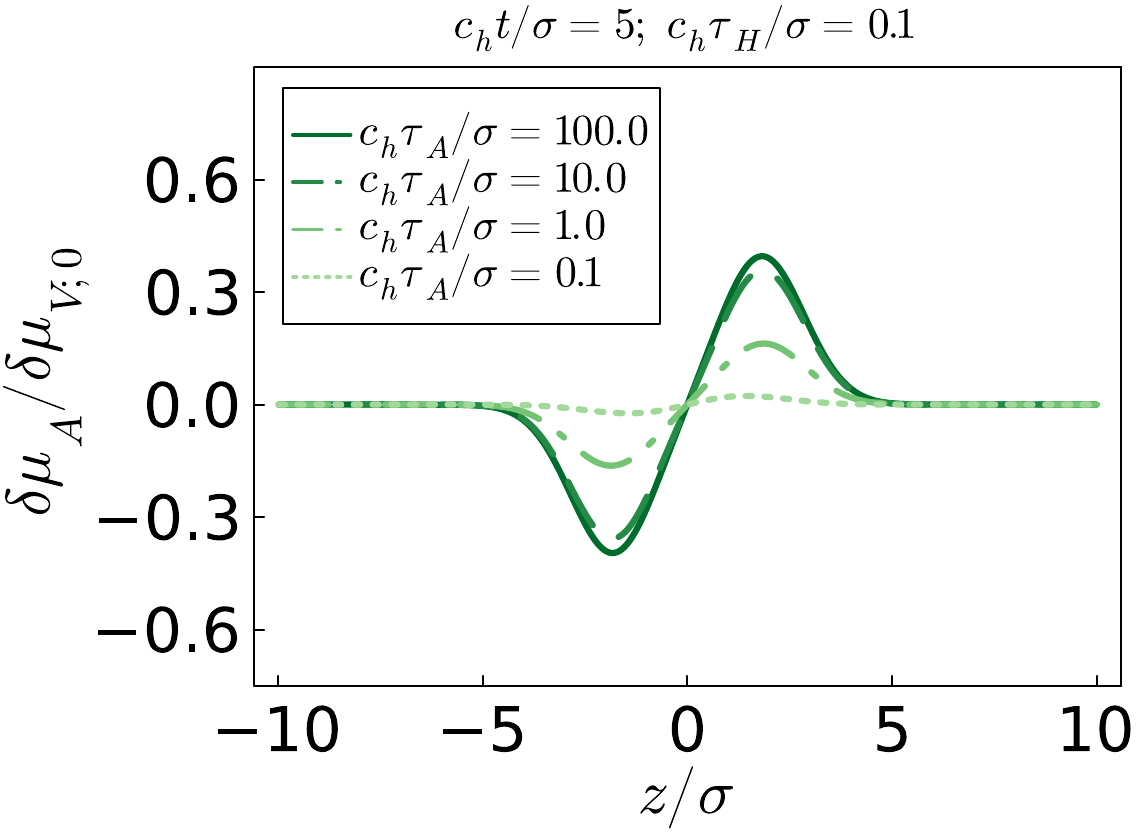}
    \end{tabular}
    \caption{
Perturbations in the axial chemical potentials at a fixed time, $c_h t/\sigma = 5$, as we vary the axial relaxation time for (top) $c_h \tau_H /\sigma=100$ (bottom) $c_h \tau_H /\sigma=0.1$. The system is initialized with a Gaussian perturbation in the vector chemical potential $\delta \mu_V$ on top of a background dimensionless vector chemical potential $\alpha_V = 13/6$. 
}
    \label{fig:QCD_tauA}
\end{figure}

The effect of the axial relaxation time on the perturbations of the axial chemical potential $\mu_A$ sourced by a Gaussian perturbation in the vector potential $\mu_V$ is exemplified in Fig.~\ref{fig:QCD_tauA}. We choose the same set of the parameters as in Fig.~\ref{fig:CVW} for two different values of the helicity relaxation time, namely $c_h \tau_H /\sigma=0.1$ and $c_h \tau_H /\sigma=100$. The first one corresponds to the traditional Chiral Vortical Wave, while the second is closer to the Helical Vortical Wave. As expected, a decrease in the lifetime of the axial charge $\tau_A$ results in a damping of the perturbations in the axial chemical potential and the variations of $\tau_A$ do not substantially affect the velocity of the axial wave propagation.

\section{Dissipative effects}
\label{sec:diss}

In our considerations so far, we have treated the V/A/H fluid as ideal, ignoring dissipative corrections to the energy-momentum tensor $T^{\mu\nu}$ and the charge currents $J^\mu_\ell$. In a realistic fluid, dissipation occurs due to the inter-particle interaction, which drives the system towards thermodynamic equilibrium. Contrary to intuition, the perfect fluid limit is achieved when the interactions are sufficiently strong, taking place on time scales much shorter than the characteristic macroscopic time scale, such that equilibration happens almost instantaneously and the deviation of the fluid from thermal equilibrium is always negligible. Since the relaxation time $\tau_H$ arises due to helicity-violating pair annihilation (HVPA) two-to-two interactions, the perfect fluid limit implies also that $\tau_H \rightarrow 0$. Similarly, the same limit implies also that the axial relaxation time $\tau_A$, describing the dissipation of axial charge, should be taken to $0$. In order to provide a more credible assessment of realistic V/A/H fluids, we will consider in this section the effect of dissipation on the spectrum of vortical waves.

\subsection{Relaxation-time approximation}\label{sec:diss:RTA}

Dissipative corrections manifest themselves whenever global thermodynamic equilibrium is perturbed. In the Landau frame, when $T^\mu{}_\nu u^\nu = E u^\mu$, they lead to the modifications \cite{Rezzolla:2013dea}
\begin{align}
 &T^{\mu\nu} \rightarrow T^{\mu\nu} = T^{\mu\nu}_0 + \pi^{\mu\nu}_d - \Pi_d \Delta^{\mu\nu}, \nonumber\\&
 J^\mu_\ell \rightarrow J^\mu_\ell = J^{\mu}_{\ell;0} + V^\mu_{\ell;d},
\end{align}
where $\Pi_d$, $V^\mu_{\ell;d}$ and $\pi^{\mu\nu}$ represent the scalar, vector and tensor dissipative degrees of freedom appearing at the level of $T^{\mu\nu}$ and $J^\mu_\ell$. By construction, $V^\mu_{\ell;d}$ and $\pi^{\mu\nu}$ are orthogonal to the fluid velocity, while $\pi^{\mu\nu}_d$ is taken to be traceless. Since we are dealing with a conformal (massless) fluid with $E = 3P$ and $T^\mu{}_\mu = E - 3P - 3\Pi_d = 0$, the bulk viscous pressure can be neglected, $\Pi_d = 0$. The remaining degrees of freedom contained in $V^\mu_{\ell;d}$ and $\pi^{\mu\nu}_d$ are {\it a priori} unconstrained and must be specified via constitutive relations.

A relativistic treatment of dissipative effects must be performed in a framework that preserves the hyperbolicity, causality and stability of the hydrodynamic equations \cite{Hiscock:1983zz,Hiscock:1985zz}. This can be achieved in the so-called M\"uller-Israel-Stewart hydrodynamics framework \cite{Muller:1967zza,Israel:1979wp}, where $V^\mu_{\ell;d}$ and $\pi^{\mu\nu}_d$ become dynamical quantities obeying equations that essentially govern their relaxation towards their Navier-Stokes (first-order) asymptotic expressions,
\begin{equation}
 V^\mu_{\ell;d} = \sum_{\ell'} \kappa_{\ell,\ell'} \nabla^\mu \alpha_{\ell'}, \qquad 
 \pi^{\mu\nu}_d = 2\eta \sigma^{\mu\nu},
 \label{eq:diss_NS}
\end{equation}
where $\eta$ is the shear viscosity, $\kappa_{\ell\ell'}$ is the so-called diffusion matrix \cite{Monnai:2014qaa,Greif:2017byw,Fotakis:2022usk}, while $\nabla^\mu = \Delta^{\mu\nu} \partial_\nu$ is the spatial gradient in the fluid rest frame, with $\Delta^{\mu\nu} = g^{\mu\nu} - u^\mu u^\nu$ being the projector on the hypersurface orthogonal to $u^\mu$. The quantities $V^\mu_{\ell;d}$ and $\pi^{\mu\nu}_d$ can be calculated from field theory via the Kubo formulas \cite{Jeon:1994if,Mallik:2016anp}, or in the frame of kinetic theory \cite{Denicol:2012cn,Fotakis:2022usk,Denicol:2022bsq}, once the inter-particle interactions have been specified. Generically, they are inversely proportional to the collision cross-section $\sigma$. The same cross-section allows one to define a characteristic relaxation time scale,
\begin{equation}
 \tau_R \propto \frac{T}{P \sigma},
\end{equation}
where we used $P/T$ instead of the particle number $n$ \cite{cercignani02}, since the latter is not conserved in a quantum fluid. We thus conclude that both $\eta$ and $\kappa_{\ell\ell'}$ are proportional to the microscopic timescale $\tau_R$. By similar arguments, the helicity and axial relaxation times $\tau_H$ and $\tau_A$ should also be proportional to $\tau_R$. One should therefore consider $\eta$ and $\kappa_{\ell\ell'}$ on an equal footing to $\tau_H$ and $\tau_A$.

In MIS hydrodynamics, the Navier-Stokes relations \eqref{eq:diss_NS} represent the asymptotic near-equilibrium, the long-wavelength limit of the dissipative quantities, and are therefore not valid if the fluid is too far from equilibrium or if the macroscopic gradients are too large compared to the microscopic relaxation time scale $\tau_R$. For $k \tau_R \ll 1$, Eq.~\eqref{eq:diss_NS} provides a reasonable approximation and second- or higher-order corrections can be neglected. Of course, the framework becomes unreliable for small wavelengths (large $k$) and we will point out this limitation when relevant in the discussion below.

The evaluation of transport coefficients in theories with vector interactions is a daunting task \cite{Carrington:2007fp,Arnold:2006fz}, and we do not pursue such calculations in what follows. We will consider the qualitative effect of dissipation on the wave spectrum in the frame of kinetic theory under the relaxation-time approximation of Anderson and Witting \cite{Anderson:1974nyl}. Denoting by $f^\sigma_{\mathbf{p},\lambda}$ the number of particles/anti-particles ($\sigma = \pm 1$) with polarization $\lambda = \pm 1/2$, we postulate a kinetic equation of the form
\begin{equation}
 p^\mu \partial_\mu f^\sigma_{\mathbf{p},\lambda} = -\frac{E_\mathbf{p} \cdot u}{\tau_R} \left(f^\sigma_{\mathbf{p},\lambda} - f^{\sigma;{\rm eq}}_{\mathbf{p},\lambda}\right),
 \label{eq_RTA}
\end{equation}
where $E_\mathbf{p} = p \cdot u$ is the particle energy in the fluid rest frame, $u^\mu$ is the fluid four-velocity and 
\begin{equation}
 f^{\sigma{\rm eq}}_{\mathbf{p},\lambda} = \left[ 
 \exp\left(\frac{p \cdot u - \mu_{\sigma,\lambda}}{T}\right) 
 + 1 \right]^{-1},
 \label{eq:feq}
\end{equation}
is the Fermi-Dirac distribution characterized by the particle four-momentum $p^\mu = (p^0, \mathbf{p})$ with $p^0 = \lvert\mathbf{p}\rvert$ for massless fermions, $\sigma = \pm 1$ distinguishes between particles and anti-particles, and $\lambda = \pm 1/2$ labels the particle helicity. 

Due to its classical nature, Eq.~\eqref{eq_RTA} cannot account for the vortical effects that modify the charge currents $J^\mu_\ell$ in the presence of finite vorticity $\omega^\mu$. In order to account for such effects, one must employ a quantum kinetic equation, such as the one derived from the Wigner equation, leading to a modified equilibrium distribution \cite{Yang:2020mtz}. An alternative approach, followed in Ref. \cite{Gorbar:2017toh} in the absence of helicity, is to make use of Chiral Kinetic Theory to describe the vorticity contributions to the currents. Similarly, Eq.~\eqref{eq_RTA} is unable to account for the non-conservation of the helical and axial charges, thus our current simplified formulation provides access only to the dissipative quantities $V^\mu_{\ell;d}$ and $\pi^{\mu\nu}_d$.

The charge currents $J^\mu_\ell$ and the stress-energy tensor $T^{\mu\nu}$ can be obtained from the distribution functions $f^\sigma_{\mathbf{p},\lambda}$ as follows:
\begin{align}
 J^\mu_\ell =& \sum_{\sigma,\lambda} q^{\ell}_{\sigma,\lambda} \int dP \, f^\sigma_{\mathbf{p},\lambda} p^\mu, \nonumber\\
 T^{\mu\nu} =& \sum_{\sigma,\lambda} \int dP \, f^\sigma_{\mathbf{p},\lambda} p^\mu p^\nu.
 \label{eq:kinetic_macro}
\end{align}
At equilibrium, when $f^\sigma_{\mathbf{p},\lambda} = f^{\sigma;{\rm eq}}_{\mathbf{p},\lambda}$, we have $J^\mu_\ell = Q_\ell u^\mu$ and $T^{\mu\nu} = (E + P) u^\mu u^\nu - P g^{\mu\nu}$. The conservation equations $\partial_\mu J^\mu_\ell = 0$ and $\partial_\mu T^{\mu\nu} = 0$ imply the Landau matching conditions $u_\mu J^{\mu}_\ell = Q_\ell$ and $u_\nu T^{\mu\nu} = E u^\mu$, valid also when the fluid is out of equilibrium, which we consider to hold in our analysis. 

\subsection{Chapman-Enskog expansion} \label{sec:diss:CE}

We now employ the Chapman-Enskog procedure to derive an approximation for $f^\sigma_{\mathbf{p}, \lambda}$, when the fluid is not too far from equilibrium \cite{cercignani02} and $\delta f^\sigma_{\mathbf{p},\lambda} = f^\sigma_{\mathbf{p},\lambda} - f^{\sigma;{\rm eq}}_{\mathbf{p},\lambda}$ can be considered small and of the order of the relaxation time, $\tau_R$. The deviation from equilibrium $\delta f^\sigma_{\mathbf{p},\lambda}$ can be obtained from Eq.~\eqref{eq_RTA},
\begin{align}
 &\delta f^\sigma_{\mathbf{p},\lambda} = -\frac{\tau_R}{E_\mathbf{p}} p^\mu \partial_\mu f^\sigma_{\mathbf{p},\lambda} \nonumber\\&= -\tau_R \left[\partial_\alpha \left(\frac{p^\alpha}{E_\mathbf{p}} f^\sigma_{\mathbf{p},\lambda}\right) + \frac{\partial_\alpha u_\beta}{E_\mathbf{p}^2} p^\alpha p^\beta f^\sigma_{\mathbf{p},\lambda}\right].
\end{align}
Due to the multiplication with $\tau_R$, the distribution on the RHS of the above equation can be approximated via $f^\sigma_{\mathbf{p},\lambda} \simeq f^{\sigma;{\rm eq}}_{\mathbf{p},\lambda}$ and eventually we obtain
\begin{align}
 V^\mu_{\ell;d} =& -\tau_R 
 \sum_{\sigma,\lambda} q^\ell_{\sigma,\lambda}
 \left(\partial_\alpha T^{{\rm eq};\mu\alpha}_{\sigma,\lambda; 1} + 
 \partial_\alpha u_\beta T^{{\rm eq};\mu\alpha\beta}_{\sigma,\lambda; 2}\right),\nonumber\\
 \pi^{\mu\nu}_d =& -\tau_R \sum_{\sigma,\lambda} 
 \left(\partial_\alpha T^{{\rm eq};\mu\nu\alpha}_{\sigma,\lambda; 1} + 
 \partial_\alpha u_\beta T^{{\rm eq};\mu\nu\alpha\beta}_{\sigma,\lambda; 2}\right),
 \label{eq:kinetic_aux}
\end{align}
where the following notation was introduced \cite{Ambrus:2016frt}:
\begin{equation}
 T^{{\rm eq};\mu_1 \mu_2 \dots \mu_m}_{\sigma,\lambda; n} =  
 \int \frac{dP}{E_\mathbf{p}^n}\, p^{\mu_1} p^{\mu_2} \cdots p^{\mu_m}
 f^{\sigma;{\rm eq}}_{\mathbf{p},\lambda}.
\end{equation}
In the massless case, when $p^2 = 0$, the integrals appearing in Eq.~\eqref{eq:kinetic_aux} are easily performed:
\begin{align}
 & \sum_{\sigma,\lambda} q^\ell_{\sigma,\lambda} T^{{\rm eq};\mu\nu}_{\sigma,\lambda;1} = \frac{1}{3} Q_\ell (4 u^\mu u^\nu - g^{\mu\nu}), \nonumber\\
 & \sum_{\sigma,\lambda} 
 \begin{pmatrix}
  T^{{\rm eq};\mu\nu\alpha}_{\sigma,\lambda;1} \\
  q^\ell_{\sigma,\lambda} T^{{\rm eq};\mu\nu\alpha}_{\sigma,\lambda;2} 
 \end{pmatrix} = \frac{1}{3} 
 \begin{pmatrix}
  E \\ Q_\ell 
 \end{pmatrix} 
 (6 u^\mu u^\nu u^\alpha 
 \nonumber\\
 & \hspace{.2\linewidth} - u^\mu g^{\nu\alpha} - u^\nu g^{\alpha\mu} - u^\alpha g^{\mu\nu}),\nonumber\\
 & \sum_{\sigma,\lambda} T^{\mu\nu\alpha\beta}_{\sigma,\lambda;2} = \frac{1}{15} E [
 48 u^\mu u^\nu u^\alpha u^\beta \nonumber\\
 &- 6(u^\mu u^\nu g^{\alpha\beta} + 
 u^\mu u^\alpha g^{\nu\beta} + u^\mu u^\beta g^{\nu\alpha} + 
 u^\nu u^\alpha g^{\mu\beta} \nonumber\\
 & + u^\nu u^\beta g^{\mu\alpha} + u^\alpha u^\beta g^{\mu\nu}) + 
 g^{\mu\nu} g^{\alpha\beta} + g^{\mu\alpha} g^{\nu\beta}\nonumber\\
 &\hspace{.6\linewidth} + g^{\mu\beta} g^{\nu\alpha}].
\end{align}
Substituting the above results into Eq.~\eqref{eq:kinetic_aux}, we obtain
\begin{align}
 V^\mu_{\ell;d} &= -\tau_R 
 \left[u^\mu \partial_\alpha(Q_\ell u^\alpha) - \frac{1}{3} \nabla^\mu Q_\ell + 
 Q_\ell D u^\mu\right],\nonumber\\
 \pi^{\mu\nu}_d &= -\tau_R \left[(3u^\mu u^\nu - \Delta^{\mu\nu})DP - u^\mu \nabla^\nu P - u^\nu \nabla^\mu P\right. \nonumber\\&\left.+ 
 4P \left(\frac{6}{5} u^\mu u^\nu \theta + u^\mu D u^\nu + u^\nu D u^\mu\right) \right.\nonumber\\
 &\left. - 
 \frac{4P}{5} (g^{\mu\nu} \theta + \nabla^\mu u^\nu + \nabla^\nu u^\mu)\right],
 \label{eq:kinetic_aux_2}
\end{align}
with $Df = u^\mu \partial_\mu f$ and $\nabla^\mu f = \Delta^{\mu\nu} \partial_\nu f$.

The relations in Eq.~\eqref{eq:kinetic_aux_2} can be simplified by taking into account the conservation relations $\partial_\mu J^\mu_\ell = 0$ and 
$\partial_\mu T^{\mu\nu} = 0$, applied to zeroth order in $\tau_R$, when $J^\mu_\ell \simeq Q_\ell u^\mu$ and $T^{\mu\nu} \simeq (E + P) u^\mu u^\nu - P g^{\mu\nu}$:
\begin{align}
 D Q_\ell + Q_\ell \theta &= O(\tau_R), \nonumber\\
 D E + (E + P) \theta &= O(\tau_R), \nonumber\\
 (E + P) D u^\mu - \nabla^\mu P &= O(\tau_R).
 \label{eq:perfect}
\end{align}
Using the above relations into Eq.~\eqref{eq:kinetic_aux_2} gives 
\begin{align}
 V^\mu_{\ell;d} &= \tau_R \left(\frac{1}{3} \nabla^\mu Q_\ell - \frac{Q_\ell \nabla^\mu P}{E + P}\right),&
 \pi^{\mu\nu}_d &= 2 \eta \sigma^{\mu\nu},
 \label{eq:kinetic_Vlpi}
\end{align}
where $\sigma^{\mu\nu} = (\frac{1}{2} \Delta^{\mu\lambda} \Delta^{\nu\kappa} + \frac{1}{2} \Delta^{\nu\lambda} \Delta^{\mu\kappa} - 
\frac{1}{3} \Delta^{\mu\nu} \Delta^{\lambda\kappa}) \partial_\lambda u_\kappa = \frac{1}{2}(\nabla^\mu u^\nu + \nabla^\nu u^\mu) - \frac{1}{3} \Delta^{\mu\nu} \theta$ is the shear stress tensor, and we identified the shear viscosity as
\begin{equation}
 \eta = \frac{4}{5} \tau_R P.\label{eq:kinetic_eta}
\end{equation}

While Eq.~\eqref{eq:kinetic_Vlpi} is sufficient for our analysis, we present below the diffusion matrix for completeness. In the massless limit, the quantities $Q_\ell / T^3$ and $P / T^4$ depend only on the ratios $\alpha_{\ell'} = \mu_{\ell'} / T$. Noting that $E + P = 4P$, it can be seen that the diffusion flux depends only on the derivatives of $\alpha_{\ell'}$. Employing $\partial P / \partial \alpha_{\ell'} = Q_{\ell'} T$, we get
\begin{align}
 &V^\mu_{\ell;d} = \sum_{\ell'} \kappa_{\ell\ell'} \nabla^\mu \alpha_{\ell'},\nonumber\\& 
 \kappa_{\ell\ell'} = 
 \tau_R \left(\frac{1}{3} \frac{\partial Q_\ell}{\partial \alpha_{\ell'}} - 
 \frac{Q_\ell Q_{\ell'} T}{E + P}\right).
 \label{eq:kinetic_Vl_alpha}
\end{align}

\subsection{Dispersion relations}

We now add the dissipative contributions $V^\mu_{\ell;d}$ and $\pi^{\mu\nu}_d$ to the Landau-frame expressions for the charge currents and energy-momentum tensor given in Eq.~\eqref{eq:Landau_currents}:
\begin{align}
 J^\mu_\ell =& Q_\ell u^\mu + \sigma^\omega_{\ell} \omega^\mu + V_{\ell;d}^\mu,\nonumber\\
 T^{\mu\nu} =& (E + P) u^\mu u^\nu - P g^{\mu\nu} + \pi_d^{\mu\nu}.
\end{align}
The dissipative quantities appearing above modify the conservation Eqs.~\eqref{eq:conservation} to 
\begin{align}
 D Q_\ell + Q_\ell \theta + \partial_\mu(\omega^\mu \sigma^\omega_\ell) + \partial_\mu V^\mu_{\ell;d} &= -\frac{T^2 \mu_\ell}{3\tau_\ell},\nonumber\\
 D E + (E + P) \theta + u_\mu \partial_\nu \pi^{\mu\nu}_d &= 0,\nonumber\\
 (E + P) Du^\mu - \nabla^\mu P + \Delta^\mu{}_\lambda \partial_\nu \pi^{\lambda\nu}_d &= 0.\label{eq_tauR_conservation_eqs}
\end{align}
We follow a similar approach as in Ref. \cite{Morales-Tejera:2024uzg} and study the waves generated by longitudinal fluctuations of the four-velocity to leading order in $\Omega$. In particular, we write
\begin{align}
    u^\mu =& \partial_t+\Omega(-y\partial_x+x\partial_y)\nonumber\\
    &+\Omega \delta u^\perp\left(-y\partial_x+x\partial_y\right) + \delta u^z\partial_z\,,
\end{align}
where the $(x,y)$-dependence of the transverse plane fluctuations is required to have a consistent solution to leading order in $\Omega$, as shown in  Ref. \cite{Morales-Tejera:2024uzg}. The fluctuations are Fourier transformed in the $(t,z)$ coordinates as in Eq. \eqref{eq_barf_Fourier}.
Then, Eqs. \eqref{eq_tauR_conservation_eqs} in Fourier space can be cast with the help of the relations \eqref{eq:kinetic_Vl_alpha} into the following form in the rotating fluid:
\begin{align}
 \pi_d^{\mu\nu} &\to  i \eta k \delta u^\perp_\omega
 \begin{pmatrix}
  0 & 0 & 0 & 0 \\ 
  0 & 0 & 0 & y \Omega \\
  0 & 0 & 0 & - x  \Omega\\
  0 & y \Omega & -x \Omega & 0
 \end{pmatrix}\nonumber\\& +\frac{2 i \eta k}{3} \delta u^z_\omega 
 \begin{pmatrix}
  0 & -y \Omega & x \Omega & 0 \\ 
  - y \Omega & 1 & 0 & -\tfrac{3\omega}{2k} y  \Omega \\
  x  \Omega & 0 & 1 & \tfrac{3\omega}{2k} x  \Omega\\
  0 & -\tfrac{3\omega}{2k} y  \Omega & \tfrac{3\omega}{2k} x  \Omega & -2
 \end{pmatrix} , \nonumber\\ \nonumber\\
 V^\mu_{\ell;d} &\to -i \tau_R
 \begin{pmatrix}
  0 \\
  y \Omega \omega\\ 
  -x \Omega \omega\\
  k
 \end{pmatrix}
 \left(\frac{1}{3} \delta Q_{\ell;\omega} - \frac{Q_\ell \delta P_\omega}{E + P}\right) .
\end{align}
Furthermore, taking into account that $u_\mu \partial_\nu \pi^{\mu\nu} = -\pi^{\mu\nu} \nabla_\nu u_\mu$ is of second order in perturbations, while
\begin{multline}
    \Delta^\mu{}_\lambda \partial_\nu \pi^{\lambda\nu} \to\frac{4\eta k^2}{3} \delta u^z_\omega\delta^\mu_z\\ + \eta k\Omega\left(k\delta u^\perp_\omega-\frac{5}{3}\omega\delta u^z_\omega\right)(x\delta^\mu_y-y\delta^\mu_x).
\end{multline}
It is clear that the energy-momentum sector remains decoupled from the charge currents sector, since it involves only the amplitudes of perturbations in the velocity $\delta u^\mu_\omega$ and in the pressure $\delta P_\omega$. We now expand the angular frequency in terms of $\Omega$ as $\omega=\omega_0+\Omega \omega_1+\dots$, and similarly we decompose the fluctuations $\delta P_\omega$ and $\delta u^z_\omega$. To leading order in $\Omega$, the conservation equations for the energy momentum tensor reduce to
\begin{equation}
 \begin{pmatrix}
  -3\omega_0 & 4 P k \\
  k & -4 P \omega_0 - \frac{4}{3} i \eta k^2
 \end{pmatrix} 
 \begin{pmatrix}
  \delta P_{\omega;0} \\ \delta u^z_{\omega;0}
 \end{pmatrix} = 0.
 \label{eq:tauR_sound_det}
\end{equation}
Imposing a non-trivial solution to the previous equation reveals that 
\begin{align}
 \omega^\pm_0 &= \pm k c_s(\eta) - \frac{i k^2 \eta}{6P}, \nonumber\\
 c_s(\eta) &= \frac{1}{\sqrt{3}} \sqrt{1 - \frac{k^2 \eta^2}{12 P^2}},
\end{align}
while the perturbations are related to each other as
\begin{equation}
    \delta P_{\pm;0} = \frac{4kP}{3\omega_0^\pm}\delta u^z_{\pm;0}.
\end{equation}
We have chosen $\delta u^{z}_{\omega;0}$ as our reference amplitude, and therefore we can set $\delta u^{z}_{\omega;1}=0$ without loss of generality. Then, the subleading contribution to the conservation equations for the energy momentum tensor take the following form:
\begin{align}
    3\omega_0^\pm\delta P_{\pm;1}+\dfrac{4kP\omega_1^\pm}{\omega_0^\pm}\delta u^z_{\pm;0}&=0\,,\nonumber\\
    k\delta P_{\pm;1}-4 P\omega_1^\pm\delta u^z_{\pm;0}&=0\,,\nonumber\\
    (12iP\omega_0^\pm - 3k^2\eta)\delta u^\perp_\pm+(4iP+\eta\omega_0^\pm)k\delta u^z_{\pm;0}&=0\,.
\end{align}
The last equation can be used to solve for the transverse perturbation $\delta u^\perp_\pm$, while the first two equations indicate that the non-trivial solution ($\delta u^z_{\pm;0}\neq 0$) requires $\omega_1^\pm =0$. Consequently, the dispersion relation for the longitudinal sound mode does not get linear corrections in $\Omega$.
The above discussion reveals that the effect of the dissipating shear is to shift the angular frequencies of the acoustic modes by an imaginary quantity, such that the non-dissipative solution shown in  Eq.~\eqref{eq:acoustic_vpm} becomes 
\begin{align}
 \omega^\pm_{R; {\rm ac.}} &= \pm k c_s(\eta) - \frac{i k^2 \eta}{6P}, \nonumber\\
 c_s(\eta) &= \frac{1}{\sqrt{3}} \sqrt{1 - \frac{k^2 \eta^2}{12 P^2}}.
\label{eq:omega_tauR}
\end{align}
The modification to the dispersion relation due to dissipation, $\omega^\pm_{R;{\rm ac.}} - \omega^\pm_{\rm ac.} = -ik^2 \eta / 6P = -i 2 k^2 \tau_R / 15$, agrees with that derived in Eq.~(58) of Ref.~\cite{Gorbar:2017toh} using Chiral Kinetic Theory.\footnote{A small discrepancy in the $O(\Omega)$ correction to $\omega^\pm_{R;{\rm ac.}}$, which can be traced back to the transverse plane dynamics ignored in Ref.~\cite{Gorbar:2017toh}, is irrelevant in the present case, as we work under the assumption of a chirally-neutral background.}
In other words, one of the effects of the presence of a finite kinetic relaxation time $\tau_R$ is to set the attenuation of the acoustic waves due to the nonvanishing kinetic shear viscosity~\eqref{eq:kinetic_eta}. In addition, Eqs.~\eqref{eq:kinetic_eta} and \eqref{eq:omega_tauR} reveal that the speed of sound in the dissipative fluid is slower than the one in its non-dissipative limit~\eqref{eq:acoustic_vpm}. The speed of the sound wave depends on the momentum $k$, which has an ultraviolet bound, implying that when $k \tau_R > 5 \sqrt{3} / 2$, the acoustic wave stops propagating. The latter effect is a feature of the first-order theory, as the inclusion of second-order terms may alter this behaviour~\cite{Sammet:2023bfo}.

What are the effects of a finite kinetic relaxation time for the propagation of perturbations in the charge sector? In the case of charge (non-)conservation, Eq.~\eqref{eq:chargecons} is modified to:
\begin{equation}
 \left[\left(\omega + \frac{i k^2 \tau_R}{3} 
 \right) \delta Q_{\ell} - k\Omega 
 \delta \sigma^\omega_\ell\right]\biggl\rvert_{\delta P = 0} = -\frac{i T^2 \delta \mu_\ell}{3\tau_\ell}.
\end{equation}
Comparing the above relations with Eq.~\eqref{eq:chargecons}, we see that, as in the case of the dissipative shear stress, the charge diffusion has the effect of shifting the angular frequencies by an imaginary quantity. Denoting by $\omega_R$ the energy dispersion relation in the presence of dissipation, we have 
\begin{equation}
 \omega_R = \omega - \frac{i k^2 \tau_R}{3}, 
 \label{eq:omega_R}
\end{equation}
where $\omega$ is the mode angular frequency. 
The modification of the dispersion relation by the amount $i k^2 \tau_R / 3$ is consistent with the results reported in Eq.~(59) of Ref.~\cite{Gorbar:2017toh}.

\subsection{Estimation of damping due to kinetic effects}\label{sec:diss:damping}

\begin{figure}[!ht]
    \centering
    \begin{tabular}{c}
    \includegraphics[width=.89\linewidth]{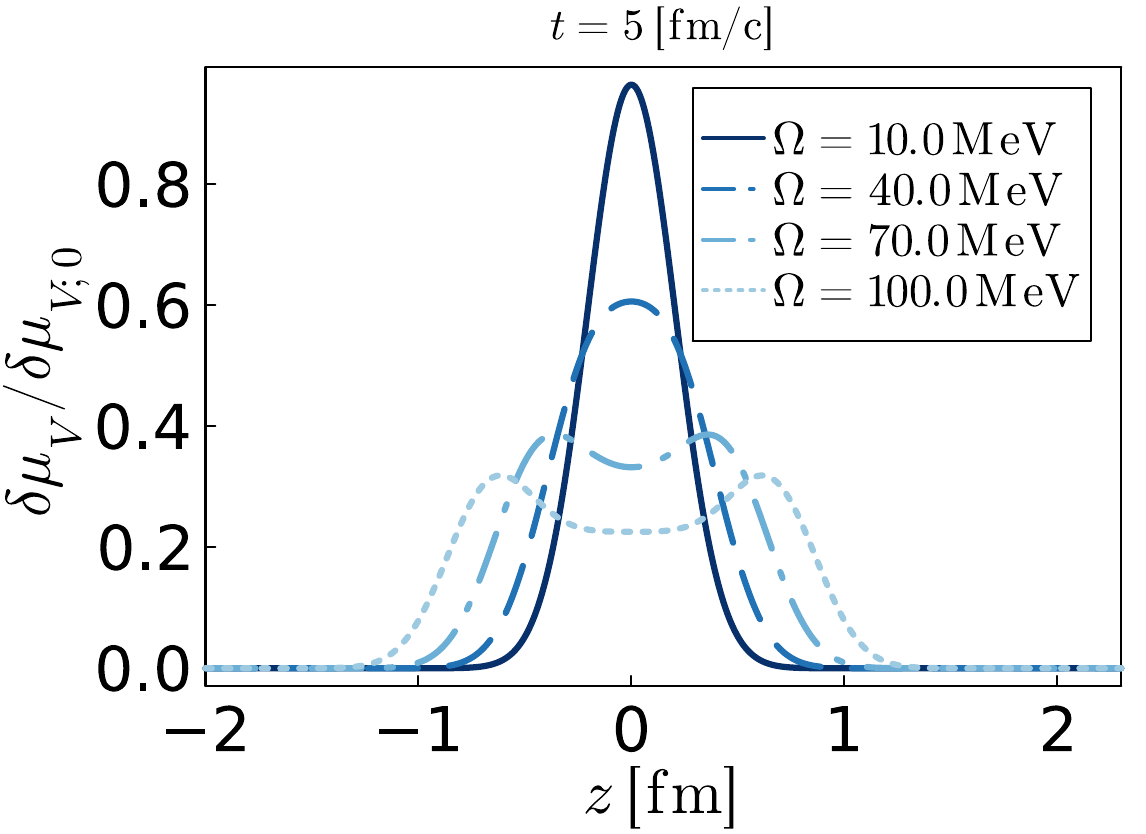}\\
    \includegraphics[width=.89\linewidth]{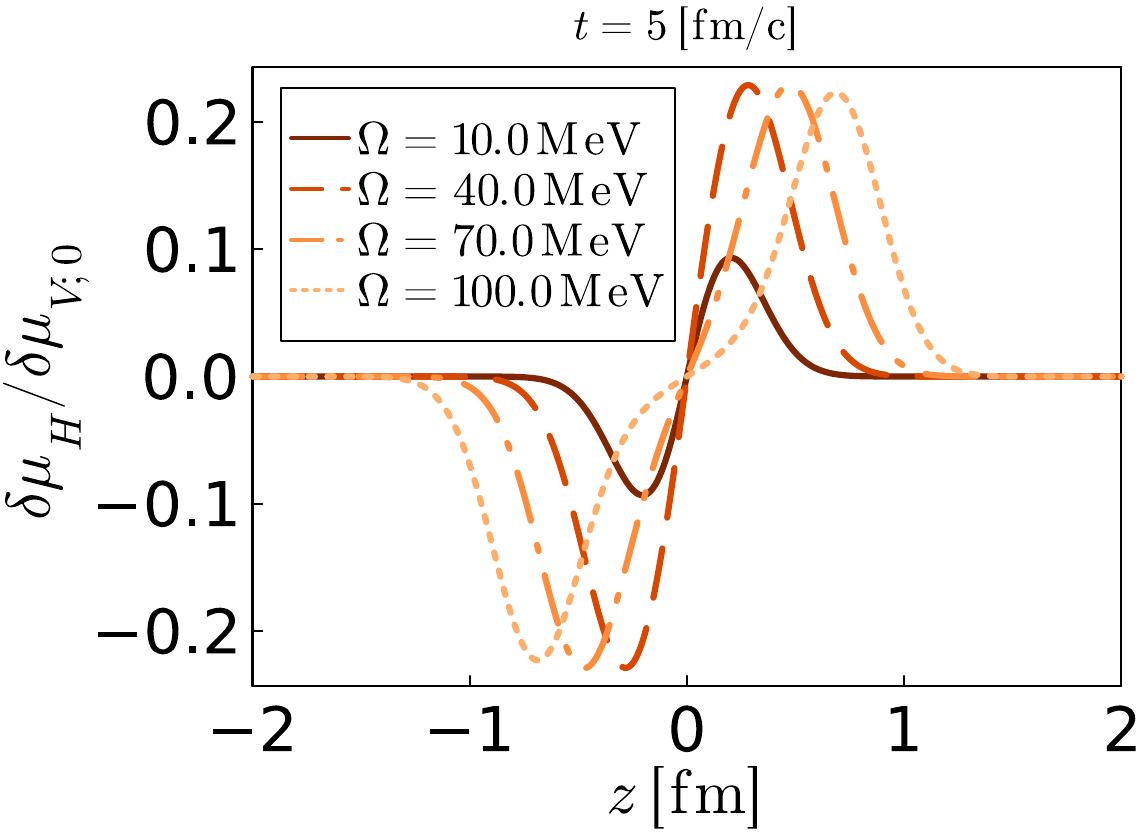}
    \end{tabular}
    \caption{
Perturbations in the chemical potentials at a fixed time $t = 5$ fm/c with finite helical ($\tau_H = 2.7$~fm/c) and axial ($\tau_A = 0.25$~fm/c) relaxation times and vanishing $\tau_R$, as we increase the angular frequency $\Omega$. The system is initialized with a Gaussian perturbation in the vector chemical potential $\delta \mu_V$ of width $\sigma=0.2$~fm on top of a neutral unpolarized background $\mu_V=\mu_A=\mu_H=0$ at temperature $T=300$~MeV. 
}
    \label{fig:QCD_phys}
\end{figure}

We now estimate the effect of damping in the setup of the quark-gluon plasma formed in heavy-ion collisions. As usual, we consider $T = 300$ MeV and $\mu_V = 0$. For definiteness, we consider a QGP with $N_f =3$ flavours, for which the helicity relaxation time evaluates to $\tau_H = 2.7$ fm/c, according to Eq.~\eqref{eq:Hell_relax}. For the axial chemical potential, we take the estimate $\tau_A \simeq 0.25\ {\rm fm}/c$, obtained from first-principle simulations in Ref.~\cite{Astrakhantsev:2019zkr}.

Ignoring for the time being the kinetic dissipation (i.e., $\tau_R \rightarrow 0$), we estimate the propagation properties of the vortical waves discussed in the previous sections at these realistic values of the relaxation times. In the case of an ideal plasma, where the axial and helical relaxation times tend to infinity, the helical vortical wave propagates with the velocity $c_h$ given in Eq.~\eqref{eq:ch}. Thus, over a time period equal to $\tau_H$, the helical vortical wave propagates a distance 
\begin{equation}
 z = c_h \tau_H \simeq 0.38 \left(\frac{\Omega}{100\ {\rm MeV}}\right) \ {\rm fm}.
\end{equation}
The above estimate implies that the wave travels a noticeable distance when $\Omega$ is of the order of $10^2\,{\rm MeV}$. At lower values of $\Omega$, the dissipative effects will lead to the damping of the fluctuations in polarization with no observable propagation. Moreover, the expected lifetime of the QGP fireball is of the order of $10$ fm/c, after which the system undergoes the phase transition towards hadronic degrees of freedom, and our considerations related to vortical effects in a conformal plasma cease to apply. 

To illustrate the effect of the rotation parameter $\Omega$ on the propagation properties of the vortical waves, we considered a system initialized with a Gaussian perturbation in the vector chemical potential $\delta\mu_V$ of width $\sigma=0.2$ fm. Figure~\ref{fig:QCD_phys} shows the vector and helical perturbations at a time $t=5$ fm/c after initialization. We do not show the axial perturbation, since in the neutral plasma, it is not excited. It can be seen that at small angular frequency $\Omega$, the dissipative effects dominate the evolution of the perturbations while increasing $\Omega$ enhances the propagating features of the waves.

\begin{figure}[!ht]
    \centering
    \begin{tabular}{c}
    \includegraphics[width=.89\linewidth]{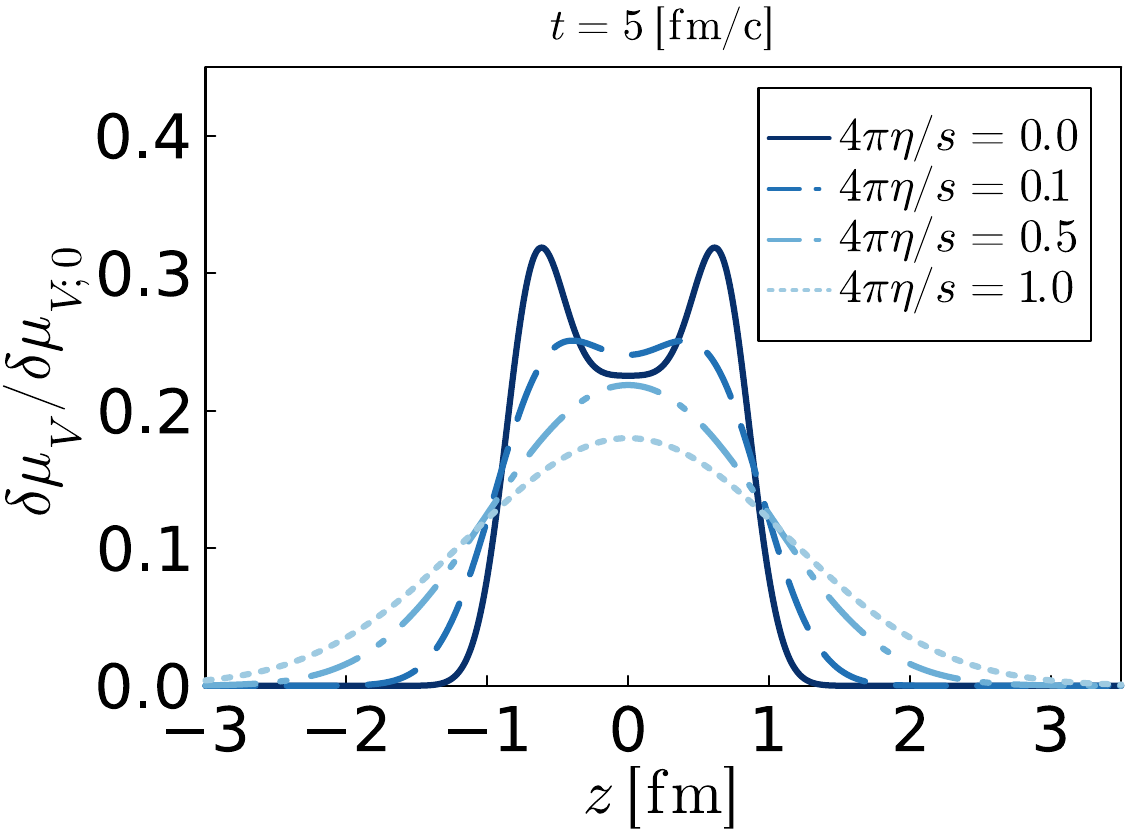}\\
    \includegraphics[width=.89\linewidth]{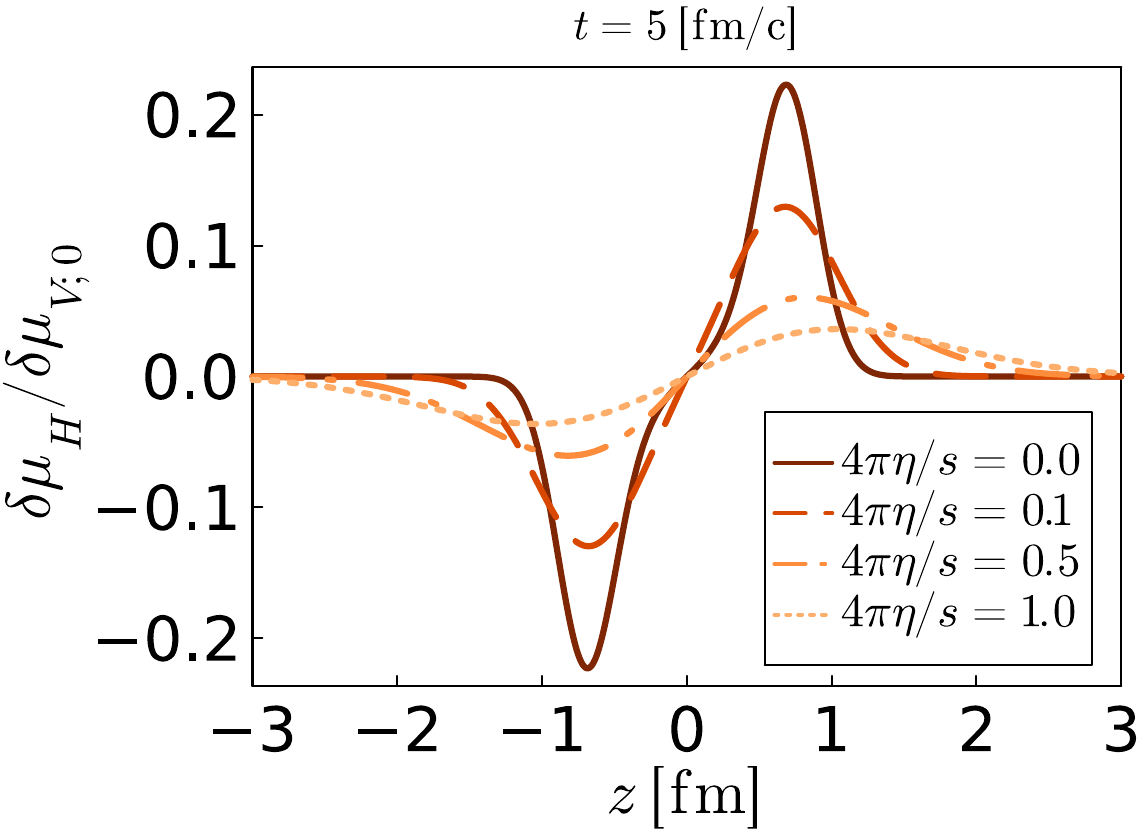}
    \end{tabular}
    \caption{
Perturbations in the chemical potentials at a fixed time $t=5$ fm/c as we increase the damping due to kinetic effects, similar to the one of Fig.~\ref{fig:QCD_phys}. We choose $\sigma=0.2$~fm for the width of initial Gaussian perturbations in all the chemical potentials $\delta \mu_\ell$ on top of a neutral background $\mu_V=0$~MeV at temperature $T=300$~MeV and the angular frequency~$\Omega=100$~MeV for the finite helical ($\tau_H = 2.7$~fm/c) and axial ($\tau_A = 0.25$~fm/c) relaxation times. 
    }
    \label{fig:physical}
\end{figure}

We now estimate the effects of kinetic dissipation, governed by the kinetic relaxation time $\tau_R$. To estimate $\tau_R$, we consider that the ratio between the shear viscosity $\eta = \frac{4}{5} \tau_R P$ and the entropy density $s = (E + P - \vec{\mu} \cdot \vec{Q}) / T$ is constant. According to Ref.~\cite{Kovtun:2004de}, the ratio $\eta / s$ should be bounded from below by $1 / 4\pi$ for strongly-interacting theories. Bayesian estimates based on matching hydrodynamics simulations to experimental data from heavy-ion collisions give $4\pi \eta / s \lesssim 1.5$ around $T = 150$ MeV \cite{Bernhard:2019bmu}, in agreement with this conjecture. Taking $4\pi \eta / s = 1$ for definiteness and considering the choice of parameters discussed above ($T = 300$ MeV, $\mu_V = 0$), we obtain
\begin{equation}
  \tau_R = \frac{5 \eta}{s T} \left(1 - \frac{\mu_\ell Q_\ell}{E + P}\right) \simeq 0.26\ {\rm fm}/c,
\end{equation}
which is very similar to the axial relaxation time $\tau_A$. Contrary to the relaxation times $\tau_A$ and $\tau_H$, which govern the dissipation of long-wavelength excitations, $\tau_R$ becomes important in the ultraviolet limit, i.e. at short wavelengths, thus leading to the smearing of the sharp features of propagating waves. The particular form of the correction to the dispersion relation due to kinetic dissipation, implied by Eq.~\eqref{eq:omega_R}, indicates that the lifetime of a mode with wavenumber $k$ is $\tau_d = 3 / (k^2 \tau_R)$. Considering now the smallest wavenumber that supports the propagation of the helical vortical wave, given in Eq.~\eqref{eq:QED_kth}, we estimate $\tau_d$ as
\begin{equation}
 \tau_d = \frac{12 \tau_H^2 c_h^2}{\tau_R} \simeq 6.64 \left(\frac{\Omega}{100\ {\rm MeV}}\right)^2\, {\rm fm}/c.
\end{equation}
Remarkably, the lifetime due to kinetic dissipation depends quadratically on the angular velocity $\Omega$ and becomes sizeable also around $\Omega \simeq 100\ {\rm MeV}$. For this purpose, we will discuss the effect of varying $\eta / s$ (or equivalently, $\tau_R$) at this particular value of $\Omega$. 

Considering the configuration and parameters shown in Fig.~\ref{fig:QCD_phys}, we now vary $\eta / s$ while keeping $\Omega = 100$ MeV fixed. The case $\tau_R = 0$ is already depicted in Fig.~\ref{fig:QCD_phys}.
Figure~\ref{fig:physical} illustrates the effect of the kinetic dissipation as $\eta/s$ is increased from $0$ to $1/(4\pi)$. 
The figure confirms that kinetic dissipation damps short wavelength perturbations, smoothing the propagating peaks and progressively erasing the propagating features of the waves as the $\eta/s$ ratio increases.

Before ending this section, we come back to our initial considerations and take into account realistic values of $\Omega$. As pointed out in Ref.~\cite{STAR:2017ckg}, the angular velocity is estimated at $\Omega \simeq 6.6\ {\rm MeV}$ for the fireball created in non-central heavy-ion collisions. At this value of $\Omega$, the propagation speed of the helical vortical wave becomes a mere
\begin{equation}
 c_h {\biggl\rvert}_{\Omega = 6.6\ {\rm MeV}} = \frac{6 \Omega \ln 2}{\pi^2 T} \simeq 0.0093\, c.
\end{equation}
Over the lifetime $\tau_{QGP} \simeq 10\ {\rm fm}/c$ of the QGP fireball, the HVW transports perturbations over a tiny distance of $c_h \tau_{QGP} \simeq 0.093\ {\rm fm}$. The threshold wavenumber over which perturbations can propagate is set by the helical relaxation time~\eqref{eq:QED_kth}: $k_{\rm th} = 1/(2 \tau_H c_h) \simeq 20\ {\rm fm}^{-1}$. At this wavenumber, the lifetime due to kinetic dissipation is decreased to $\tau_d \simeq 0.029$ fm/c, which gives an infinitesimal value for the distance over which the wave can propagate: $c_h \tau_d \simeq 2.7 \times 10^{-4}$ fm. Clearly, in such conditions, no signs of propagation due to the vortical effects can be seen: at small wavelengths, restriction of the wave propagation appears as a result of kinetic dissipation, while at large wavelengths, the wave overdamping occurs due to the non-conservation of polarization-related charges.

\section{Summary and Conclusions} \label{sec:conc}

In our paper, we further developed an alternative approach to chiral fermionic systems, put forward in Refs.~\cite{Ambrus:2019ayb,Ambrus:2019khr,Morales-Tejera:2024uzg}, which extends the traditional premise that chiral fluids are fully described by the pair of vector and axial local charges. In the previous works, we have demonstrated that the incorporation of the helical degree of freedom of non-interacting relativistic fermions -- which is a conserved degree of freedom related but not equal to the axial charge in general -- enriches the hydrodynamic spectrum of the system via the interplay of Axial and Helical Vortical Effects, visualized in an intuitively appealing pictorial form in Fig.~1 of Ref.~\cite{Morales-Tejera:2024uzg}.

\paragraph{Helical and Axial Waves.} In Ref.~\cite{Ambrus:2019khr}, we uncovered a new type of gapless mode, the Helical Vortical Wave, which represents a coherent propagating excitation of vector and helical charge densities and their currents. We also found the Axial Vortical Wave~\cite{Morales-Tejera:2024uzg}, which is driven by fluctuations in the axial charge density that propagates in the background of a large axial chemical potential $\mu_A$. Moreover, a background with non-zero $\mu_A$ induces non-reciprocity in the spectrum, namely that waves travel with different velocities along and opposite to the direction of the angular velocity. Contrary to the Chiral Vortical Wave~\cite{Jiang:2015cva, Gorbar:2017toh}, which operates in vector and axial sectors, the Axial and Helical Vortical Waves emerge in an electrically neutral regime characterized by a vanishing vector chemical potential. In the opposite, high-density limit, these waves merge and form a mixed Axial-Helical-Vortical Wave~\cite{Morales-Tejera:2024uzg}. In this work, we questioned whether these hydrodynamic modes could withstand realistic conditions that appear in the interacting vortical quark-gluon plasmas.

\paragraph{Relaxation time approximation.} In our analysis, we used the relaxation time approximation in which interactions can phenomenologically be described via the introduction of the helical ($\tau_H$), axial ($\tau_A$) and kinetic ($\tau_R$) relaxation times. The first two quantities control the time scale of the non-conservation of the corresponding charges, while the third one determines the dissipative evolution of the fluid towards thermal equilibrium. The relaxation time of the vector charge is taken to be infinite as the electric charge is a conserved quantity. This property allowed us to study finite-density systems at non-zero vector chemical potential. Accordingly, in this paper, we do not consider the linear hydrodynamic waves at a finite background axial density since this case is not consistent with a finite axial relaxation time. 

\paragraph{Helical charge relaxation.} We first analyzed the genuine effects of finite helical relaxation time $\tau_H$ in the limit when other relaxation effects are absent, i.e. $\tau_A =\infty$ and $\tau_R=0$ (Section~\ref{sec:QED}). In the dilute plasma regime, where the density of the vector chemical potential is small compared to temperature, $\mu_V \ll T$, and the other chemical potentials are zero, $\mu_A = \mu_H = 0$, we get an expected effect: in addition to the damping due to the finite $\tau_H$, the Helical Vortical Wave evolves to a purely diffusive mode when the half-period of the wave exceeds the helical relaxation time $\tau_H$. In other words, the helical relaxation cuts off the low-frequency ``helical'' modes, corresponding predominantly to the helical-vortical charge fluctuations of the Helical Vortical Wave. The ``axial'' mode, a leftover of the non-reciprocal Axial Vortical Wave branch, is purely diffusive in this regime. 

In the dense plasma, at $\mu_V \gtrsim T$, the magnitude of the helical relaxation time determines the qualitative properties of the hydrodynamic spectrum. For slowly relaxing helical plasma with a large $\tau_H$, the oscillations of helical and axial charge densities -- which are indistinguishable in this limit\footnote{The helical and axial charges of an ensemble of fermions carrying the same vector charge (i.e. either fermions or anti-fermions) are identical up to a sign since the same statement is valid for the helical and axial charges of a single given particle (or anti-particle).} -- and the vector charge oscillations merge into a common hydrodynamic excitation, the Axial-Helical Vortical Wave.

The Chiral Vortical Wave emerges in the vanishing-$\tau_H$ limit where the helical degree of freedom disappears from the spectrum (it freezes). It is worth stressing that the Chiral Vortical Wave and the mixed Axial-Helical Vortical Wave, both existing in the high-density limit at finite $\tau_H$, are different excitations, as is reflected in their different propagation velocities and the chemical content of these waves. 

Notice also that for a degenerate ($\mu_V \gg T$) axially conserved ($\tau_A \to \infty$) chiral fluid, the limit of the fast helical relaxation, $\tau_H \to 0$, is accompanied by the redistribution of the degrees of freedom. At large but finite $\tau_H$ (approximately conserved helicity), the spectrum is represented by two non-reciprocal modes of the Axial-Helical Wave that mixes almost indistinguishable helical-axial oscillations with fluctuations in the vector charge density. However, at small $\tau_H$, these modes transform into a non-propagating helical mode (dissipation of all charges driven by the helical charge relaxation) and two bi-directionally propagating axial-vector modes (the Chiral Vortical Wave).

At a finite (non-zero and non-infinite) helical relaxation time, the interplay of the axial and helical modes in dense fermionic matter becomes rather non-trivial, as summarized in Fig.~\ref{fig:k*}. For the chemical potential below a certain limiting value, the wave spectrum contains three regions: (i) two propagating waves with a small wavenumber (large wavelength) are the generalization of the Chiral Vortical Wave (axial-vector oscillations) that includes an accompanying helical mode; (ii) two propagating waves with a large wavenumber which is essentially a generalization of the Helical Vortical Wave (helical-vector oscillations) to the finite-density case; (iii) an intermediate region with a moderate wavelength in which both mentioned modes are diffusive. Thus, we have two branches of hydrodynamic excitations. (iii) When the vector chemical potential exceeds the mentioned critical value, the diffusive region disappears, and the (i) CVW and (ii) HVW domains merge, constituting a single branch of solutions. In all mentioned regions, these two helical modes are accompanied by a single diffusive axial non-propagating mode, a remainder of the Axial Vortical Wave. 

\paragraph{Simultaneous helical and axial charge relaxations.} The accounting for a finite axial relaxation time $\mu_A$ makes the whole picture more realistic and, somewhat expectedly, more intricate. While we refer the interested reader to Section~5 for more details, here we describe the main features of the hydrodynamic waves with finite helical and axial relaxation times. 

For a non-conserved axial charge, there are no propagating modes for small wavenumbers (large wavelengths). This property contrasts with the picture at the conserved axial charge (and non-conserved helical charge), where two modes propagate for all wavelengths at high vector chemical potential. At the same time, the relaxation time of the axial and/or helical charges does not affect qualitatively the wave spectrum in the large wavenumber limit compared to the previously studied case of the conserved axial charge.

If the helical charge dissipates rapidly (faster than the axial charge), then a single non-propagating helical mode dissipates rather quickly, while the two axial modes propagate at small wavevectors and are purely diffusive otherwise. In the case of conserved helicity, $\tau_H \to \infty$ (and at finite $\tau_A$), two non-dissipative helical modes co-exist along with the purely dissipative axial mode, for all wavelengths.

The ratio of the relaxation times, $\tau_H/\tau_A$, determines whether we have two branches of thermodynamic excitations (large-wavenumber Helical and low-wavenumber Chiral Vortical Waves separated by a non-propagating domain of wavelengths) or a single branch that merges these excitations. The corresponding diagram is shown in Fig.~8.

\paragraph{Kinetic dissipation.} 
We also considered the effects of kinetic dissipation produced by the corrections to the energy-momentum tensor and charge currents due to interactions in the fluid. In our work, these corrections are also taken into account in the relaxation time approximation governed by the kinetic relaxation time $\tau_R$ within the relativistically consistent framework of M\"uller-Israel-Stewart hydrodynamics. The presence of the kinetic dissipation leads to the damping of the high-wavenumber modes. This property, together with the fact that for a non-conserved axial charge, there are no propagating modes for small wavenumbers, poses a stringent restriction to the existence of the vortical waves in realistic plasmas. 

\paragraph{Consequences for quark-gluon plasma.} 
For the environment of the quark-gluon plasma produced in the current experiments in relativistic heavy-ion collisions, the estimations made in Section 5 do not favour the observation of the Helical Vortical Waves. It appears that the angular velocity of the plasma is too small, so the wave propagates too slowly (about 1\% of the speed of light), implying that during the lifetime of the plasma fireball, the wave advances over a phenomenologically irrelevant distance of a small fraction of a fermi. However, it appears that even this estimation is too optimistic because of the damping effects. The helical relaxation time sets an upper threshold of the wavelength of the propagating wave because when the wavelengths are too large, wave propagation is prohibited as a result of the non-conservation of the helical charge. However, even this threshold wavelength appears to be too small as the kinetic dissipation effects -- that increase quadratically with the decrease of the wavelength -- appear to be too strong so that they damp the wave at small wavelengths. Therefore, we conclude that helical vortical waves cannot propagate in the realistic environment of quark-gluon plasmas.

It is worth mentioning that we do not consider the effect of finite electric conductivity that promotes a diffusive electromagnetic backreaction of the oscillating electric charge density generated by the waves. 
This electromagnetic effect can be relevant for all hydrodynamic excitations that involve oscillations of the vector chemical potential. An analysis of the electromagnetic backreaction has been performed for the Chiral Magnetic Wave in the background magnetic field in Ref.~\cite{Shovkovy:2018tks} and for the longitudinal sound wave, a circularly polarized vortical wave and diffusive modes, among others, in a vortical background in Ref.~\cite{Rybalka:2018uzh}. It appears that these waves are overdamped in quark-gluon plasma, which aligns well with our conclusions made for the helical and axial hydrodynamic excitations. Moreover, a finite electric conductivity will further limit the propagation of helical and axial waves as they also contain a vector component. Therefore, our theoretical analysis gives an upper bound on the dissolution time of these modes.

{\bf Acknowledgments.}
We thank P. Aasha for fruitful discussions. We also thank an anonymous referee for helping us improve the connections to existing literature.
This work was funded by the EU’s NextGenerationEU instrument through the National Recovery and Resilience Plan of Romania - Pillar III-C9-I8, managed by the Ministry of Research, Innovation and Digitization, within the project entitled ``Facets of Rotating Quark-Gluon Plasma'' (FORQ), contract no. 760079/23.05.2023 code CF 103/15.11.2022.

\appendix

\section{Dissipation basis: charge vs. chemical potential}\label{apa}




In section \ref{sec:QED}, we discussed the collective excitations of the unpolarized plasma when the non-conservation of helical charge is included through $\dot{Q}_H\propto-\mu_H/\tau_H$. In this Appendix, we review the results for the inclusion of dissipation as $\dot{Q}_H\propto-Q_H/\tau_H$ and show that it unavoidably gives rise to instability.

In order to see that the two approaches are inequivalent, let us consider the variation of $Q_H$ with respect to small perturbations in $T$ and $\mu_{V/A/H}$. For simplicity, we restrict ourselves to the large temperature expression for $Q_H$. From Eq.~\eqref{eq:Charges_Conds} with $\ell=H$ (see also Eq.~(64b) in Ref.~\cite{Morales-Tejera:2024uzg}), we find that 
\begin{multline}
    Q_H = \frac{\mu_H T^2}{3} + \dfrac{4 T \ln 2 }{\pi^2}\mu_A \mu_V \\
    + \dfrac{\mu_H(3 \mu_V^2 + 3\mu_A^2+\mu_H^2)}{3 \pi^2} + O(T^{-1})
\end{multline}
The fluctuations of the helicity charge $\delta Q_H$ receive non-trivial contributions from fluctuations in both the helical and the axial charges, even in the case of an unpolarized (but not necessarily neutral, $\mu_V \neq 0$) background,
\begin{multline}
 \delta Q_H{\biggl\rvert}_{\mu_A = \mu_H = 0} = \left(\frac{T^2}{3} + \frac{\mu_V^2 }{\pi^2}\right) \delta \mu_H\\ 
 + \frac{4 T \ln 2}{\pi^2} \mu_V \delta \mu_A  + O(T^{-1}).
\end{multline}
Therefore, in the prescription for relaxation given by $\partial_\mu J^\mu_H = -Q_H / \tau_H$ there is an extra term coming from the crossed susceptibility $\chi_{HA} = \partial Q_H/\partial \mu_A$ which is responsible for a relaxation towards a state with $Q_H = 0$, which does not necessarily enforce $\mu_H = 0$. On the contrary, the prescription used in the main text $\partial_\mu J^\mu_H = -T^2 \mu_H/ 3\tau_H$ does lead to a relaxation towards $\mu_H=0$. We proceed now to discuss the consequences of introducing dissipation on the basis of charge.

The (non)-conservation equations for the charge sector become
\begin{equation}\label{eq:diss_QH}
    \partial_{\mu}J^{\mu}_V =0\,,\hspace{0.3cm} \partial_{\mu}J^{\mu}_A =0\,,\hspace{0.3cm} \partial_{\mu}J^{\mu}_H =-\dfrac{Q_H}{\tau_H}\,.
\end{equation}

The matrix $\mathbb{M}$ is now slightly different from that in Eq.~\eqref{eq:M_aux2}. In particular, taking the $\tau_A \to \infty$ limit, we obtain:
\begin{multline}
    \frac{1}{T^2} \mathbb{M}_{\ell\ell'} = \omega \mathbb{M}^\omega_{\ell\ell'} - \kappa_\Omega \mathbb{M}^\Omega_{\ell\ell'}  + \dfrac{2 i \sigma_A^\omega}{\tau_H T^2} \delta_{\ell H} \delta_{\ell'H} \\
    + \dfrac{2i \sigma_H^\omega}{\tau_H T^2} (\delta_{\ell H}\delta_{\ell'A} ),
\end{multline}
where the last term is due to the cross-susceptibility $\chi_{HA}$.
The determinant of $\mathbb{M}$ can be evaluated as follows:
\begin{multline}
 {\rm det} \left(\frac{1}{T^2} \mathbb{M}\right) = {\rm det}\left(\omega \mathbb{M}_\omega - \kappa_\Omega \mathbb{M}_\Omega+ \dfrac{2 i \sigma_A^\omega}{\tau_H T^2} \mathbb{I}_H \right) \\
 - \frac{2i\sigma_H^\omega}{\tau_H T^2} 
 \left\lvert\begin{array}{cc}
    \frac{2\omega}{T^2} \left(\sigma_A^\omega - \frac{T^2}{3} \Delta H\right)  &  -\frac{\kappa_\Omega}{H} B \\
    -\kappa_\Omega A & \frac{2\omega}{T^2} \sigma_H^\omega
 \end{array}\right\rvert,
\end{multline}
where the first term can be obtained by substituting $\tau_H \rightarrow \tau_H \times T^2 / (6 \sigma^\omega_A)$ in Eq.~\eqref{eq:QED_detM}.

\begin{figure}[!ht]
 \centering
 \includegraphics[width=0.89\linewidth]{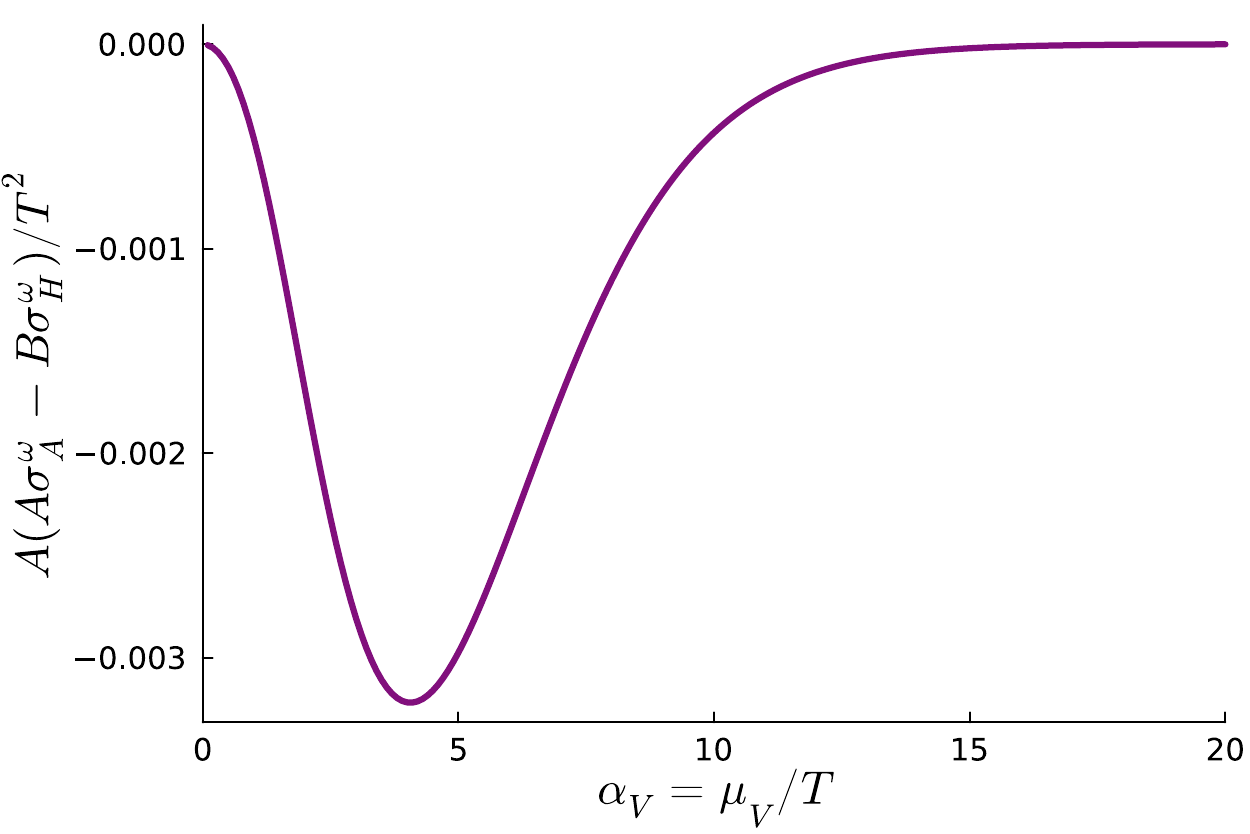}
 \caption{ Explicit demonstration that the left-hand side of Eq.~\eqref{eq:inst} is negative and therefore there is always one unstable mode at small wavenumber when charge dissipation is implemented on the basis of charge fluctuations (see Appendix~\ref{apa}).
 \label{fig:inst}}
\end{figure}

Accordingly, the master equation ${\rm det}(\mathbb{M}) = 0$, shown in Eq.~\eqref{eq:QCD_detM}, gets modified to 
\begin{multline}
 \frac{2\omega}{T^2} \left\{\left(\frac{2 \omega}{T^2}\right)^2 \left(\sigma^\omega_A - \frac{T^2}{3} \Delta H \right) [(\sigma^\omega_A)^2 - (\sigma^\omega_H)^2] \right. \\ \left. -
 \frac{\kappa_\Omega^2}{H} [(A^2 + B^2) \sigma_A^\omega - 2 A B \sigma^\omega_H]\right\}\\
 + \frac{i}{3\tau_H } \left[\left(\frac{2\omega}{T^2}\right)^2 \dfrac{6}{T^2}\left(\sigma_A^\omega - \frac{T^2}{3} \Delta H\right)[(\sigma_A^\omega)^2-(\sigma_H^\omega)^2 ]\right. \\ - \left.  \frac{\kappa_\Omega^2 A}{H}\dfrac{6}{T^2}(A\sigma_A^\omega-B\sigma_H^\omega)\right]=0\,.
 \label{eq:QEDII_detM}
\end{multline}
We can solve the previous equation perturbatively in $\kappa_\Omega$, considering the following expansion for $\omega$:
\begin{equation}
    \omega = \omega_0 + \omega_1 \kappa_\Omega+ \dots
\end{equation}
The master equation to zeroth order in $\kappa_\Omega$ reads:
\begin{multline}
 \left(\frac{2 \omega_0}{T^2}\right)^2 \left(\sigma^\omega_A - \frac{T^2}{3} \Delta H \right) [(\sigma^\omega_A)^2 - (\sigma^\omega_H)^2]\\ 
 \times \left(\frac{2 \omega_0}{T^2}+\dfrac{2i}{\tau_H T^2}\right)
 =0\,.
\end{multline}
There is a non-trivial and purely dissipating solution $\omega_0 = -i/\tau_H$ and a double trivial root $\omega_0=0$. We now compute the corrections to the trivial root from the next-to-leading order for the master equation:
\begin{align}
    &\dfrac{2i}{\tau_H T^2} \left[\left(\frac{2 \omega_1}{T^2}\right)^2 \left(\sigma_A^\omega - \frac{T^2}{3} \Delta H\right)[(\sigma_A^\omega)^2-(\sigma_H^\omega)^2 ] \right.\nonumber\\&\left. - \frac{ A}{H}(A\sigma_A^\omega-B\sigma_H^\omega)\right]\kappa_\Omega^2=0
\end{align}
Therefore, we find
\begin{equation}
    \omega_1^{\pm} = \pm\dfrac{T^2}{2}\sqrt{\dfrac{A (A\sigma_A^\omega-B\sigma_H^\omega)}{H\left(\sigma_A^\omega - \frac{T^2}{3} \Delta H\right)[(\sigma_A^\omega)^2-(\sigma_H^\omega)^2 ]}}\,.
\end{equation}
The quantity $A(A\sigma_A^\omega-B \sigma_H^\omega)$ is negative semi-definite, whereas the remaining factor inside the square root is positive semi-definite. Specifically, from the definitions in Sec.~\ref{sec:setup:unpol}, we find $\sigma_A^\omega>\sigma_H^\omega$ and 
\begin{align}
  H = \frac{15 \alpha_V^4 + 30\pi^2 \alpha_V^2+7\pi^4}{\pi^2(15 \alpha_V^2+7\pi^2)} &> 0, \nonumber \\  
   \frac{1}{T^2} \left(\sigma_A^\omega - \frac{T^2}{3} \Delta H\right) =\frac{15 \alpha_V^4+6 \alpha_V^2\pi^2+7\pi^4}{6\pi^2 (15 \alpha_V^2+7\pi^2)} &> 0,
\end{align}
where we have written $\alpha_V=\mu_V/T\,$. The behaviour of $A(A\sigma_A^\omega-B \sigma_H^\omega)$ at small chemical potential is given by
\begin{equation}
    \dfrac{1}{T^2}A(A\sigma_A^\omega-B \sigma_H^\omega)\simeq  -0.000432 \alpha_V^2  <0.
    \label{eq:inst}
\end{equation}
As a result, $\omega_1^{\pm}$ come in complex conjugate pairs, and there is one unstable mode at finite vector chemical potential. While the previous result is derived for small chemical potential, it can be verified (see Fig. \ref{fig:inst}) that $A(A\sigma_A^\omega-B \sigma_H^\omega)$ is indeed negative definite for any non-zero chemical potential, and thus, the instability always appears for small wavenumber.

\section{Helicity relaxation time at finite chemical potential}\label{app:tauH}

In this section of the appendix, we extend the calculation of the helicity relaxation time $\tau_H$ presented in Ref.~\cite{Ambrus:2019khr} to the case of an unpolarized, charged plasma, i.e. at finite vector chemical potential $\mu_V = \alpha_V T$. We start from Eqs.~(101), (102) and (104) of Ref.~\cite{Ambrus:2019khr}:
\begin{multline}
 \frac{d Q_H}{dt} = g \sum_{\lambda,\sigma} 2\sigma \lambda \int dP\, C_{\rm HVPA}[f],\\
 C_{\rm HVPA}[f] = \int dK dP' dK' \delta^4(p + k - p' - k') s (2\pi)^6 \\ 
 \times [f^\sigma_{\bp', -\lambda} f^{-\sigma}_{\bk',\lambda} \tilde{f}^\sigma_{\bp,\lambda} \tilde{f}^{-\sigma}_{\bk, -\lambda} - 
 f^\sigma_{\bp, \lambda} f^{-\sigma}_{\bk,-\lambda} \tilde{f}^\sigma_{\bp',-\lambda} \tilde{f}^{-\sigma}_{\bk', \lambda}] \\\times 
 \frac{g \alpha^2_{\rm QCD}}{18 E_{\rm cm}^2} (1 - \cos \theta_{\rm cm}),
 \label{eq:dQH_dt_gen}
\end{multline}
with $\tilde{f}^\sigma_{\bp,\lambda} = 1 - f^\sigma_{\bp,\lambda}$
and $s = E^2_{\rm cm} = (p + k)^2 = (p' + k')^2$ being the center-of-mass energy, while $\theta_{\rm cm}$ is the angle between $\bp'$ and $\bp$, measured in the center of mass frame, satisfying
\begin{equation}
 1 - \cos\theta_{\rm cm} = \frac{4 p \cdot p'}{E_{\rm cm}^2}.
 \label{eq:costhcm}
\end{equation}

We now consider that the plasma is near thermal equilibrium at finite $T$ and $\mu_V$. A small helicity imbalance can be modelled via the Fermi-Dirac distribution,
\begin{equation}
 f^{\sigma}_{\bp,\lambda} \simeq f^{{\rm eq};\sigma}_{\bp,\lambda} \simeq f^\sigma_{0\bp} + 2\lambda \sigma \beta \mu_H f^\sigma_{0\bp} \tilde{f}^\sigma_{0\bp},
\end{equation}
where $f^\sigma_{0\bp} = [e^{\beta E_\bp - \sigma \alpha_V} + 1]^{-1}$ is the equilibrium distribution for a charged, unpolarized fluid, thus extending Eq.~(105) of Ref.~\cite{Ambrus:2019khr} for the neutral fluid to the case of a charged fluid. 
Note that in Eq. (107) of Ref.~\cite{Ambrus:2019khr}, Eq.~\eqref{eq:dQH_dt_gen} was replaced by $dQ_H / dt = -Q_H /\tau_H$, which is valid under the assumption of a neutral background with a small helicity imbalance, i.e. $\mu_V = \mu_A = 0$ and $\lvert \mu_H \rvert \ll T$, when $Q_H \simeq \mu_H T^2 / 3$. In this work, we consider the case when $\mu_V$ is arbitrary, such that instead Eq.~\eqref{eq:dQH_dt_gen} becomes 
\begin{equation}
 \frac{dQ_H}{dt} = -\frac{\mu_H T^2}{3\tau_H},
\end{equation}
with
\begin{multline}
 \tau_H^{-1} = \frac{8}{3} (2\pi)^6 g \alpha_{\rm QCD}^2 \beta^3 \\ \int dP dK dP' dK' 
 (1 - \cos \theta_{\rm cm})^2 \delta^4(p + k - p' - k') \\\times
 \frac{1}{4} \sum_{\sigma = \pm 1} f^\sigma_{0\bp} f^{-\sigma}_{0\bk} \tilde{f}^\sigma_{0\bp'} \tilde{f}^{-\sigma}_{0\bk'} (\tilde{f}^\sigma_{0\bp} + \tilde{f}^{-\sigma}_{0\bk} + f^{\sigma}_{0\bp'} + f^{-\sigma}_{0\bk'}).
\end{multline}
Noting that, under the conservation of total four-momentum, $f^\sigma_{0\bp} f^{-\sigma}_{0\bk} \tilde{f}^\sigma_{0\bp'} \tilde{f}^{-\sigma}_{0\bk'} (f^{\sigma}_{0\bp'} + f^{-\sigma}_{0\bk'}) = f^\sigma_{0\bp} f^{-\sigma}_{0\bk} \tilde{f}^\sigma_{0\bp'} \tilde{f}^{-\sigma}_{0\bk'} (f^{\sigma}_{0\bp} + f^{-\sigma}_{0\bk})$, the expression between the parentheses above evaluates to 
\begin{equation}
 \tilde{f}^\sigma_{0\bp} + \tilde{f}^{-\sigma}_{0\bk} + f^{\sigma}_{0\bp'} + f^{-\sigma}_{0\bk'} \rightarrow \tilde{f}^\sigma_{0\bp} + \tilde{f}^{-\sigma}_{0\bk} + f^{\sigma}_{0\bp} + f^{-\sigma}_{0\bk} = 2,
\end{equation}
where we used the property $\tilde{f}^\sigma_{0\bp} = 1 - f^\sigma_{0\bp}$. Writing now as in Eq. (159) of Ref.~\cite{Ambrus:2019khr},
\begin{equation}
 \tau_H^{-1} = \frac{8}{3} (2\pi)^6 g \alpha_{\rm QCD}^2 \beta^3 \int dP dK 
 \frac{1}{2} \sum_\sigma f^\sigma_{0\bp} f^{-\sigma}_{0\bk} I^\sigma_\Omega,
\end{equation}
with
\begin{multline}
 I^\sigma_\Omega = \int dP' dK' (1 - \cos\theta_{\rm cm})^2 \delta^4(p + k - p' - k') \\\times 
 \tilde{f}^\sigma_{0\bp'} \tilde{f}^{-\sigma}_{0\bk'},
\end{multline}
the $k'$ and $p'$ outgoing momenta can be boosted in the center of mass frame, where the delta function reads $\delta(E_{\rm cm} - p'_0 - k'_0) \delta^3(-\bp' - \bk')$. The $dK'$ integral can be performed automatically, leading to
\begin{align}
 &I^\sigma_\Omega = \frac{1}{(2\pi)^6} \int_0^\infty dp'\nonumber \\ & \int d\Omega_{\bp'} \delta(E_{\rm cm} - 2p'_0) (1 - \cos\theta_{\rm cm})^2 \tilde{f}^{\sigma}_{0 \bp'} \tilde{f}^{-\sigma}_{0\bk'}
\end{align}
The integration with respect to $p'$ yields a factor of $1/2$ (see Ref.~\cite{Helicity:2024err}), leading to:
\begin{equation}
 I_\Omega^\sigma = \frac{1}{2(2\pi)^5} \int dx \tilde{f}^{\sigma}_{0 \bp'} \tilde{f}^{-\sigma}_{0\bk'} \int_0^{2\pi} \frac{d\varphi_{\bp'}}{2\pi} (1 - \cos\theta_{\rm cm})^2.
\end{equation}
In the above, $x = \cos \theta_{\bp'}$ represents the angle between $\bp'$ and $\bp + \bk$, while $\varphi_{\bp'}$ measures an angle in the plane perpendicular to $\bp + \bk$. 

\begin{figure}[!ht]
 \centering
 \includegraphics[width=0.9\linewidth]{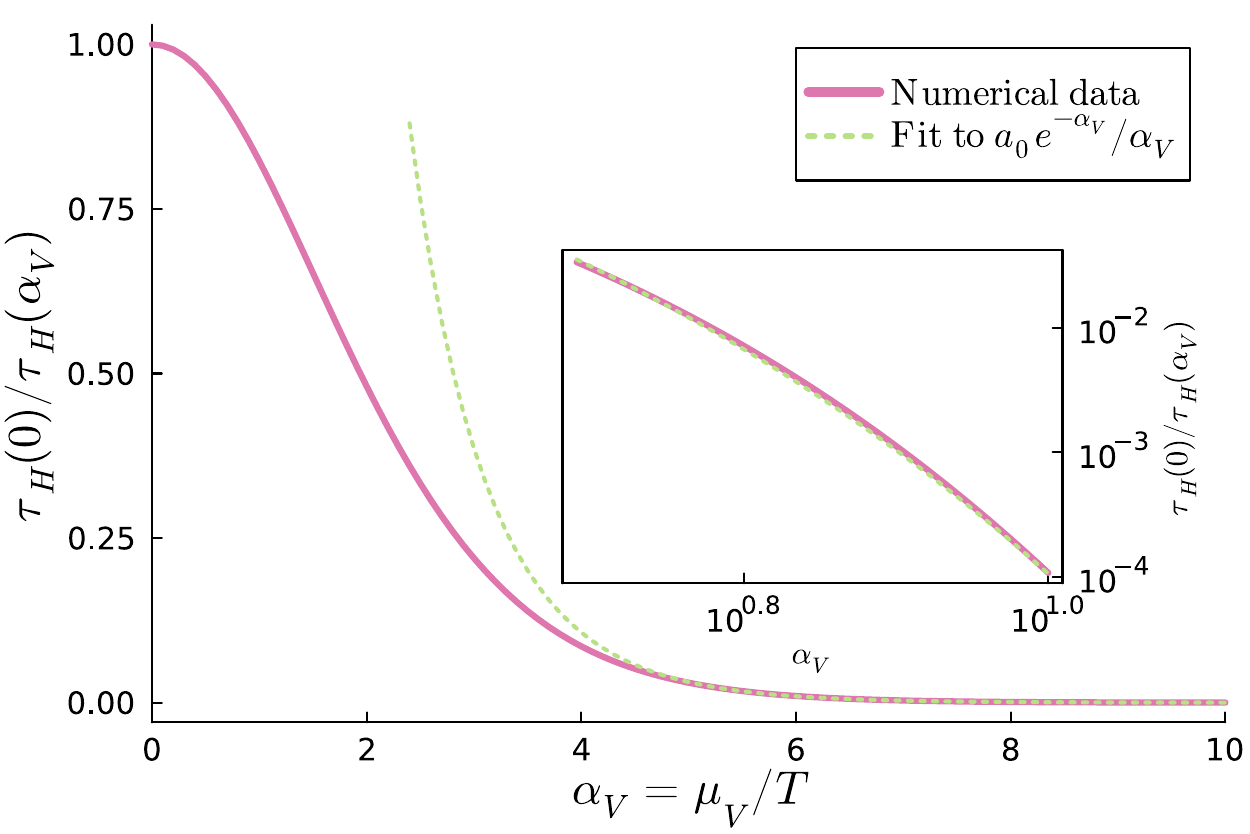}
 \caption{Ratio $\tau_H(0)/\tau_H(\alpha_V)$ between the helicity relaxation time in a neutral plasma and at dimensionless vector chemical potential $\alpha_V = \mu_V / T$, in an unpolarized plasma. The inset plot shows the agreement between the numerical integration and the fitted function in a log-log scale. 
 \label{fig:tauH}} 
\end{figure}

The integral with respect to $\varphi_{\bp'}$ can be performed using Eq.~\eqref{eq:costhcm} to replace $1 - \cos\theta_{\rm cm}$.
After boosting $p'$ to the center of mass frame, $p^\mu$ is replaced by $p^\mu_{\rm cm} = (E_{\rm cm} / 2, \bp_{\rm cm})$, with 
\begin{equation}
 \bp_{\rm cm} = \bp - \frac{p^0 + E_{\rm cm} / 2}{p^0 + k^0 + E_{\rm cm}} (\bp + \bk).
\end{equation}
Splitting $\bp_{\rm cm} = \bp_{\lvert \rvert} + \bp_\perp$, where $\bp_{\vert \rvert}$ is oriented along $\bk + \bp$ and $\bp_{\perp}$ is perpendicular to this vector, we have 
\begin{equation}
 p_{\lvert\rvert} = \frac{\bp_{\rm cm} \cdot (\bp + \bk)}{\lvert\bp + \bk \rvert} = \frac{E_{\rm cm} (p^0 - k^0)}{2\lvert \bp + \bk \rvert},
\end{equation}
same as Eq.~(168) in Ref.~\cite{Helicity:2024err}. The calculation proceeds further exactly as shown in Ref.~\cite{Ambrus:2019khr}, leading to 
\begin{equation}
 \tau_H = \frac{6\pi^3 \beta}{g \alpha_{\rm QCD}^2 \mathcal{I}},
\end{equation}
with 
\begin{multline}
 \mathcal{I} = \int_0^\infty dz\, z^3 \int_{-1}^1 dx \int_{-1}^1 d\delta \int_{\lvert \delta \rvert}^1 d\xi \\\times \left[\frac{3 - x^2}{2} \xi -2x \delta + \frac{3x^2 - 1}{2\xi} \delta^2\right] \\\times 
 \frac{1}{2} \sum_{\sigma = \pm 1} [e^{\frac{z}{2}(1 + \delta) - \sigma \alpha_V} + 1]^{-1} [e^{\frac{z}{2}(1 - \delta) + \sigma \alpha_V} + 1]^{-1}\\ \times[e^{-\frac{z}{2}(1 + x\xi) + \sigma \alpha_V} + 1]^{-1}   [e^{-\frac{z}{2}(1 - x\xi) - \sigma \alpha_V} + 1]^{-1}.
\end{multline}
At vanishing chemical potential, $\mathcal{I} = 5.09434$ \cite{Helicity:2024err} and the helicity relaxation time in the neutral plasma reads 
\begin{align}
 \tau_H &= 0.392 \times \frac{\pi^3 \beta}{N_f \alpha_{\rm QCD}^2} \nonumber \\ &\simeq 
 \left(\frac{250\ {\rm MeV}}{k_B T}\right) \left(\frac{1}{\alpha_{\rm QCD}}\right)^2 \left(\frac{2}{N_f}\right) \times 4.80\ {\rm fm}/c.
\end{align}
As the vector chemical potential is increased, the relaxation time also increases, as shown in Fig.~\ref{fig:tauH}. The large chemical potential behaviour of the relaxation time can be well fitted by the following function:
\begin{equation}
    \tau_H^{-1}(\alpha_V\gg 1) = \tau^{-1}_{\rm deg.} \frac{e^{-\lvert \alpha_V \rvert}}{\lvert \alpha_V \rvert},
\end{equation}
where the best fit was found to be given by $\tau^{-1}_{\rm deg.} = 118.6\ c / {\rm fm}$, or $\tau_{\rm deg.} \simeq 8.432 \times 10^{-3}\,{\rm fm}/c$.

\bibliography{hvw}

\end{document}